\documentclass[11pt,twoside]{article}
\usepackage{atmp}
\usepackage{amsmath,amsfonts,amssymb,amsthm,amstext,eucal}
\usepackage{array}
\usepackage[latin1]{inputenc}

\copyrightnotice{2002}{6}{703}{793}
\newcommand{\half}{\tfrac{1}{2}}
\renewcommand{\d}{\partial}
\newcommand{\fg}{\mathfrak{g}}
\newcommand{\fh}{\mathfrak{h}}
\newcommand{\fk}{\mathfrak{k}}
\newcommand{\fp}{\mathfrak{p}}
\newcommand{\fso}{\mathfrak{so}}
\newcommand{\fsp}{\mathfrak{sp}}

\newcommand{\fsu}{\mathfrak{su}}

\newcommand{\ISO}{\mathrm{ISO}}
\newcommand{\SO}{\mathrm{SO}}
\newcommand{\Cl}{\mathrm{C}\ell}
\newcommand{\Spin}{\mathrm{Spin}}
\def\Sp{\mathrm{Sp}}
\newcommand{\SL}{\mathrm{SL}}
\newcommand{\SU}{\mathrm{SU}}

\newcommand{\EE}{\mathbb{E}}
\newcommand{\RR}{\mathbb{R}}
\newcommand{\PP}{\mathbb{P}}
\newcommand{\CC}{\mathbb{C}}
\newcommand{\HH}{\mathbb{H}}
\newcommand{\ZZ}{\mathbb{Z}}

\newcommand{\eD}{\mathcal{D}}
\newcommand{\eH}{\mathcal{H}}
\newcommand{\eL}{\mathcal{L}}
\newcommand{\eM}{\mathcal{M}}
\newcommand{\eT}{\mathcal{T}}
\newcommand{\eV}{\mathcal{V}}
\DeclareMathOperator{\AdS}{AdS}
\DeclareMathOperator{\dvol}{dvol}
\DeclareMathOperator{\Mat}{Mat}
\newcommand{\bx}{\boldsymbol{x}}
\newcommand{\by}{\boldsymbol{y}}
\newcommand{\btheta}{\boldsymbol{\theta}}
\newcommand{\bC}{\boldsymbol{C}}
\newcommand{\eP}{\mathcal{P}}
\newcommand{\eQ}{\mathcal{Q}}
\newcommand{\eK}{\mathcal{K}}

\newcommand{\1}{\mathbb{1}}
\allowdisplaybreaks[1]
\numberwithin{equation}{section}
\begin{document}
\title{Supersymmetric Kaluza--Klein reductions of M2 and M5-branes}

\url{hep-th/0208107}

\author{José Figueroa-O'Farrill}
\address{School of Mathematics, The University of Edinburgh, Scotland,
  United Kingdom}
\addressemail{j.m.figueroa@ed.ac.uk}

\author{Joan Simón}
\address{The Weizmann Institute of Physical Sciences, Department of
  Particle Physics, Rehovot, Israel}
\addressemail{jsimon@weizmann.ac.il}

\pagestyle{myheadings}
\markboth
{SUPERSYMMETRIC KALUZA-KLEIN REDUCTIONS}
{FIGUEROA-O'FARRILL and SIM\'ON}

\begin{abstract}
  We classify and construct all the smooth Kaluza--Klein reductions to
  ten dimensions of the M2- and M5-brane configurations which preserve
  some of the supersymmetry.  In this way we obtain a wealth of new
  supersymmetric IIA backgrounds describing composite configurations
  of D-branes, NS-branes and flux/nullbranes; bound states of
  D2-branes and strings, D4-branes and NS5-branes, as well as some
  novel configurations in which the quotient involves
  nowhere-vanishing transverse rotations to the brane twisted by a
  timelike or lightlike translation.  From these results there also
  follow novel M-theory backgrounds locally isometric to the M-branes,
  some of which are time-dependent and all of which are asymptotic to
  discrete quotients of eleven-dimensional Minkowski spacetime.  We
  emphasise the universality of the formalism by briefly discussing
  analogous analyses in type IIA/IIB dual to the ones mentioned above.
  Some comments on the dual gauge theory description of some of our
  configurations are also included.
\end{abstract}
\cutpage

\section{Introduction and conclusions}
\label{sec:introconc}

New sectors in string theory have emerged by the embedding of the
Melvin universe \cite{Melvin} into string theory
\cite{GibbonsWiltshire, GibbonsMaeda, DGGH1, DGGH2, DGGH3, CGS, CG,
  GSflux, Saffin, CHC, EmparanFlux, BrecherSaffin, Uranga, EmpGut}.
These are the so-called fluxbranes.  Their supergravity description is
in terms of the Kaluza--Klein reduction of eleven-dimensional
Minkowski spacetime along the orbits of suitable one-parameter
subgroups of the Poincaré group.

In our previous work \cite{FigSimFlat} we investigated and classified
all smooth Kaluza--Klein reductions of eleven-dimensional Minkowski
spacetime by one-parameter subgroups of the Poincaré group, paying
special attention to those preserving some amount of supersymmetry.
These give rise to a wealth of supersymmetric IIA backgrounds
including fluxbranes, the then novel nullbranes and combinations
thereof.  This analysis teaches us that besides fluxbrane sectors,
there are new supersymmetric sectors in string theory associated with
nullbranes.  We will summarise the results of \cite{FigSimFlat} in
Section~\ref{sec:flat}, but let us comment here briefly on them.  The
constraint that the IIA background be smooth imposes constraints on
the one-parameter subgroup $\Gamma$ by which we reduce.  In
particular, it follows that $\Gamma$ cannot be compact; in other
words, it is not a circle subgroup as in the usual Kaluza--Klein
reduction but is diffeomorphic to the real line.  Since Kaluza--Klein
reduction is usually phrased in terms of circle subgroups, one can
still do that here provided one performs the reduction by $\Gamma$ in
two steps.

It will be convenient to generalise the discussion and consider not
just the reduction of Minkowski spacetime but that of an arbitrary
M-theory background $\eM = (M,g,F_4)$ by a noncompact one-parameter
Lie subgroup $\Gamma$ of the symmetry group of $\eM$.  The first step
consists in performing a discrete quotient to obtain another M-theory
background.  To this effect one first chooses a co-compact discrete
subgroup $\Gamma_0 \subset \Gamma$ (that is, a subgroup such that the
quotient group $\Gamma/\Gamma_0$ is compact, e.g., if $\Gamma \cong
\RR$ then $\Gamma_0 \cong \ZZ$) and performs the quotient by
$\Gamma_0$.  The resulting background $\eM/\Gamma_0$ is a smooth
supersymmetric M-theory background which is locally isometric to
$\eM$: it consists of making identifications along the orbits of the
Killing vector generating $\Gamma$ at equal intervals in the parameter
space.  The choice of this interval implies a choice of scale relative
to which the (possibly varying) size of the M-theory circle is
measured.

Having quotiented by $\Gamma_0$, it is the circle subgroup
$\Gamma/\Gamma_0$ which acts nontrivially on $\eM/\Gamma_0$.  Its
orbits are the M-theory circles, which may be of varying size.  The
second step in the reduction is then the standard Kaluza--Klein
reduction of $\eM/\Gamma_0$ by the circle subgroup $\Gamma/\Gamma_0$.

The advantage of performing the reduction along $\Gamma$ in two steps,
apart from placing the dimensional reduction squarely in the familiar
context of circle reductions, is that the result of the first step is
often interesting in its own right.

In fact, in \cite{FigSimFlat} we discovered a novel reduction of
eleven-dimensional Minkowski spacetime: a generalised fluxbrane in
type IIA which we dubbed a nullbrane, since the RR 2-form field
strength is null and moreover the corresponding Killing vector $\xi$
has a component which is a null rotation.  The subgroup $\Gamma$
generated by this Killing vector is noncompact, parametrised by a real
number $t$.  The isomorphism $\RR \to \Gamma$ is given explicitly by
$t \mapsto g(t) := \exp t \xi$.  This map is a group homomorphism:
$g(t) g(t') = g(t+t')$, whence if we choose a scale $R$, the elements
$\Gamma_0 = \left\{g(nR)\mid n\in\ZZ \right\}$, form a co-compact
discrete subgroup isomorphic to $\ZZ$.  The quotient
$\RR^{1,10}/\Gamma_0$ of Minkowski spacetime by $\Gamma_0$ is a smooth
supersymmetric M-theory background, locally isometric to Minkowski
spacetime.  It is moreover time-dependent, since the null rotation
involves time nontrivially.  It has been dubbed the ``nullbrane'' in
\cite{LMS2} because the circle reduction to IIA along the orbits of
$\Gamma/\Gamma_0$ gives rise to the nullbrane of \cite{FigSimFlat}.
We will call it here the ``eleven-dimensional nullbrane'' to avoid
confusion.

Similarly, there are ``eleven-dimensional fluxbranes'' which are
obtained by performing a discrete quotient of $\RR^{1,10}$ by
$\Gamma_0 \subset \Gamma$, where $\Gamma$ is now such that
$\RR^{1,10}/\Gamma$ is a IIA fluxbrane.

The above discussion holds in any number $D{+}1$ of dimensions---in
particular in ten dimensions.  Ten-dimensional fluxbranes, that is,
$\RR^{1,9}/\Gamma_0$, allow a conformal field theory (CFT) \cite{RT1,
  RT, TseytlinFlux, Suyama, TU, RTFlux} description in terms of
resolutions of the ordinary euclidean orbifolds.  Analogously,
ten-dimensional nullbranes resolve null (or parabolic) orbifolds
\cite{HorowitzSteif, TseytlinExact}.  The geometry of these null
orbifolds has been recently discussed in \cite{JoanNull,LMS1}.  It was
further noticed in \cite{LMS1} that, by going to the light cone gauge
such spacetimes allow an interpretation in terms of a Big Crunch phase
connected to a Big Bang phase through a point where our spacetime
manifold is non-Hausdorff.  The nullbrane not only resolves the fixed
points of the null orbifold and its non-Hausdorff nature, but it is
also free of closed causal curves and it is stable against the
formation of black holes when probed by particles at least at low
energies and for a large enough number of spacetime dimensions
\cite{HorPol,FabMcG}.  The geometry of the eleven-dimensional
nullbrane is a metric product of a flat seven-dimensional euclidean
space (which can be further compactified) and a four-dimensional
locally minkowskian spacetime which can be viewed as the total space
of a circle fibration over a 3-dimensional manifold, where the radius
of the circle varies with time and reaches a minimum nonzero size $R$,
which is nothing but the scale in the previous discussion.  These
properties seem to allow the use of string perturbation theory in
these discrete quotients \cite{LMS2,FabMcG}, extending the original
formulation for the null orbifold given in \cite{LMS1}.

The existence of these backgrounds with cosmological interpretations
motivates and suggests a natural mathematical problem, namely the
classification of smooth supersymmetric M-theory or type IIA/IIB
backgrounds which are obtained by discrete quotients (often termed
\emph{orbifolds} even if the resulting quotient is smooth) of a given
one.  This is a much harder problem already for the case of Minkowski
spacetime and it definitely does not follow from the results of
\cite{FigSimFlat}.  Its solution would require classifying all the
discrete subgroups $\Gamma_0$ of the corresponding Poincaré group
which act freely and properly discontinuously on Minkowski spacetime
and which preserve some supersymmetry.  As discussed in
\cite{JMWaves,Bryant-ricciflat} supersymmetry in eleven dimensions
requires that $\Gamma_0$ be contained in the subgroup $\Spin(7)
\ltimes \RR^9$ of the Poincaré group.  The classification of smooth
flat supersymmetric M-theory backgrounds would include, in particular,
the classification of crystallographic subgroups of $\Spin(7) \subset
\SO(8)$: a hard problem in practice despite the existence of a
powerful algorithm.  Some comments and partial results concerning
smooth time-dependent backgrounds constructed in this way can be found
in Section~7 of \cite{LMS2}.

In this paper, as in \cite{FigSimFlat}, we will not emphasise the
discrete quotients of the M-theory backgrounds, but it is worth
remarking that from our results there follows a classification of
those discrete quotients by (infinite) cyclic subgroups.  As some of
the subgroups we will exhibit act nontrivially in the time direction,
some of the M-theory backgrounds resulting from the corresponding
discrete identifications will be time-dependent.  There has been much
recent interest in this class of backgrounds, as evidenced by the
recent work on spacelike branes \cite{GSSbranes, CGGSbranes,
  KMPSbranes, DKSbranes}, Sen's proposal \cite{SenRT,SenTE} that
dynamical rolling of the tachyon of open string field theory can lead
to interesting cosmologies, double Wick rotations of stationary
configurations \cite{AFHSWick, BRWick, CaiWick, GMWick, BLWWick},
coset models \cite{EGKRCoset, CKRCoset} and related orbifold constructions
\cite{KOSST, SeibergBC, BHKVN, CorCos, Nekrasov}.  The time-dependent
backgrounds implicit in our results hence add to the existing classes
of examples of such backgrounds.  It should be remarked that unlike
the ones found in \cite{FigSimFlat}, these will not be flat, but they
will be asymptotic to $\RR^{1,10}/\Gamma_0$, as corresponds to
excitations of these new M-theory vacua.  We will call them
\emph{asymptotically locally flat} spaces.

The purpose of the present work is twofold.  Firstly, it is natural to
ask about which D-branes, NS-branes and other dynamical objects in
string theory exist in flux/nullbrane sectors.  This question has been
partially addressed for the fluxbrane sector using CFT techniques
\cite{TU2,DudasMourad}.  In the following, we shall adopt a low-energy
effective closed string description to fully answer this point.  In
this way, we shall not only classify all the allowed supersymmetric
configurations, but we shall find their explicit supergravity
realisation at the same time.  If one is interested in classifying
composite configurations of branes, waves, monopoles,... and
flux/nullbranes, it is natural to consider the same configuration at
strong coupling, where they should be described in terms of the
uplifted known M-theory backgrounds satisfying certain nontrivial
global identifications reminiscent of the twisted identifications
associated with flux/nullbranes in flat spacetime.  Thus, our
classification problem of composite configurations in type IIA is
equivalent to a classification of certain one-parameter subgroups
$\Gamma$ of the symmetry group of an M-brane background $\eM$.  The
Kaluza--Klein reduction $\eM/\Gamma$ will then give rise to the
desired composite configurations.  This has already been shown for a
particular configuration in \cite{JoanD0F5} through a probe analysis
of D0-branes in flux 5-brane (F5-brane) backgrounds.  Similar ideas
were also pointed out in the appendix of \cite{RTFlux}.  On the other
hand, performing the reduction in steps: first quotienting by a
discrete subgroup $\Gamma_0 \subset \Gamma$ and then quotienting the
resulting M-theory background by the circle subgroup
$\Gamma/\Gamma_0$, will give rise, after the first step, to new
eleven-dimensional backgrounds $\eM/\Gamma_0$, some of which will be
time-dependent and all of which will be asymptotic to
$\RR^{1,10}/\Gamma_0$.

The second goal of this work is to make progress in the classification
of supersymmetric M-theory and type IIA supergravity backgrounds.  Due
to the equivalence stated above, it is natural to make a thorough
analysis of all smooth Kaluza-Klein reductions of a given fixed
M-theory background $\eM$.  It will turn out that for curved
spacetimes, such as the ones describing M-branes, new reductions arise
which were not taken into account in the past and are not possible in
Minkowski spacetime.

For example, it is possible to reduce along the orbits generated by
Killing vectors involving time translations and rotations transverse
to the brane, but which nevertheless are everywhere spacelike, so as
to avoid the existence of manifest closed timelike
curves.\footnote{Although such a restriction is perfectly legitimate
  from a physical point of view, one cannot discard the possibility of
  obtaining novel string backgrounds by reducing along the orbits of a
  freely-acting Killing vector $\xi$ after excising from the spacetime
  those regions where $\xi$ fails to be spacelike, with the usual
  completeness caveats.}  These novel reductions appear whenever the
unit sphere transverse to the brane is odd-dimensional, because such
spheres will admit nowhere-vanishing Killing vectors.  This is the
case for M2-branes or delocalised M5-branes, but it is certainly a
much more general phenomenon, present in the delocalised M-wave
\cite{FigSimGrav} and in some intersecting brane backgrounds
\cite{FigRajSim}, for instance.

A second example of novel reductions results whenever the brane is
delocalised.  In such cases one can consider Killing vectors involving
linear combinations of a translation tangent to the worldvolume of the
brane and a translation transverse to it.  If both translations are
spacelike, the type IIA configuration describes non-threshold bound
states.  If one of them is lightlike, the IIA observer no longer
measures an asymptotically Minkowski spacetime, but a wave propagating
at the speed of light.  Whenever one of the translations is timelike,
the physical interpretation remains unclear to us.

It is a natural question to ask about the global causal structure of
the spacetimes obtained by reducing along Killing vectors which act
nontrivially on time.  This question remains open for the former set
of novel reductions discussed above, but we will show that for the
latter set there are no closed causal curves whenever the Killing
vector is everywhere spacelike, even if it contains timelike and
lightlike translations.

It should be clear that any of the techniques mentioned above also
applies to any on-shell background in type IIA/IIB supergravity.  In
this context, our analysis must be understood as the classification of
all inequivalent quotients (discrete or not) of such a given
background by a one-parameter subgroup of symmetries.  As will be
discussed in the last section of this paper, both constructions, the
one obtained from M-theory through Kaluza--Klein reduction and the one
directly constructed in type IIA are equivalent through a TST chain of
dualities, T standing for T-duality and S for S-duality
transformations, respectively.

In the context of the AdS/CFT correspondence \cite{AdSCFTReview}, some
of our configurations (composites of D-branes and fluxbranes), when
studied in certain decoupling limits, provide us with supergravity
duals of the dipole theories introduced in \cite{BerGanDipoles, DGR,
  BDGKR, DSJDipole, AliYav}.  Indeed, our configurations depend on
some set of parameters (related to the charge of the fluxbranes) which
can be held fixed in a Maldacena-type limit \cite{Malda,IMSY}.  They
can thus be viewed as some sort of deformation parameters of the
corresponding near horizon geometries of D-branes.  Our analysis shows
the existence of further supersymmetric configurations both in the
flux- and nullbrane sectors.  Indeed, the identifications underlying
dipole theories involve the R-symmetry group transverse to the brane.
We point out the possibility of making identifications involving
transverse and longitudinal (either spacelike or lightlike) directions
to the brane at the same time.  Their dual gauge theory description
involves supersymmetric Yang--Mills with the adjoint matrices
satisfying twisted conditions generalising the ones written before
\cite{BerGanDipoles,MotlMelvin}.  Similar techniques were used also in
\cite{MorTri}.

We would like to emphasise the universality of the construction
presented in this paper.  Even though we shall restrict ourselves to
M2 and M5-brane configurations in this work, it is clear that any
supersymmetric configuration, with enough symmetry, of a $D{+}1$
supergravity theory (ordinary or gauged) for which a consistent
Kaluza--Klein truncation to a $D$-dimensional supergravity theory is
known, would generate new $D$-dimensional configurations by using the
same methods developed below.  In particular, extensions to M-waves
and MKK-monopoles \cite{FigSimGrav}, intersections of M-branes
\cite{FigRajSim}, supersymmetric wrapped branes, antibranes,
multicentred configurations or supergravity duals of non-commutative
gauge theories, among others, are conceptually straightforward.  It is
particularly interesting to address these questions in supersymmetric
backgrounds of the form $\AdS_p\times S^q$ both in M-theory and in
type IIB.  The BTZ black hole \cite{BTZ} construction out of $\AdS_3$
\cite{BHTZ} and the dual description of string theory in these
backgrounds in terms of supersymmetric Yang--Mills theories opens many
interesting questions both in the gravity and gauge theory sides.
This is currently under investigation and some results will appear
elsewhere \cite{JoanNullAdS,FigSimAdS}.

In what remains of this introductory section, we provide a
self-contained and hopefully comprehensive description of the
geometrical set-up underlying Kaluza--Klein reductions.  It also fixes
the notation and conventions used in the rest of the paper.
Section~\ref{sec:geometry} deals with an arbitrary bosonic background
having certain group of isometries, whereas in
Section~\ref{sec:susyKK} we explain which is the criterion for
preservation of supersymmetry.  In this way, we are naturally led to
introduce the notion of \emph{moduli space of supersymmetric
  Kaluza--Klein reductions}.  In Section~\ref{sec:branes}, we present
the argument that will allow us to determine this moduli for the M2-
and M5-brane configurations in the body of the paper.  Finally,
Section~\ref{sec:summary} details the organisation of the rest of the
paper and a brief summary of its results.

\subsection{Supersymmetric M-theory backgrounds}
\label{sec:susyMback}

The bosonic fields of eleven-dimensional supergravity \cite{Nahm,CJS}
are a lorentzian metric $g$ and a closed four-form $F_4$ defined on an
eleven-dimensional spin manifold $M$.  The bosonic equations of motion
are a generalisation of the Einstein--Maxwell equations in four and
five dimensions.  They consist of a nonlinear Maxwell-type equation
for $F_4$:
\begin{equation*}
  d\star F_4 = \half F_4 \wedge F_4~,
\end{equation*}
and an Einstein-type equation for $g$ whose explicit form need not
concern us here, but which relates the Einstein tensor of $g$ to the
energy momentum tensor of $F_4$---an algebraic tensor depending on $g$
and quadratically on $F_4$.

By hypothesis $M$ has a spin structure and hence a bundle of spinors,
which is a (symplectic) real bundle of rank $32$.  The supersymmetry
of the supergravity action (with fermions included) induces a
covariant derivative operator $\eD$ on spinors.  It is defined as the
supersymmetry variation of the gravitino $\Psi_M$ restricted to the
subspace of the configuration space where the gravitino vanishes.  In
other words, if $\varepsilon$ is a spinor, then
\begin{equation*}
  \eD_M \varepsilon = \delta_\varepsilon \Psi_M\bigr|_{\Psi=0}~.
\end{equation*}
We will not need the explicit expression of $\eD$; suffice it to say
that it can be written in the form $\eD_M = \nabla_M + \Omega_M$,
where $\nabla$ is the spin connection and $\Omega_M$ is an
endomorphism of the spinor bundle depending algebraically in $g$ and
linearly in $F_4$.  A nonzero spinor $\varepsilon$ which obeys $\eD_M
\varepsilon = 0$ is called a \emph{Killing spinor}, as they are the
``square roots'' of Killing vectors.  A pair $(g,F_4)$ satisfying the
field equations (with fermions set to zero) is called a
\emph{(bosonic) background}.  A background is \emph{supersymmetric} if
it possesses Killing spinors.  Since the Killing spinor equation is
linear, the space of Killing spinors is a vector space, and since the
equation is first order, its dimension is not bigger than the rank of
the spinor bundle, here $32$.  A background $(g,F_4)$ is said to
preserve a fraction $\nu$ of the supersymmetry if the dimension of the
space of Killing spinors is $32\nu$.  The space of backgrounds is
stratified by the possible values
$\{0,\frac1{32},\frac1{16},\dots,1\}$ of $\nu$.  At the time of
writing it is not known whether all strata are nonempty.

\subsection{The geometry of Kaluza--Klein reductions}
\label{sec:geometry}

Ten-dimensional type IIA supergravity can be defined as the
Kaluza--Klein reduction of eleven-dimensional supergravity along a
spacelike direction \cite{CampbellWestIIa, GianiPerniciIIA,
  HuqNamazieIIA}.  This means that if the Kaluza--Klein reduction of
an M-theory background exists, it is automatically a type IIA
background and moreover every IIA background arises in this way (at
least locally).  Of course, not every M-theory background can be
reduced, as this requires the existence of a symmetry.

By an \emph{(infinitesimal) symmetry} of a background $(g,F_4)$ we
mean a vector field $\xi$ on $M$ such that
\begin{itemize}
\item $\xi$ is Killing: $\eL_\xi g = 0$; and
\item $\xi$ preserves $F_4$: $\eL_\xi F_4 = 0$---equivalently $d i_\xi
      F_4 = 0$, since $F_4$ is closed.
\end{itemize}

Now suppose that an infinitesimal symmetry $\xi$ integrates to an
action of a one-dimensional group $\Gamma$ (either $\RR$ or $S^1$)
obeying the following two properties:
\begin{enumerate}
\item the action is free; and
\item the norm of $\xi$ never vanishes.
\end{enumerate}
The first property involves two separate conditions: one
infinitesimal, that the vector field $\xi$ never vanishes; and one of
a more global nature, that every point in $M$ should have trivial
stabiliser.  This guarantees that the space $N=M/\Gamma$ of
$\Gamma$-orbits is a smooth manifold.  The second property means that
$N$ inherits a metric: lorentzian if $\xi$ is spacelike or riemannian
if it is timelike.  As in \cite{FigSimFlat}, we will concentrate
solely on the case of a spacelike $\xi$.

The condition that $\xi$ does not vanish is a necessary condition for
the smoothness of the quotient.  In many of the examples we will
consider in this paper, $\xi$ will tend to zero as we approach the
horizon.  Such a zero need not have geometric meaning, as the
coordinate system in which our ansätze are written are singular in the
horizon.  One way to determine whether or not this zero of the vector
field yields a singularity in the quotient would be to extend the
solution beyond the horizon, as the brane ansätze we will write down
only cover the spacetime exterior to the brane.  In order to keep this
paper down to a reasonable length, we will choose to ignore these
potential singularities and take the conservative approach that our
solutions only describe the spacetime exterior to the brane, even
though the physics of these potential singularities in the throats of
the different branes and their resolutions, if any, are very
interesting questions for future research.

Since it plays an important role in our approach, we now describe in
some detail the geometric underpinning of the Kaluza--Klein ansatz.
We think of the original spacetime $M$ as the total space of a
principal $\Gamma$-bundle $\pi:M \to N = M/\Gamma$, where $\pi$ is the
map taking a point in $M$ to the $\Gamma$-orbit on which it lies.  At
every point $p$ in $M$, the tangent space $T_pM$ of $M$ at $p$
decomposes into two orthogonal subspaces: $T_p M = \eV_p \oplus
\eH_p$, where the \emph{vertical subspace} $\eV_p = \ker \pi_*$
consists of those vectors tangent to the $\Gamma$-orbit through $p$,
and the \emph{horizontal subspace} $\eH_p=\eV_p^\perp$ is its
orthogonal complement relative to the metric $g$.  The resulting
decomposition is indeed a direct sum by virtue of the
nowhere-vanishing of the norm of $\xi$, whose value at $p$ spans
$\eV_p$ for all $p$.  The derivative map $\pi_*$ sets up an
isomorphism between $T_p M$ and $T_q N$, where $\pi(p) = q$.  There is
a unique metric on $N$ for which this isomorphism is also an isometry
and it is defined as follows.  We choose a point $p$ in the fibre above
$q$.  Then if $X,Y \in T_qN$, we define their inner
product\footnote{The reason for the prime in the notation will become
  obvious below.}  $h'(X,Y) = g(\tilde X, \tilde Y)$, where $\tilde
X,\tilde Y \in \eH_p$ obeying $\pi_*\tilde X =X$ and $\pi_*\tilde Y
=Y$ are the \emph{horizontal lifts} of $X$ and $Y$, respectively.
This does not actually depend on the choice of $p$ because $\Gamma$
acts by isometries.

The horizontal sub-bundle $\eH$ gives rise to a connection one-form
$\alpha$ on $M$ such that $\eH = \ker \alpha$ and such that
$\alpha(\xi) = 1$, where $\xi$ is the Killing vector generating the
$\Gamma$-action.  Relative to a local trivialisation, and letting $z$
be a coordinate along the $\Gamma$-orbits, we can write $\alpha = dz +
A$, where $A$ is a locally-defined one-form on $N$.  In terms of this
data (and omitting pullbacks by $\pi$) the metric $g$ on $M$ can be
written as
\begin{equation*}
  g = h' + e^{4\Phi/3} (dz + A)^2~,
\end{equation*}
where $e^{2\Phi/3}$ is the norm of the Killing vector $\xi$, which in
this trivialisation is given by $\xi = \d_z$.  The function
$\Phi:N\to\RR$ is the dilaton.  To make contact with the metric which
appears in the effective action of the type IIA string, it is
convenient to conformally rescale the metric on $N$ by a function of
the dilaton.  Doing this we finally arrive at the familiar
string-frame Kaluza-Klein ansatz for the dimensional reduction of
eleven-dimensional supergravity to type IIA supergravity:
\begin{equation}
  \label{eq:kkg}
  g = e^{-2\Phi/3} h + e^{4\Phi/3} (dz + A)^2~.
\end{equation}

The four-form $F_4$ also reduces and gives rise to two forms on $N$:
the NSNS three-form $H_3$ and the RR four-form $H_4$.  To see how this
comes about, let us first decompose $F_4$ as follows
\begin{equation*}
  F_4 = G_4 - \alpha \wedge G_3~,
\end{equation*}
where $\alpha$ is the connection one-form defined above.  The
curvature two-form $d\alpha$ is both \emph{horizontal}, so that
$\imath_\xi d\alpha = 0$ and \emph{invariant}, so that $\eL_\xi
d\alpha = 0$.  (In this case, invariance follows from horizontality
since $d\alpha$ is closed.)  To prove horizontality, we first observe
that $\alpha = \xi^\flat/\|\xi\|^2$, where $\xi^\flat$ is the one-form
defined by $\xi^\flat(X) = g(\xi, X)$ for all vectors field $X$ on
$M$, is invariant\footnote{This is equivalent to the physical
  statement that the photon carries no charge.}.  This means that
$\imath_\xi d \alpha + d\imath_\xi \alpha = 0$, but since
$\alpha(\xi)=1$ and hence constant, this is precisely horizontality of
$d\alpha$.

Forms which are both horizontal and invariant are called \emph{basic},
since they are pull-backs of forms on the base $N$.  The curvature
two-form $d\alpha$ is the pull-back of the RR two-form field-strength
$H_2 = dA$.  Let us remark in passing that in coordinates adapted to
$\xi$, so that $\xi = \d_z$, a form is basic if and only if it has no
$dz$ component (horizontality) and does not depend explicitly on $z$
(invariance).

We now claim that both $G_4$ and $G_3$ are basic.  It is clear from
the above expression that $G_3 = -\imath_\xi F_4$, so that it is
manifestly horizontal.  Invariance of $F_4$ means that $G_3$ is
closed, whence it is basic.  It is therefore the pull-back of a
three-form on $N$: the NSNS three-form field-strength $H_3=dB_2$.
Finally, we observe that $G_4$ is also basic.  It is manifestly
horizontal, and invariance follows by a simple calculation
\begin{equation*}
  \begin{split}
    \eL_\xi G_4 &= d \imath_\xi G_4 + \imath_\xi d G_4\\
    &= \imath_\xi (-d\alpha \wedge G_3)\\
    &= 0~,
  \end{split}
\end{equation*}
where we have used that $G_3$, $G_4$ and $d\alpha$ are horizontal.
This means that $G_4$ is the pull-back of a four-form on $N$.  It is
convenient to write that four-form as $H_4 - H_3 \wedge A$, where
$H_4$ is the RR four-form field-strength.

In the local trivialisation used above, we can rewrite the above
decomposition of $F_4$ in a more familiar form (omitting pullbacks)
\begin{equation}
  \label{eq:kkf}
  F_4 = H_4 - dz \wedge H_3~.
\end{equation}

In summary, if $(M,g,F_4)$ is an M-theory background admitting a free
$\Gamma$-action ($\Gamma=\RR$ or $\Gamma=S^1$) with spacelike orbits,
then $(N=M/\Gamma,h,\Phi,H_2=dA,H_3,H_4)$ is a type IIA background.
Conversely any IIA background $(N,h,\Phi,H_2,H_3,H_4)$ can be lifted
locally to an M-theory background $(M,g,F_4)$ possessing a spacelike
symmetry in such a way that the Kaluza--Klein reduction of $(M,g,F_4)$
along that symmetry reproduces the IIA background we started out with.

Some of the backgrounds we will be studying admit a simpler
description in terms of the dual seven-form $F_7 = \star F_4$.  The
Kaluza--Klein reduction of $F_7$ is very similar to that of $F_4$.  We
start by decomposing $F_7$ as
\begin{equation*}
  F_7 = G_7 + \alpha \wedge G_6~,
\end{equation*}
where $G_7$ and $G_6$ are horizontal.  This means that $G_6 =
\imath_\xi F_7$.  We claim that $G_6$ and $G_7$ are also invariant,
whence basic.  To see this we compute their Lie derivative with
respect to $\xi$:
\begin{equation*}
  L_\xi G_6 =   L_\xi \imath_\xi F_7 = \imath_\xi L_\xi F_7 = 0~,
\end{equation*}
since $F_7$ is invariant.  Similarly, using that $G_7 = F_7 - \alpha
\wedge G_6$, we calculate
\begin{equation*}
  L_\xi G_7 = L_\xi (F_7 - \alpha \wedge G_6) = - L_\xi \alpha
  \wedge G_6 =0~,
\end{equation*}
where we have used that $\alpha$ is invariant.  In summary, since they
are basic, $G_6$ and $G_7$ are pullbacks of forms on the base.  In
adapted coordinates where $\xi = \d_z$ and $\alpha = dz + A$, we can
write (omitting pullbacks)
\begin{equation}
  \label{eq:kkfdual}
  F_7 = H_7 + dz \wedge H_6~,
\end{equation}
where $H_6$ pulls back to $G_6$ and $H_7$ pulls back to $G_7 - A\wedge
G_6$.

It is possible to trace the Hodge-dual across the reduction and to
relate $H_7$ and $H_6$ to $H_4$ and $H_3$ and the fields to which the
eleven-dimensional metric reduces.  One finds
\begin{equation}
  \label{eq:duals}
  \begin{aligned}[m]
    H_6 &= \star_{10} \left( H_4 + A \wedge H_3 \right)\\
    H_7 &= A \wedge \star_{10} H_4 + \star_{10} \imath_{A^\sharp} (A
    \wedge H_3) - e^{-2\Phi} \star_{10} H_3~,
  \end{aligned}
\end{equation}
where $\star_{10}$ is the ten-dimensional Hodge star operator relative
to the metric $h$ and $A^\sharp$ is the vector field dual to the RR
one-form $A$.

One can also invert the above relation and express $H_3$ and $H_4$ in
terms of $H_6$ and $H_7$.  With the same notation as above, the result
is the following
\begin{equation}
  \label{eq:slaud}
  \begin{aligned}[m]
    H_3 &= e^{2\Phi} \star_{10} \left( H_7 - A \wedge H_6 \right)\\
    H_4 &= \star_{10} H_6 - e^{2\Phi} A \wedge \star_{10} H_7 +
    e^{2\Phi} \star_{10} \imath_{A^\sharp} (A \wedge H_6)~,
  \end{aligned}
\end{equation}
which we record here for future reference.

\subsection{Supersymmetric Kaluza--Klein reductions}
\label{sec:susyKK}

Closely tied to the symmetries of a supergravity background are
its supersymmetries, which manifest themselves through Killing
spinors.  In this section we review what happens to supersymmetry
under Kaluza--Klein reduction.

Every Killing vector $\xi$ acts naturally on a spinor $\varepsilon$
via the \emph{spinorial Lie derivative} introduced by Lichnerowicz
(see, e.g. \cite{Kosmann} and more recently \cite{JMFKilling} for
applications closer to the present context) and defined by
\begin{equation*}
  \eL_\xi \varepsilon = \nabla_\xi \varepsilon + \tfrac14 d\xi^\flat
  \cdot \varepsilon~,
\end{equation*}
where $\nabla$ is the spin connection and where $\cdot$ means the
Clifford action of forms on spinors.  Choosing a local frame, we can
write
\begin{equation*}
  \eL_\xi \varepsilon = \nabla_\xi \varepsilon + \tfrac14 \nabla_a
  \xi_b \Gamma^{ab} \varepsilon~.
\end{equation*}

The spinorial Lie derivative enjoys many of the properties that we
expect of a Lie derivative; in particular, it is a derivation
\begin{equation}
  \label{eq:derivation}
  \eL_\xi (f \varepsilon) = (\xi f) \varepsilon + f \eL_\xi
  \varepsilon~,
\end{equation}
for any function $f$.  Furthermore one can show that for any vector
field $X$,
\begin{equation*}
  [\eL_\xi, \nabla_X] = \nabla_{[\xi,X]}~,
\end{equation*}
and if in addition $\xi$ preserves $F_4$, then
\begin{equation*}
  [\eL_\xi, \eD_X] = \eD_{[\xi,X]}~.
\end{equation*}
This implies that the Lie derivative along $\xi$ of a Killing spinor
is again a Killing spinor, and hence the space of Killing spinors
becomes a linear representation of the group $\Gamma$ generated by
$\xi$.

It follows from the definition of type IIA supergravity as the
Kaluza--Klein reduction of eleven-dimensional supergravity, that the
supersymmetries of the IIA background are precisely the
$\Gamma$-invariant Killing spinors of the M-theory background.  Notice
that the condition $\eL_\xi\varepsilon =0$ for $\varepsilon$ a Killing
spinor is actually an algebraic condition, since $\eL_\xi - \eD_\xi$
is a zeroth order operator.  This is nothing but the condition which
arises from the supersymmetry variation of the IIA dilatino fields.
Thus if an M-theory background admits a Kaluza--Klein reduction along
the orbits of a group $\Gamma$ such that there are $\Gamma$-invariant
Killing spinors, the resulting IIA background will be supersymmetric.
We will call this a \emph{supersymmetric Kaluza--Klein reduction}.
The IIA background will preserve a fraction $\nu$ of the
supersymmetry, where $32\nu$ is the dimension of the space of
$\Gamma$-invariant Killing spinors of the corresponding M-theory
background.

Given a supersymmetric M-theory background $(M,g,F_4)$ with
symmetries, the problem of finding supersymmetric Kaluza--Klein
reductions can be phrased in the following terms.  Let $G$ be the Lie
group of symmetries of the background and let $\fg$ denote its Lie
algebra, consisting of infinitesimal symmetries, namely those Killing
vectors on $M$ which also preserve $F_4$.  Let $\eT$ be the subset of
$\fg$ corresponding to those spacelike Killing vectors which integrate
to a free group action preserving some Killing spinors.  The subset
$\eT$ parametrises the supersymmetric Kaluza--Klein reductions of the
background $(M,g,F_4)$; however not all points in $\eT$ correspond to
physically distinct reductions.  Indeed, the action of $G$ on $M$
induces the adjoint action on $\fg$, and this action preserves the
subset $\eT$.  Two points in $\eT$ which are $G$-related, correspond
to two physically indistinguishable supersymmetric reductions, as they
are related by a change of variables corresponding to a symmetry.  The
space $\eT$ is also preserved by rescaling the Killing vectors.
Indeed, notice that if $\xi\in\eT$, then $R\xi\in\eT$ for every
nonzero real number $R\in\RR^\times$.  Although the scale $R$ has
physical meaning---for example, it is related to the radius of the
M-theory circle in the standard Kaluza--Klein reduction---it is
natural, from a mathematical perspective at least, to identify
collinear elements in $\eT$ and define the \emph{moduli space $\eM$ of
  supersymmetric Kaluza--Klein reductions} as the set of orbits of the
action of $G \times \RR^\times$ on $\eT$; in other words, as the set
of equivalence classes $[X]$ where $X\in\eT$, where $X \sim R g X
g^{-1}$ for all $g\in G$ and $R\in\RR^\times$.  The number of Killing
spinors left invariant by $X$ and by $R g X g^{-1}$ is clearly the
same, whence the space $\eM$ is stratified by the value of $\nu$, the
fraction of the supersymmetry preserved by the IIA background, which
will be a further fraction of that preserved by the M-theory
background from which we reduce.  In \cite{FigSimFlat} we determined
$\eM$ for the flat eleven-dimensional M-theory vacuum and in this
paper we determine $\eM$ for the M2 and M5-brane solutions by mapping
the problem essentially (but not quite) to the flat case, albeit with
a restricted symmetry group.  A similar method also allows us to
determine $\eM$ for the purely gravitational M-wave and Kaluza--Klein
monopole \cite{FigSimGrav}.

We should remark that the failure of the problem to map
\emph{precisely} to the flat case, far from being a problem, is
actually responsible for the existence of novel reductions by vector
fields which would have Killing horizons in flat space but which are
spacelike relative to the brane metric.

\subsection{Classifying supersymmetric brane reductions}
\label{sec:branes}

The classification problem of supersymmetric reductions has two parts.
One is the classification proper: determining the moduli space $\eM$
of the given background.  Then given $\eM$, a second part of the
problem is to perform the reduction explicitly.  Although one can
address this problem for an arbitrary supersymmetric background, in
practice only the most symmetric backgrounds admit an explicit
solution.

Since Killing spinors square to Killing vectors, one way to generate
backgrounds with sufficient symmetry is to look for backgrounds
preserving a large fraction of the supersymmetry.  In
\cite{FigSimFlat} we considered the case of flat space.  Due to the
high degree of symmetry present in the flat background, we were able
to employ group-theoretical methods to solve the first part of the
problem and classify all the supersymmetric (generalised) fluxbranes.
We were then able to write down explicit formulae for the resulting
IIA backgrounds.

In the present paper we extend these results to the M2 and M5-brane
half-BPS backgrounds.  These brane-like solutions we consider in this
paper are asymptotically flat and crucial to our approach is that they
inherit isometries and supersymmetries from the asymptotic geometry.
Indeed, both Killing vectors and Killing spinors are determined
uniquely by their asymptotic values.  Moreover, as the following
simple argument will show, this correspondence is equivariant with
respect to the action of the Killing vectors on the Killing spinors,
reducing the problem in essence to the flat space case, albeit with a
restricted Poincaré group.

Consider for simplicity a typical electric $p$-brane solution in $d$
dimensions:
\begin{equation}
  \label{eq:elemint}
  \begin{aligned}[m]
    g &= e^{2A} ds^2(\EE^{1,p}) + e^{2B} ds^2(\EE^{d{-}p{-}1}) \\
    F_4 &= \dvol(\EE^{1,p}) \wedge dC~,
  \end{aligned}
\end{equation}
where $A,B,C$ are functions of the radial distance $r$ in the
transverse space $\EE^{d{-}p{-}1}$ approaching $0$ as $r\to\infty$.
(In the magnetic ansatz we would replace $F_4$ with $\star F_4$.)  The
asymptotic geometry is therefore flat and invariant under
$\ISO(1,d-1)$, whereas the brane solution is only invariant under a
subgroup $G=\ISO(1,p) \times \SO(d{-}p{-}1)$.  The Killing spinors for
such a background are of the form
\begin{equation*}
  \varepsilon = e^D \varepsilon_\infty~,
\end{equation*}
where $D$ is another function of the transverse radius approaching $0$
at infinity and where the asymptotic value $\varepsilon_\infty$ is a
(covariantly) constant spinor in flat space subject to a condition of
the form
\begin{equation}
  \label{eq:plane}
  \dvol(\EE^{1,p}) \cdot \varepsilon_\infty  = \varepsilon_\infty~.
\end{equation}
We claim that the action of $G$ on the Killing spinors is induced by
the action of $\Spin(1,p) \times \Spin(d{-}p{-}1)$ on the asymptotic
spinors $\varepsilon_\infty$.  Indeed, consider the action of a
Killing vector $\xi$ in the Lie algebra $\fg$ of $G$ on a Killing
spinor $\varepsilon = e^D \varepsilon_\infty$.  Since $D$ is constant
along the orbits of $\xi$,
\begin{equation*}
  \eL_\xi \varepsilon = e^D \eL_\xi \varepsilon_\infty~,
\end{equation*}
where we have used the derivation property \eqref{eq:derivation} of
the Lie derivative.  On the other hand, since $\xi$ also preserves
$F_4$, the Lie derivative of a Killing spinor is again a Killing
spinor, whence
\begin{equation*}
  \eL_\xi \varepsilon = e^D \varepsilon'_\infty~,
\end{equation*}
for some constant spinor $\varepsilon'_\infty$.  Comparing the two
expressions we see that
\begin{equation*}
  \eL_\xi \varepsilon_\infty = \varepsilon'_\infty~,
\end{equation*}
which is manifestly independent of $r$.  We can therefore compute it
in the asymptotic limit $r\to\infty$ where the spacetime is flat; but
in flat space, the action of $\eL_\xi$ on spinors coincides with the
restriction to $\fg$ of the spinor representation of $\Spin(1,10)$.
In other words, translations act trivially and a Lorentz
transformation $\xi \in \fg$ of the form $\xi = \lambda^{MN} x_M \d_N$
acts as
\begin{equation}
   \label{eq:susycons}
   \eL_\xi \varepsilon_\infty = \half \lambda^{MN} \Sigma_{MN}
   \varepsilon_\infty~,
\end{equation}
where $\Sigma_{MN}$ are the spin generators in the Clifford algebra.
Notice that since $\dvol(\EE^{1,p})$ is $G$-invariant, the action of
$G$ on spinors restricts to the subspace defined by \eqref{eq:plane}.

Essentially the same argument works for delocalised branes, where the
functions $A$, $B$ and $C$ entering the solution are invariant under
translations in the transverse space.  This will allow us to also use
the methods of \cite{FigSimFlat} in order to study the supersymmetric
reductions of delocalised brane solutions.  There are however two
fundamental differences between flat space and the brane backgrounds.
One is the obvious fact that the brane metric is not flat and hence
the norm of vector fields will differ.  We will see that there exist
Killing vector fields which in flat space would have a Killing horizon
but which in the spacetime exterior to the brane are everywhere
spacelike (except perhaps at the brane horizon).  The second
difference is reflected in the fact that the symmetries of the brane
are a \emph{proper} subgroup of the symmetries of the flat asymptotic
geometry.

These very differences make this problem more interesting than the
flat space problem.  The existence of spacelike Killing vectors which
would have Killing horizons in flat space underlies some of the novel
reductions described in this paper.  Similarly, in flat space any two
(spacelike, say) translations are equivalent under conjugation.  This
means that we can choose coordinates so that any translation is along
one of the coordinates.  In contrast, for a (delocalised) brane
solution, the symmetry is not large enough to conjugate a translation
tangent to the brane into a translation perpendicular to the brane.
This means that we can consider translations $a \d_\parallel + b
\d_\perp$ where $\d_\parallel$ and $\d_\perp$ are the components
tangent and perpendicular to the brane, respectively.  The
Kaluza--Klein reduction along such translations gives a ``pencil'' of
supersymmetric reductions interpolating smoothly between the extremes
$a=0$ and $b=0$.

\subsection{Contents and summary of results}
\label{sec:summary}

Let us now outline the results and the organisation of the paper.

Section~\ref{sec:KK} starts with a brief review of the results of
\cite{FigSimFlat}, and continues with the main theoretical results
which will be applied in the remainder of the paper to study the set
of M-theory backgrounds obtained by reducing the M2- and M5-brane
configurations along the orbits of a one-parameter subgroup of their
isometry groups.  We discuss the action of Killing vectors on Killing
spinors and two equivalent methods which can be used to determine the
loci of supersymmetric reductions.  We also derive general formulae
for the type IIA fields within the ansätze considered in this paper,
which are applied in later sections when we study specific
configurations.

Section~\ref{sec:M2} applies the technology developed in
Section~\ref{sec:KK} to the M2-brane both for the localised solution
in Section~\ref{sec:M2l} and for the solution which has been
delocalised along one transverse direction in Section~\ref{sec:M2d}.
As a consequence of our analysis, we find composite configurations of
strings, D2-branes and bound states of strings and D2-branes with both
fluxbranes and nullbranes.  The allowed supersymmetric configurations
are summarised in Tables~\ref{tab:M2}, \ref{tab:M2d2},
\ref{tab:M2(p,q)}.  In addition to these configurations, we find new
backgrounds obtained through reductions involving timelike and
lightlike translations.  When the M2-branes are localised, one can use
a linear combination of a timelike translation and a nowhere-vanishing
transverse rotation to the brane.  This is case~(A) in
Section~\ref{sec:M2l}.  If the M2-branes are delocalised, one can
reduce by the action of Killing vectors involving a linear combination
of the delocalised direction and a timelike or lightlike direction
along the brane.  This is case~(B) in Section~\ref{sec:M2d}.  In
Section~\ref{sec:M2dCCC} we prove that in this case there are no
closed causal curves in the quotient manifold.

Section~\ref{sec:M5} does for the M5-brane what Section~\ref{sec:M2}
did for the M2-brane.  Composite configurations of D4-branes,
NS5-branes and bound states D4-NS5 with both fluxbranes and nullbranes
are found and classified.  The results concerning supersymmetric
configurations are summarised in Tables~\ref{tab:M5susy1},
\ref{tab:M5susy2} and \ref{tab:M5susy3}.  As in the M2-brane
reductions, we find new backgrounds under the same circumstances as
stated above, but this time both occur for the delocalised M5-brane:
both cases (A) and (B) in Section~\ref{sec:M5d} include novel
reductions.  In Section~\ref{sec:M5dCCC} we prove that for some of
these reductions, the quotient manifold has no closed causal curves.

Section~\ref{sec:dual} argues why our technology can be
straightforwardly applied to many other configurations directly in
type IIA/IIB.  The duality relation among the configurations obtained
in this way and the ones discussed extensively in this paper is
explained.  We end up with some remarks concerning the gauge theory
dual description of some our backgrounds in the context of the AdS/CFT
correspondence.

Some facts about the decomposition of the spinors representation of
$\Spin(1,10)$ under certain subgroups, needed to determine the amount
of supersymmetry preserved by the backgrounds described in the body of
the paper, are discussed in Appendix~\ref{sec:groups}.

\section{Kaluza--Klein reductions of brane solutions}
\label{sec:KK}

In this section we describe some general features of the problem and
of the method used in this paper to solve it.  The problem of
determining the admissible supersymmetric Kaluza--Klein reductions can
be essentially mapped to a problem in flat space but with a restricted
group of isometries.  We will outline our method to classify the
supersymmetric Kaluza--Klein reductions of these backgrounds and we
will then derive general expressions for the IIA fields obtained by
Kaluza--Klein reduction.  The results of this section will be used
repeatedly in the rest of the paper.  In order to ease comparison with
the flat case and to make this paper self-contained, we will start the
section with a brief review of the flat case.

\subsection{Brief review of flat reductions}
\label{sec:flat}

In \cite{FigSimFlat} we classified those Kaluza--Klein reductions of
the eleven-dimensional vacuum of M-theory which give rise to smooth
ten-dimensional geometries and determined which of those were
supersymmetric.  Let us rephrase this problem in a way that will
facilitate the comparison with the problems treated in this paper.
Let $P = \ISO(1,10)$ be the eleven-dimensional Poincaré group and let
$\fp$ be the its Lie algebra.  We will identify $\fp$ with the Lie
algebra of Killing vectors in Minkowski space.  We denote $\eT \subset
\fp$ the \emph{subset} of $\fp$ consisting of Killing vectors $\xi$
which obey the following properties:
\begin{enumerate}
\item $\xi$ is everywhere spacelike; and
\item $\xi$ preserves some nonzero (covariantly) constant spinor.
\end{enumerate}
The subset $\eT$ is preserved by the adjoint action of $P$ on $\fp$,
which is the action induced on $\fp$ by the geometric action of $P$ on
Minkowski space.  Similarly, $\eT$ is preserved by rescalings $\xi
\mapsto s\xi$ for any nonzero real $s \in \RR^\times$.  In
\cite{FigSimFlat}, we determined the moduli space $\eM = \eT/(P \times
\RR^\times)$.  There are two families of solutions with three
parameters each:
\begin{align*}
  \xi &= \d_z + \theta_1 R_{12} + \theta_2 R_{34} +
  \theta_3 R_{56} + \theta_4 R_{78}~,\\
  \intertext{and} \xi &= \d_z + N_{+1} + \theta'_1 R_{34} +
  \theta'_2 R_{56} + \theta'_3 R_{78}~,
\end{align*}
with $\theta_1 + \theta_2 + \theta_3 + \theta_4 = 0$ and $\theta'_1 +
\theta'_2 + \theta'_3 = 0$.  The notation is that $z$ is the tenth
spacelike direction, $R_{ij}$ is the generator of infinitesimal
rotations in the $(ij)$-plane, and $N_{+i}$ the generator of an
infinitesimal null rotation in the $i$th direction, where the
light-cone coordinates are given by $x^\pm = (x^9 \pm x^0)/\sqrt{2}$.
The former family gives rise to fluxbranes, whereas the latter gives
rise to nullbranes and solutions interpolating between nullbranes and
fluxbranes.  The resulting IIA objects are summarised in
Table~\ref{tab:flat} together with the fraction $\nu$ of the
supersymmetry of the vacuum which the solution preserves and the
spinor isotropy subalgebra to which the Killing vector belongs.  The
notation F$p$ and N stands for a flux $p$-brane and a nullbrane,
respectively; and the objects labelled F$p$/N interpolate between
them.  In these cases, the fraction of the supersymmetry displayed in
the table is the one at generic points in the moduli space: there are
lower-dimensional loci where supersymmetry is enhanced.

\begin{table}[h!]
  \begin{center}
    \setlength{\extrarowheight}{3pt}
    \begin{tabular}{|>{$}c<{$}|c|>{$}l<{$}|}
      \hline
      \nu & Object & \multicolumn{1}{c|}{Subalgebra}\\
      \hline
      \hline
      \frac12 & F5 & \fsu(2)\\
      \frac12 & N & \RR\\
      \frac14 & F3 & \fsu(3)\\
      \frac14 & F1 & \fsp(2)\\
      \frac14 & F5/N & \fsu(2) \times \RR\\
      \frac18 & F1 & \fsu(4)\\
      \frac18 & F3/N & \fsu(3) \times \RR\\[3pt]
      \hline
    \end{tabular}
    \vspace{8pt}
    \caption{Supersymmetric IIA fluxbranes and nullbranes}
    \label{tab:flat}
  \end{center}
\end{table}

Now let $G \subset P$ be a subgroup of the Poincaré group and $\fg
\subset \fp$ be the corresponding Lie subalgebra.  The subspace $\eT
\cap \fg$ is now preserved by scaling and by conjugation by $G$, and
this prompts us to define the moduli space $\eM(G) = (\eT \cap
\fg)/(G\times \RR^\times)$.  The problem of determining $\eM(G)$ is a
restriction of the flat space problem, obtained by restricting the
Poincaré group to $G$.  As we will see presently, the problem of
determining the possible physically distinct supersymmetric
Kaluza--Klein reductions of many brane-like solutions, reduces morally
to determining $\eM(G)$ where $G$ is the symmetry group of the
brane-like solution.  We say `morally' because there will be a small
subtlety in that we will have to enlarge the space $\eT \cap \fg$ to
include Killing vectors in $\fg$ which are spacelike relative to the
brane metric and which may vanish at the horizon.  These vector fields
will not belong to $\eT$ but to some larger subspace of the Poincaré
algebra which will depend on the particular brane solution (e.g., on
the charge) and not simply on the symmetry group.  It is precisely
this fact which prevents us from claiming that the problem of
determining the possible supersymmetric Kaluza--Klein reductions of
M-branes reduces \emph{precisely} to the flat case; although it does
so to a large extent.

\subsection{Methodology}
\label{sec:method}

Let us recall our aims and outline the method we have followed to
achieve them.

Given a subgroup $G \subset \ISO(1,10)$ of symmetries of an M-theory
background $(M,g,F_4)$ of type \eqref{eq:elemint}, we would like
\begin{enumerate}
\item to classify the one-dimensional (connected) subgroups $\Gamma
  \subset G$ acting freely on $M$ with spacelike orbits and leaving
  invariant some nonzero Killing spinor; and
\item for every such subgroup, to write down explicitly the reduced
  IIA background.
\end{enumerate}
In the following section we will address (2) by giving general
formulas for the IIA fields, but let us first restate (1) using the
observations made above.

Let $\fg$ be the Lie algebra of $G$, which we identify with the
Killing vectors generating its action on $M$.  As a first step we will
determine those vectors $\xi \in \fg$ which are spacelike (at least
outside the brane horizon).  This in effect results in the
classification of reductions which are not necessarily supersymmetric
or even smooth.  Doing so requires a careful yet elementary analysis
of the norm of a Killing vector relative to the brane metric.  A
perhaps surprising result born out of this analysis is the existence
of everywhere spacelike Killing vectors which on flat space would not
have this property.

Having determined the everywhere spacelike Killing vectors we must
ensure that the action they generate is free: being spacelike means
that they never vanish, whence the action is locally free.  To ensure
that every point has trivial stabiliser is equivalent to demanding the
presence of the translation component in the Killing vector.  Since we
are interested in the moduli space $\eM$ of smooth supersymmetric
Kaluza--Klein reductions, we are free to conjugate by the action of
$G$; that is, to choose appropriate coordinates by performing a
symmetry transformation.

Finally we must impose that $\xi$ preserve some nonzero Killing spinor
in the background.  The same argument as in Section \ref{sec:branes}
shows that the action of $\xi$ on spinors coincides with the action of
$\fg$ on the spinor representation of $\Spin(1,10)$.  This reduces the
problem of finding supersymmetric Kaluza--Klein reductions of
solutions of the type \eqref{eq:elemint} to a flat space problem with
a restricted Poincaré group.  Furthermore, we must ensure that the
invariant spinors satisfy equation \eqref{eq:plane}.  Since
translations act trivially on spinors, only the Lorentz component of
$\xi$ is constrained.  In all cases, this component is either a
rotation or a commuting linear combination of a rotation and a null
rotation, and as we will now outline, supersymmetry will only
constrain the rotation.  Two methods can be used to determine the
supersymmetric loci: a representation-theoretical method based on the
weight decomposition of various spinorial representations, and a
direct computational method based on the Clifford algebra.  Since both
have their merits, we will describe them both.

\subsubsection{Representation-theoretical method}
\label{sec:reptheory}

Let $S$ be the unique half-spin representation of the
eleven-dimensional spin group.  (In Appendix~\ref{sec:groups} this
representation will be denoted $S_{11}$ to emphasise its
eleven-dimensional origin.)  The Clifford algebra can act on $S$ in
one of two inequivalent ways, but as mentioned above we will assume
that a choice has been made once and for all.  (This choice is
dictated by the supersymmetry transformation laws, equivalently by the
Killing spinor equation.)  The Killing spinors of the background are
in one-to-one correspondence with a linear subspace $S_0$ of $S$
determined by some $G$-equivariant equations of the form
\eqref{eq:plane}.  Equivariance simply means that the action of $G$,
and hence that of $\xi$, restricts to $S_0$.  We would like to
determine for which $\xi$ are there $\xi$-invariant spinors in $S_0$.
As mentioned in the previous paragraph, only the Lorentz component
$\lambda$ acts nontrivially on spinors, and in all the cases we will
consider this will be of the form $\lambda = \nu + \rho$, where the
null rotation $\nu$ and the rotation $\rho$ commute.  Since $\nu$ is
nilpotent and $\rho$ semisimple, $\lambda$ annihilates a spinor if and
only if it is annihilated by both $\nu$ and $\rho$ separately (cf.,
the argument in \cite[Section~2.2]{FigSimFlat}).  Since $\nu$ and
$\rho$ commute, the subspace of $\rho$-invariant spinors in $S_0$ is
preserved by $\nu$.  Now, without loss of generality, $\nu$ acts on
spinors as a multiple of $\Gamma_+\Gamma_1$, say, and both $\Gamma_+$
and $\Gamma_-$ commute with $\rho$.  Because $\Gamma_+\Gamma_- +
\Gamma_-\Gamma_+$ is proportional to the identity, the subspace of
$\rho$-invariant spinors in $S_0$ is even-dimensional and breaks up
into two equidimensional subspaces consisting of spinors annihilated
by $\Gamma_+$ (hence by $\nu$) and by $\Gamma_-$.  Therefore only
$\rho$ is constrained by the existence of invariant spinors and the
presence of the null rotation is only reflected in a further halving
of the fraction of supersymmetry which the reduction preserves.

Let us now discuss the constraints on $\rho$ arising out of the
existence of $\rho$-invariant spinors in $S_0$.  The rotation $\rho$
belongs to the maximal compact subalgebra $\fk$ of $\fg$ and hence in
some fixed Cartan subalgebra $\fh$ of $\fk$.  Up to isomorphism, the
action of $\fh$ on $S_0$ is completely specified by giving the weights
together with their multiplicities.  Weights are linear functionals on
$\fh$ hence the existence of $\rho$-invariant spinors is equivalent to
there being some weight $\mu$ which annihilates $\rho$.  The rotation
$\rho$ defines a point in $\fh$ with coordinates $\btheta
=(\theta_1,\dots,\theta_\ell)$ and the condition that a weight $\mu$
annihilates $\rho$ translates into a homogeneous linear equation on
$\btheta$.  The solutions of this equation will define a hyperplane in
$\fh$.  Doing this for all the weights\footnote{Due to the freedom to
  conjugate by $G$, we need only consider one weight in each Weyl
  orbit.} we obtain a family of hyperplanes defining the locus of
supersymmetric reductions.  The bigger the number of these hyperplanes
a rotation $\rho$ belongs to, the more supersymmetry will the
reduction preserve.

It is then a simple matter to determine the weight decomposition of
$S_0$ for each of the M-theory backgrounds discussed in this paper and
determine the locus of supersymmetry reductions for each one.  The
necessary group theory is outlined in Appendix~\ref{sec:groups}.

\subsubsection{Direct method}
\label{eq:projectors}

Alternatively one can obtain the same result by directly solving the
algebraic equation \eqref{eq:susycons}.  The most general form of the
latter, that is going to be used in this paper, can be written as
\begin{equation}
  \left(\alpha\eP_+ +
    \sum_{i=1}^n\beta_i\eP_i\right)\varepsilon_\infty = 0~,
 \label{eq:susyabs}
\end{equation}
where $\eP_+$ and $\eP_i$ are commuting linear transformations on
spinors obeying $\eP_+^2=0$ and $\eP_i^2 = -\1$ for all
$i$.\footnote{Notice that when one considers such an expression, the
  freedom under conjugation by the isometry group has already been
  taken into account, as explained in more detail for the different
  configurations studied in this paper.  In particular, one can think
  of $\eP_+=\Gamma_{+1}$ and $\eP_i = \Gamma_{i, i+1}$.}

The above equation defines two linear transformations $N=\alpha\eP_+$
and $S=-\sum_{i=1}^n\beta_i\eP_i$,
\begin{equation*}
  N\varepsilon_\infty = S \varepsilon_\infty~,
\end{equation*}
$N$ being nilpotent and $S$ semisimple.  Using the same argument of
\cite[Section~2.2]{FigSimFlat}, the original equation decomposes into
\begin{equation}
   \label{eq:susysol}
    N\varepsilon_\infty = 0 \qquad\text{and}\qquad
    S\varepsilon_\infty = 0~.
\end{equation}

The first equation breaks one half of the supersymmetry, and it is
associated with null rotations.  Let us now discuss the solutions to
the second equation in \eqref{eq:susysol}.  By squaring it, one
derives the further constraint
\begin{equation*}
  \left(\sum_{i<j} \alpha_{ij}\eQ_{ij}\right)\varepsilon_\infty = 
  \eK\varepsilon_\infty~,
\end{equation*}
where $\eQ_{ij}=\eP_i\eP_j$, $\alpha_{ij}=2\beta_i\beta_j$ and
$\eK=-\sum_{i=1}^n \left(\beta_i\right)^2$.  Notice that the
$\eQ_{ij}$ are also a commuting family of linear transformations.

Assuming that the parameter $\beta_i$ do not satisfy any relations,
the general solution to the above equation is given by
\begin{equation*}
  \eQ_{ij}\varepsilon_\infty = \eta_{ij}\varepsilon_\infty \qquad
  \eK = \sum_{i<j}\eta_{ij}\alpha_{ij}~,
\end{equation*}
for some signs $\eta_i$.  This is easily shown by induction on the
number of pairs $m= \binom{n}{2}$ determining the number of matrices
$\eQ_{ij}$.  Indeed, for $m=1$ (equivalently, $n=2$) it is trivial to
prove that $\eK=\eta_{12}\alpha_{12}$ and $\eQ_{12}\varepsilon_\infty=
\eta_{12}\varepsilon_\infty$ is the solution.  Assuming, our result is
true for a given $m$, we shall now show the result also holds for
$m+1$.  Indeed, the starting equation can be decomposed as
\begin{equation*}
  \left(\left(\sum_{i=1}^m\alpha_{(i)}\eQ_{(i)}\right) 
  + \alpha_{(m+1)}\eQ_{(m+1)}\right)\varepsilon_\infty = 
  \eK_{(m+1)}\varepsilon_\infty~.
\end{equation*}
By assumption, it can just be rewritten as
\begin{equation*}
  \alpha_{(m+1)}\eQ_{(m+1)} \varepsilon_\infty = 
  \left(\eK_{(m+1)}-\eK_{(m)}\right)\varepsilon_\infty~,
\end{equation*}
which is solved by
\begin{gather*}
  \eQ_{(m+1)} \varepsilon_\infty = \eta_{(m+1)}
  \varepsilon_\infty~,\quad \text{and}\\
  \eK_{(m+1)}=\eK_{(m)} + \eta_{(m+1)}\alpha_{(m+1)}~.
\end{gather*}
Thus we obtain the full solution for $m+1$.

It is important to stress that not all conditions on the asymptotic
Killing spinors are independent.  Indeed, there are only $n-1$
independent conditions
\begin{equation*}
  \eP_1\varepsilon_\infty = -\eta_{1j}\eP_j\varepsilon_\infty \quad
  , \quad j=2,\dots ,n
\end{equation*}
since it is trivial to show that $\eta_{jj'}=-\eta_{1j}\eta_{1j'}$
for all $2 \leq j\neq j' \leq n$.  Inserting these relations into the
solution for the eigenvalue $\eK$, and solving for $\beta_i$, one
derives the constraint on the parameters determining the
infinitesimal isometry:
\begin{equation}
  \beta_1 = \sum_{j=2}^n\eta_{1j}\beta_j~.
 \label{eq:susyfinal}
\end{equation} 
With all this information, it is straightforward to show that the
original second equation in \eqref{eq:susysol} is satisfied.

As emphasised before, it is not claimed that the above solution is the
most general one.  Indeed, whenever the coefficients $\beta_i$ satisfy
certain relations among them, there may be an enhancement of
supersymmetry.  It is actually very simple to argue when such a
phenomena is going to happen.  Indeed, whenever we have a linear
combination of matrices annihilating a vector, the annihilation still
holds if a subset of the matrices already annihilate the vector by
themselves.  Thus, there are as many solutions as different ways of
decomposing the original linear combination compatible with the
existence of a solution.  Instead of developing the general theory for
arbitrary $n$, we shall just state the results needed in the bulk of
this paper.  For our purposes, $n\leq 5$ (since there are ten
spacelike directions).  It is obvious that $n=1$ breaks supersymmetry
completely.  For $n=2$, the unique solution is the one specified by
\eqref{eq:susysol}.  It corresponds to an $\fsu(2)$ subalgebra.  For
$n=3$, equation \eqref{eq:susysol} is still the general solution,
since all the decompositions into different subsets break
supersymmetry.  This case corresponds to the $\fsu(3)$ subalgebra.
For $n=4$, there is the generic solution given in \eqref{eq:susysol},
corresponding to $\fsu(4)$, but there is also the possibility of
decomposing $n=n_1+n_2=2+2$, which corresponds to the
$\fsu(2)\times\fsu(2)$ subalgebra.  This case arises when
$\beta_1=\eta_2\beta_2$ and $\beta_3=\eta_4\beta_4$.  Finally, for
$n=5$, there is the standard $\fsu(5)$ subalgebra solution
\eqref{eq:susysol}, but also the $\fsu(2)\times\fsu(3)$ one.  The
latter takes place whenever $\beta_1=\eta_2\beta_2$ and
$\beta_3=\eta_4\beta_4 + \eta_5\beta_5$.  It is understood that
depending on the isometry group of the background under consideration,
given a decomposition $n=\sum_i n_i$, there might be more than one
inequivalent configuration associated with it.

\subsection{Explicit formulas for the Kaluza--Klein reduction}
\label{sec:explicitkk}

We now give a general formula for the Kaluza--Klein reduction of the
metric and the four-form.  This is facilitated by going to adapted
coordinates where the Killing vector $\xi$ is simply a translation.
As in \cite{FigSimFlat} we will exhibit $\xi$ as a ``dressed''
translation, which yields at once the required change of coordinates.

\subsubsection{Kaluza--Klein reduction of the metric}
\label{sec:metric}

In this section we give a general formula for the Kaluza--Klein
reduction of the metric.  Let us assume that the eleven-dimensional
metric in the coordinate system $(z,\by)$, where the Killing vector
being used in the reduction is $\xi=\partial_z + \alpha$, can be
written as \footnote{In the bulk of the paper, we shall also deal with
  configurations in which the $z$ coordinate is also either a timelike
  or a lightlike coordinate.  It is straightforward to extend the
  formalism developed below to these cases.}
\begin{equation}
   \label{eq:metans}
   g = \sum_i V_i(\by)^{\alpha_i}ds^2(\EE^i) + V^\gamma(\by)(dz)^2~.
\end{equation}
Ansatz \eqref{eq:metans} includes the backgrounds discussed in this
paper.  Furthermore, $\alpha$ is an affine transformation of the
Cartesian coordinates $(\by)$.  It is important to stress that in
later applications, $\alpha$ will always be a linear combination of
commuting infinitesimal transformations commuting with $\partial_z$.

It is useful to introduce some set of projectors $\PP_i$ satisfying,
for all $i$,
\begin{equation*}
  ds^2(\EE^i) = (d\by)^t\PP_i\eta\,d\by~,
\end{equation*}
where $\eta$ stands for the ten-dimensional Minkowski metric.  As in
\cite{FigSimFlat}, the description of the explicit geometry obtained
through the reduction along the orbits of the Killing vectors is
obtained by working in coordinates $(z,\bx)$ adapted to the Killing
vector, $\xi=\partial_z$.  The explicit change of coordinates is
obtained by noticing that $\xi$ is simply a dressed version of its
translation component
\begin{equation*}
  \xi = U\partial_z U^{-1} \quad \text{where} \quad U=\exp
  (-z\alpha)~. 
\end{equation*}

Thus, defining
\begin{equation}
  \bx = U\by~,
 \label{eq:adapted}
\end{equation}
it follows that $\xi\,\bx=0$, so that $\bx$ are good coordinates for
the space of orbits.  Since $\alpha$ is a linear combination of
infinitesimal Lorentz transformations and translations, its action on
$\by$ can be defined by
\begin{equation}
  \alpha \by = B\by + \bC~,
 \label{eq:data}
\end{equation}
where $B$ is, generically, a $10\times 10$ constant matrix, whereas
$\bC$ is a constant 10-vector taking care of the inhomogeneous part of
the infinitesimal transformation generated by $\alpha$.  Thus,
$\bx(z,\by) = e^{-zB} (\by+ B^{-1}\bC) - B^{-1}\bC$, so that
\begin{equation*}
  d\by = e^{zB} [d\bx + (B\bx+\bC) dz]~.
\end{equation*}

We can now rewrite the metric \eqref{eq:metans} in the adapted
coordinate system $(z,\bx)$, obtaining
\begin{equation*}
  g = \Lambda (dz + A)^2 - \Lambda A^2 + \sum_i
  V_i(\by)^{\alpha_i}(\bx) (d\bx)^t\PP_i\eta\,d\bx~,
\end{equation*}
where
\begin{equation}
   \label{eq:RR1}
   \begin{aligned}[m]
     \Lambda &= V^\gamma(\bx) + \sum_i V_i(\by)^{\alpha_i}(\bx)
     (B\bx + \bC)^t\PP_i\eta (B\bx + \bC) \\
     A &= \Lambda^{-1}\sum_i V_i(\by)^{\alpha_i}(\bx) (B\bx +
     \bC)^t\PP_i\eta d\bx~.
   \end{aligned}
\end{equation}

Using the Kaluza--Klein ansatz \eqref{eq:kkg} we can read off the
dilaton $\Phi = \tfrac{3}{4} \log\Lambda$ and the IIA metric
\begin{equation}
  \label{eq:IIAmetric}
  g = \Lambda^{1/2}\left\{\sum_i
    V_i(\by)^{\alpha_i}(\bx)(d\bx)^t\PP_i\eta \,d\bx - \Lambda
    A^2\right\}~,
\end{equation}
whereas the RR 1-form is given by \eqref{eq:RR1}.

\subsubsection{Kaluza--Klein reduction of the four-form and its dual}
\label{sec:fourform}

We now give a general formula for the Kaluza--Klein reduction of the
four-form $F_4$ and its dual.  In the $(z,\by)$ coordinates we can
write the four-form $F_4$ uniquely as
\begin{equation}
  \label{eq:F4zy}
  F_4(z,\by) = K(z,\by) - dz \wedge L(z,\by)~,
\end{equation}
where $K$ and $L$ are a four-form and three-form, respectively,
without $dz$ components.  Let us change coordinates to $(z,\bx)$.  By
the results of Section~\ref{sec:geometry} we know that the resulting
expression for $F_4$ is given by equation \eqref{eq:kkf}, where the
forms $H_4$ and $H_3$ are basic.  This means that in the adapted
coordinate system $(z,\bx)$ they do not depend explicitly on $z$ nor
do they have any component in $dz$.  We can exploit this fact in order
to give an explicit expression for $H_3$ and $H_4$ in terms of the
forms $K$ and $L$ in \eqref{eq:F4zy}.  The idea is simple: we perform
the explicit change of coordinates in \eqref{eq:F4zy} and write the
result in the form \eqref{eq:kkf}.  To simplify the calculation we set
$z=0$, since we know \emph{a priori} that the resulting forms do not
depend on $z$.  We find that
\begin{equation*}
  K(z,\by(z,\bx))\bigr|_{z=0} = K(0,\bx) + dz \wedge \imath_{B\bx+\bC}
  K(0,\bx)~,
\end{equation*}
and similarly for $L$, where we have used that at $z=0$, $\by(0,\bx)=
\bx$.  Inserting this into \eqref{eq:F4zy} and comparing with
\eqref{eq:kkf} we find
\begin{equation}
  \label{eq:kkf4}
  H_4(\bx) = K(0,\bx) \qquad\text{and}\qquad
  H_3(\bx) = L(0,\bx) - \imath_{B\bx+\bC} K(0,\bx)~.
\end{equation}

The same method also works \emph{mutatis mutandis} for the seven-form
$F_7$ dual to $F_4$, and indeed for any invariant $p$-form.  In some
backgrounds it is more convenient to work with $F_7$ and use the above
method to determine the forms $H_7$ and $H_6$ in \eqref{eq:kkfdual},
from which we can then recover the forms $H_3$ and $H_4$ using the
duality relations \eqref{eq:slaud}.

\section{Kaluza--Klein reductions of the M2-brane}
\label{sec:M2}

In this section we classify the set of M-theory backgrounds obtained
by modding out the M2-brane background by a one-parameter subgroup of
its isometry group and study the smooth supersymmetric Kaluza--Klein
reductions along the orbits of the Killing vectors generating such
subgroups.  We shall first consider the standard M2-brane
configuration in section~\ref{sec:M2l}.  Afterwards, we shall discuss
the M2-brane delocalised along one transverse direction in
section~\ref{sec:M2d}.

\subsection{Supersymmetric reductions of the M2-brane}
\label{sec:M2l}

The M-theory membrane \cite{DS2brane} is described by a metric of the
type \eqref{eq:elemint} with two factors,
\begin{equation}
  \label{eq:mM2}
  g = V^{-2/3} ds^2(\EE^{1,2}) + V^{1/3} ds^2(\EE^8)~,
\end{equation}
where $V = 1 + |Q|/r^6$ with $|Q|$ some positive constant and $r$ the
radial distance in the transverse $\EE^8$.  The $4$-form is given by
\begin{equation}
  \label{eq:FM2}
  F_4 = \dvol(\EE^{1,2}) \wedge d V^{-1}~,
\end{equation}
up to a constant of proportionality.  The Killing spinors are of the
form
\begin{equation}
  \label{eq:SM2}
  \varepsilon = V^{-1/6} \varepsilon_\infty~,
\end{equation}
where $\varepsilon_\infty$ is a constant spinor satisfying
\begin{equation}
  \label{eq:PM2}
  \dvol(\EE^{1,2}) \cdot \varepsilon_\infty = \varepsilon_\infty~.
\end{equation}
The symmetry group is
\begin{equation}
  \label{eq:GM2}
  G = \ISO(1,2) \times \SO(8) \subset \ISO(1,10)~,
\end{equation}
with Lie algebra
\begin{equation}
  \label{eq:gG2}
  \fg = \left(\RR^{1,2} \rtimes \fso(1,2)\right) \times \fso(8)~,
\end{equation}
whence any Killing vector $\xi$ can be decomposed as
\begin{equation}
  \label{eq:KVM2}
  \xi = \tau_\parallel + \lambda_\parallel + \rho_\perp~,
\end{equation}
where, mnemonically, $\tau$, $\lambda$ and $\rho$ denote a
translation, a Lorentz transformation and a rotation, respectively,
and where the subscripts $\parallel$ and $\perp$ refer to vector
fields tangent and perpendicular to the brane worldvolume,
respectively.  We will often omit these subscripts if doing so does
not result in ambiguity.

\subsubsection{Freely-acting spacelike isometries}

The geometrical action of $G$ on the coordinates induces an action on
the Killing spinors which translates into conjugation by $G$ on the
Lie algebra $\fg$.  Using this freedom, we may bring $\lambda$ into a
normal form.  Nontrivial Lorentz transformations in $\fso(1,2)$ come
in three flavours depending on the type of vector in $\EE^{1,2}$ that
they leave invariant.  Therefore either $\lambda=0$ or, via a Lorentz
transformation in the worldvolume of the membrane, it can be brought
to one of the following three normal forms:
\begin{enumerate}
\item $\lambda$ fixes a timelike vector:
  \begin{equation}
    \label{eq:LM2t}
    \lambda = \theta R_{12}\qquad \theta \neq 0~,
  \end{equation}
\item $\lambda$ fixes a spacelike vector:
  \begin{equation}
    \label{eq:LM2s}
    \lambda = \beta B_{02} \qquad \beta \neq 0~,
  \end{equation}
\item $\lambda$ fixes a null vector:
  \begin{equation}
    \label{eq:LM2n}
    \lambda = N_{+2}~,
  \end{equation}
\end{enumerate}
where $R_{12}$ is the generator of infinitesimal rotations in the
$(12)$-plane, $N_{+2}$ is the generator of infinitesimal null rotation
in the $x^2$ direction with light-cone coordinates $x^\pm = (x^1 \pm
x^0)/\sqrt{2}$, and where $B_{02}$ is the generator of infinitesimal
boosts along the $x^2$ direction.

We can now use the freedom to change origin in the worldvolume of the
brane---equivalently, to conjugate by the translation subgroup in
$\ISO(1,2)$---in order to bring $\tau$ to a normal form.  If
$\lambda=0$, $\tau$ does not change; but in the other normal forms we
can bring $\tau$ to the following: (1) $\tau \propto \d_0$, (2) $\tau
\propto \d_1$, and (3) $\tau \propto \d_-$.  It is easy to see that in
case (2) there are points outside the brane horizon where $\xi$ is
timelike, hence this case is ruled out.  It is also easy to see that
in case (3) we must have $\tau=0$ for precisely the same reasons.
This narrows down the possibilities to three cases:
\begin{enumerate}
\item[(a)] $\xi = \tau + \rho_\perp$, with $\tau$ so far
  unconstrained;
\item[(b)] $\xi = \tau + \rho_\parallel + \rho_\perp$, with $\tau$
  timelike and orthogonal to $\rho_\parallel$; and
\item[(c)] $\xi = \nu_\parallel + \rho_\perp$, with $\nu_\parallel$ a
  null rotation.
\end{enumerate}
In the above expressions $\rho_\perp$ is an infinitesimal rotation in
$\fso(8)$ and hence can be brought to a normal form
\begin{equation}
  \label{eq:rhoperpnormalform}
  \rho_\perp = \theta_1 R_{34} + \theta_2 R_{56} + \theta_3 R_{78}
  + \theta_4 R_{9\natural}~.
\end{equation}
We must distinguish between two cases: either one or more of the
$\theta$s vanish or none does.  If some $\theta$s vanish, only case
(a) above gives rise to a spacelike $\xi$ and in that case we must
take $\tau$ to be spacelike.  If none of the $\theta$s vanish, we
have more possibilities: case (c) can occur, and so can cases (a) and
(b) provided that $\|\tau\|^2$ is not too negative.  To understand
this, let us compute the norm of $\xi$ in these cases.

Consider a Killing vector of the form $\xi = \tau + \rho_\parallel +
\rho_\perp$, where $\tau$ is orthogonal to $\rho_\parallel$ and where
we allow $\rho_\parallel$ to be zero.  In this way we can discuss
cases (a) and (b) simultaneously.  The norm of this vector field
relative to the membrane metric is given by
\begin{equation*}
  \|\xi\|^2 = V^{-2/3}\left( \|\tau\|^2_\infty +
  \|\rho_\parallel\|^2_\infty \right) + V^{1/3}
  \|\rho_\perp\|^2_\infty~,
\end{equation*}
where $\|\cdot\|_\infty$ is the norm relative to the flat metric.  The
tangent vector $\rho_\perp$ at a point a distance $r>0$ away from the
membrane is tangent to the transverse sphere of radius $r$ through
that point.  This means that the norm at that point is given by
\begin{equation*}
  \|\rho_\perp\|^2_\infty = r^2 \|\rho_\perp\|^2_S~,
\end{equation*}
where $\|\cdot\|_S$ is the norm relative to the round metric on the
sphere of unit radius.  Since the sphere is compact,
$\|\rho_\perp\|_S$ acquires a maximum and a minimum, whence
\begin{equation}
  \label{eq:spherenorminequalities}
  m^2 r^2 \leq \|\rho_\perp\|^2_\infty \leq M^2 r^2~,
\end{equation}
for some non-negative real numbers $m \leq M$.  In fact, it is easy to
see that for $\rho_\perp$ given in \eqref{eq:rhoperpnormalform}, these
numbers are given by
\begin{equation*}
  m^2 = \min_i \theta_i^2 \qquad\text{and}\qquad M^2 = \max_i
  \theta_i^2~.
\end{equation*}
It should be stressed that both inequalities in
\eqref{eq:spherenorminequalities} are sharp, since there are
directions (i.e., points in the sphere) where the inequalities are
saturated.  The lower bound $m$ is positive if and only if
$\rho_\perp$ does not leave any directions invariant.  This is
possible for the M2-brane since the transverse sphere is
odd-dimensional, for on an even-dimensional sphere every (continuous)
vector field has a zero (in fact, two) and hence $m=0$ in those cases.
The norm of $\xi$ is then bounded below by
\begin{equation*}
  \|\xi\|^2 \geq V^{-2/3} \|\tau\|_\infty^2 + V^{-2/3}
  \|\rho_\parallel\|^2 + V^{1/3} r^2 m^2~,
\end{equation*}
and again this inequality is sharp.  The rotation $\rho_\parallel$ has
a zero at the `origin' of the membrane worldvolume.  We can thus
simplify the above bound even further:
\begin{equation*}
  \|\xi\|^2 \geq V^{-2/3} \|\tau\|_\infty^2 + V^{1/3} r^2 m^2~,
\end{equation*}
which is still sharp.  The right-hand side of the above equation is a
function of $r$ and it will attain a minimum at a critical radius
$r_0$, which is zero if $\|\tau\|_\infty^2 \geq 0$ and positive if
$\|\tau\|_\infty^2 < 0$.  Indeed, the lower bound for the norm of
$\xi$ is given by the function
\begin{equation*}
  f(r) = V^{-2/3} \|\tau\|_\infty^2 + V^{1/3} r^2 m^2~,
\end{equation*}
whose derivative is given by
\begin{equation*}
  f'(r) = V^{-5/3} \frac{2 |Q|}{r^7} \left( 2 \|\tau\|_\infty^2 + m^2
  r^2 \left( 1 + \frac{r^6}{|Q|} \right) \right)~.
\end{equation*}
This function has a critical point at $r=0$ and at the positive root
$r_0$ of the equation
\begin{equation*}
  V(r) r^8 = - \frac{2 |Q|}{m^2} \|\tau\|_\infty^2~,
\end{equation*}
should such a root exist.  For $\|\tau\|_\infty^2\geq 0$, no such root
exists and the minimum of $f$ is at $r=0$, whereas for
$\|\tau\|_\infty^2<0$, the minimum is at $r_0$.  In any case we have
the bound
\begin{equation*}
  \|\xi\|^2 \geq V^{-2/3}(r_0) \|\tau\|_\infty^2 + V^{1/3}(r_0) r_0^2
  m^2~,
\end{equation*}
which is still sharp.  The right-hand side is positive for all
\begin{equation}
  \label{eq:muM2}
  \|\tau\|_\infty^2 > - \tfrac32 m^2 (2|Q|)^{1/3}~.
\end{equation}
This means that we can allow for $\tau$ to be timelike (but not too
much) and still obtain a spacelike Killing vector.  If $\rho_\perp$
fixes some directions, so that $m=0$, we see that $\tau$ must be
spacelike.

In summary, the following Killing vectors in $\fg$ are spacelike:
\begin{enumerate}
\item[(A)] $\xi = \tau_\parallel + \rho_\parallel + \rho_\perp$, with
  $\rho_\perp$ without fixed directions and $\tau$ obeying a
  constraint on the norm: $\|\tau\|_\infty^2 > - \mu^2$, where $\mu$
  can be read from equation \eqref{eq:muM2}, and where we can also
  allow for $\rho_\parallel = 0$ in this case;
\item[(B)] $\xi = \tau_\parallel + \rho_\perp$, with $\rho_\perp$
  fixing some directions and $\tau$ spacelike;
\item[(C)] $\xi = \nu_\parallel + \rho_\perp$, with $\rho_\perp$
  without fixed directions.
\end{enumerate}

It is clear that in cases (A) (with $\tau \neq 0$) and (B), $\xi$
integrates to a free action of a subgroup $\RR \subset G$, since $\xi$
contains a translation.  The absence of translations in case (C) makes
it different from (A) and (B). Even though the action is locally free
(for $r>0$), one can prove that there are points with nontrivial
stabilisers, so that the action is not free and the quotient is
therefore singular.  Consider, for example, the point $P$ with
coordinates $x^\pm = x^2 = x^3 = \cdots = x^9 = 0$ and $x^\natural =
1$.  The orbit of this point under the action generated by the Killing
vector $\xi= \nu_\parallel + \rho_\perp$, with $\rho_\perp$ given by
\eqref{eq:rhoperpnormalform} with all $\theta$s different from zero,
is $x^\pm = x^2 = x^3 = \cdots x^8 = 0$, $x^9(t) = \cos \theta_4 t$
and $x^\natural(t) = \sin \theta_4 t$.  As a result, the point $P$ is
mapped to itself by those points $t = 2\pi n/\theta_4$, for any
$n\in\ZZ$.  This defines a subgroup isomorphic to $\ZZ$ in the $\RR$
subgroup generated by $\xi$.  Reducing by $\xi$ would therefore result
in singularities outside the horizon and so we discard it and with it
case (C).  A very similar argument would allow us to discard case (A)
with $\tau=0$.  This leaves us with two possibilities for a
freely-acting spacelike vector field: case (A) with $\tau \neq 0$ and
case (B).

\subsubsection{Moduli space of smooth reductions}

Let us now identify more precisely the moduli space of smooth
reductions in each of these cases.

In case (A) we must distinguish between two cases, depending on
whether or not $\rho_\parallel$ vanishes.  If $\rho_\parallel \neq 0$,
then by changing the origin we can put $\tau = a \d_0$, where $0< a^2
< \mu^2$, with $\mu$ given in \eqref{eq:muM2}.  In summary, $\xi$ can
be written in the following form
\begin{equation}
  \label{eq:xiM2(A)}
  \xi = a \d_0 + \theta_1 R_{12} + \theta_2 R_{34} +
  \theta_3 R_{56} + \theta_4 R_{78} + \theta_5 R_{9\natural}~,
\end{equation}
where none of the $\theta$s vanish and $0<|a|<\mu$.  The moduli
space is obtained from this space by projectivising and by some
discrete identifications coming from the action of the Weyl group.
Its dimension is therefore five-dimensional, and we will see below
that supersymmetry will select a four-dimensional locus.

If $\rho_\parallel = 0$, the causal character of the translation is
not fixed, although the norm constraint \eqref{eq:muM2} is still in
force.  We must distinguish between three cases, depending on whether
$\tau$ is timelike, spacelike or null.  If $\tau$ is timelike we can
choose coordinates such that $\tau = a \d_0$ with $0<|a|<\mu$.
Similarly, if $\tau$ is spacelike, it can be arranged that $\tau =
a\d_1$ with $a\neq 0$ but otherwise unconstrained.  In either case we
have a four-dimensional moduli space.  Finally if $\tau$ is null,
coordinates can be chosen where $\tau = \d_+$ (no free parameter!)
whence the moduli space is three-dimensional.  In all cases,
supersymmetry will select a codimension-one locus.

Finally, in case (B) we can arrange for $\tau = a \d_1$ and hence
\begin{equation}
  \label{eq:xiM2(B)}
  \xi = a \d_1 + \theta_2 R_{34} + \theta_3 R_{56} +
  \theta_4 R_{78}~,
\end{equation}
where $a\neq 0$ is not otherwise constrained.  The moduli space in
this case is only three-dimensional, with supersymmetry selecting
a two-dimensional locus.

\subsubsection{Supersymmetry}

Now we impose the condition that the reduction preserves
supersymmetry.  As explained in Section~\ref{sec:method},
supersymmetry only constrains the rotation component of the Killing
vector $\xi$ to lie in the isotropy of some spinor satisfying
\eqref{eq:PM2}.  In all the cases above, the rotation component of
$\xi$ takes the general form
\begin{equation}
  \label{eq:so10rotation}
  \rho = \theta_1 R_{12} + \theta_2 R_{34} + \theta_3 R_{56} +
  \theta_4 R_{78} + \theta_5 R_{9\natural}~,
\end{equation}
with perhaps some of the $\theta$s vanishing.  We find it convenient
to treat the general case first, which will give some relations
between the $\theta$s and then impose any further conditions.

As discussed in Appendix~\ref{sec:groups}, the condition that $\rho$
annihilates a Killing spinor is equivalent to $\rho$ being annihilated
by some weight in the subspace $S_0$ of the half-spin representation
defined by \eqref{eq:PM2}.  These weights are given in equation
\eqref{eq:weightsM2} with the negative sign, according to our
conventions.  Therefore a weight will annihilate $\rho$ if and only if
the $\theta_i$ belong to the union of the following eight hyperplanes
\begin{equation}
  \label{eq:hyperplanesM2}
  \sum_{i=1}^5 \mu_i \theta_i = 0\qquad\text{where $\mu_i^2=1$ with
    $\mu_2\cdots\mu_5 = -1$.}
\end{equation}
Notice that if $\rho$ is annihilated by $\mu$ it is also annihilated
by $-\mu$ which is also a weight in the representation $S_0$: this
explains why there are only eight hyperplanes in the above family.  If
$\rho$ belongs to one and only one such hyperplane the amount of
supersymmetry preserved by such a reduction is $\nu= \frac1{16}$.
This corresponds to $\rho$ belonging to an $\fsu(5)$ subalgebra of
$\fso(1,10)$.\footnote{More precisely, the phrase ``$\rho$ belongs to
  an $\fsu(5)$ subalgebra'' is to be interpreted as meaning that
  $\rho$ belongs to the intersection of an $\fsu(5)$ subalgebra of
  $\fso(1,10)$ with $\fg$.  In general, the rotations which give rise
  to supersymmetric reductions belong to the intersection of a spinor
  isotropy subalgebra of $\fso(1,10)$ with $\fg$, but we choose to
  organise the results in terms of the spinor isotropy subalgebra,
  e.g., $\fsu(5)$ in this case.}

There is enhancement of supersymmetry if $\rho$ belongs to the
intersection of two or more hyperplanes in \eqref{eq:hyperplanesM2}.  
Assuming that none of the $\theta$s are
zero, we can only have simple intersections between two hyperplanes,
e.g., $\theta_4 = \theta_5$ and $\theta_1 + \theta_2 + \theta_3 = 0$.
This corresponds to a $\rho$ which belongs to an $\fsu(2) \times
\fsu(3)$ subalgebra.  In this case there are four weights which
annihilate $\rho$ and hence the fraction of the supersymmetry
preserved by the reduction is enhanced to $\nu = \frac18$.

If $\rho_\parallel = 0$ then $\theta_1=0$, and if the remaining
$\theta$s do not vanish, the supersymmetric locus is given by the
intersection of the hyperplane $\theta_1 = 0$ with the hyperplanes in
\eqref{eq:hyperplanesM2}.  This is a family of four hyperplanes in the
$\theta_1=0$ subspace, given by the equations
\begin{equation*}
  \sum_{i=2}^5 \mu_i \theta_i = 0 \qquad \text{where $\mu_i^2=1$ and
    $\mu_2\cdots\mu_5 = -1$.}
\end{equation*}
A generic $\rho$ in one of these hyperplanes is annihilated by four
weights, hence the reduction preserves a fraction $\nu = \frac18$ of
the supersymmetry, corresponding to $\rho$ in an $\fsu(4)$ subalgebra.
There is again supersymmetry enhancement at the intersection of these
hyperplanes, for example, $\theta_4 = \theta_5$ and $\theta_1 +
\theta_2 = 0$, which corresponds to $\rho$ in an $\fsp(1) \times
\fsp(1)$ subalgebra.  The reduction now preserves a fraction $\nu =
\frac14$ of the supersymmetry.

Finally, in case (B) with $\rho_\parallel =0$ and $\rho_\perp$ fixing
some directions, we can choose coordinates so that $\theta_1 = 0 =
\theta_2$.  The supersymmetric locus consists of the intersection of
the hyperplanes \eqref{eq:hyperplanesM2} with the hyperplanes
$\theta_1=0$ and $\theta_2=0$.  The resulting four hyperplanes are
described by the equations
\begin{equation*}
  \sum_{i=3}^5 \mu_i \theta_i = 0 \qquad \text{where $\mu_i^2=1$.}
\end{equation*}
A generic $\rho$ in this locus is annihilated by four weights, hence
the reduction preserves a fraction $\nu = \frac18$ of the
supersymmetry, and $\rho$ belongs to an $\fsu(3)$ subalgebra.  We are
free to specialise to any intersection of the above planes: simple
intersections correspond to setting another one of the $\theta_i$ to
zero, or equivalently to $\rho$ lying in an $\fsu(2)$ subalgebra.
This enhances the supersymmetry to a fraction $\nu=\frac14$.  Finally,
the only point in more than two hyperplanes is the origin, whence
$\rho=0$ and hence the reduction preserves all the supersymmetry of
the membrane, namely a fraction $\nu = \half$.  These results are
summarised in Table~\ref{tab:M2susy}.

As mentioned in the introduction, we could consider a discrete
subgroup $\Gamma_0\subset\Gamma$ such that $\Gamma/\Gamma_0$ is
compact.  The corresponding eleven-dimensional configurations
classified above, $\eM_{\text{M2}}/\Gamma_0$, correspond to new smooth
supersymmetric vacua which are asymptotic to $\RR^{1,10}/\Gamma_0$.
In the particular case in which $\xi = \tau_\parallel + \rho$,
$\tau_\parallel$ being an spacelike translation,
$\eM_{\text{M2}}/\Gamma_0$ corresponds to a stack of M2-branes in an
eleven-dimensional fluxbrane vacua.  We shall next concentrate on the
Kaluza--Klein reductions.

\begin{table}[h!]
  \begin{center}
    \setlength{\extrarowheight}{5pt}
    \begin{tabular}{|>{$}c<{$}|>{$}c<{$}|>{$}c<{$}|>{$}c<{$}|}
      \hline
      \multicolumn{1}{|c|}{Translation} & \text{Subalgebra} & \nu &
      \dim\\
      \hline
      \hline
      & \fsu(4) & \frac1{16} & 3\\
      & \fsu(3) & \frac18 & 2\\
      a \d_1 & \fsp(1) \times \fsp(1) & \frac18 & 2\\
      a\neq 0 & \fsu(2) & \frac14 & 1\\
      & \{0\} & \frac12& 0 \\[3pt]
      \hline
      \d_+ & \fsu(4) & \frac1{16} & 2\\
      & \fsp(1) \times \fsp(1) & \frac18 & 1\\[3pt]
      \hline
      & \fsu(5) & \frac1{32} & 4\\
      a\d_0& \fsu(2) \times \fsu(3) & \frac1{16} & 3\\
      0<|a|<\mu& \fsu(4) & \frac1{16} & 3\\
      & \fsp(1) \times \fsp(1) & \frac18 & 2 \\[3pt]
      \hline
    \end{tabular}
    \vspace{8pt}
    \caption{Supersymmetric reductions of the M2 brane.  We indicate
      the form of the translation, the spinor isotropy subalgebra to
      which the rotation belongs, the fraction $\nu$ of the
      supersymmetry preserved and the dimension of the corresponding
      stratum of the moduli space $\eM$ of supersymmetric reductions.}
    \label{tab:M2susy}
  \end{center}
\end{table}

\subsubsection{Explicit reductions}

We shall start by studying the reductions not involving timelike
translations.  The Killing vector can thus be written as $\xi =
\partial_z + \lambda$, where $z$ stands for a longitudinal direction,
i.e. $y^2$, and $\lambda$ stands for the infinitesimal rotation in the
space transverse to the brane
\begin{equation*}
  \begin{aligned}[m]
    \lambda &= \theta_1 (y^3\partial_4
    -y^4\partial_3) + \theta_2 (y^5\partial_6-y^6\partial_5) \\
    & + \theta_3 (y^7\partial_8-y^8\partial_7) 
    + \theta_4 (y^9\partial_\natural-y^\natural\partial_9)~.
  \end{aligned}
\end{equation*} 

The constant matrix $B$ introduced in \eqref{eq:data} is an $8\times
8$ matrix which can be written as
\begin{equation}
  \label{eq:Bmatrixm2a}
  B= 
  \begin{pmatrix}
    0 & -\theta_1 & 0 & 0 & 0 & 0 & 0 & 0 \\
    \theta_1 & 0 & 0 & 0 & 0 & 0 & 0 & 0 \\
    0 & 0 & 0 & -\theta_2 & 0 & 0 & 0 & 0 \\
    0 & 0 & \theta_2 & 0 & 0 & 0 & 0 & 0 \\
    0 & 0 & 0 & 0 & 0 & -\theta_3 & 0 & 0 \\
    0 & 0 & 0 & 0 & \theta_3 & 0 & 0 & 0 \\
    0 & 0 & 0 & 0 & 0 & 0 & 0 & -\theta_4 \\
    0 & 0 & 0 & 0 & 0 & 0 & \theta_4 & 0 \\
  \end{pmatrix}~,
\end{equation}
in the basis $\{x^3,x^4,\dots\,x^9,x^\natural\}$ spanned by the
adapted coordinates defined in \eqref{eq:adapted}.  It is
straightforward to derive the ten-dimensional metric
\begin{multline*}
  g = \tilde{\Lambda}^{1/2}\left\{V^{-1}ds^2(\EE^{1,1}) 
  + ds^2(\EE^8)\right\} \\
  - \tilde{\Lambda}^{-1/2}\,V\left(\theta_1 \omega^{34} + 
  \theta_2 \omega^{56} + \theta_3 \omega^{78} + \theta_4 \omega^{9\natural}
  \right)^2~,
\end{multline*}
where we have introduced the notation $\omega^{ij} := x^i dx^j - x^j dx^i$.
The RR 1-form $A_{1}$, NS-NS 3-form field strength $H_{3}$ and dilaton 
$\Phi$ are listed below:
\begin{equation*}
  \begin{aligned}[m]
    A_1 &= \tilde{\Lambda}^{-1}\,V\,\left(\theta_1 \omega^{34} + 
    \theta_2 \omega^{56} + \theta_3 \omega^{78} + \theta_4 \omega^{9\natural}
    \right) \\
    H_3 &= \dvol\EE^{1,1}\wedge dV^{-1} \\
    \Phi &= \tfrac{1}{2}\log \left(\tilde{\Lambda}^{3/2}\cdot
    V^{-1}\right)~,
  \end{aligned}
\end{equation*}
whereas the RR 4-form $H_4$ field strength vanishes.  The
configuration depends on an scalar function $\tilde{\Lambda}$ which is
defined in terms of the scalar function $\Lambda$ appearing in
Section~\ref{sec:explicitkk}, by
\begin{equation*}
\Lambda = V^{-2/3}\cdot \tilde{\Lambda}~,
\end{equation*}
and equals
\begin{multline*}
  \tilde{\Lambda} = 1 + V\left\{\left(\theta_1\right)^2\left[(x^3)^2 +
      (x^4)^2\right] + \left(\theta_2\right)^2\left[(x^5)^2
      +(x^6)^2\right]
  \right. \\
  \left. +\left(\theta_3\right)^2\left[(x^7)^2 +(x^8)^2\right] +
    \left(\theta_4\right)^2\left[(x^9)^2
      +(x^\natural)^2\right]\right\}~.
\end{multline*}

Notice that for arbitrary values of the angles $\theta_i$, the string
coupling constant blows up, irrespectively of the direction, at large
distances, whereas it is bounded from above by the constant
$\left(\sum_i (\theta_i)^2 \right)^{3/4} Q^{1/4}$ at $r\to 0$.  Thus,
it is always possible to have a weakly coupled region close to the
origin, whereas as we move away from it, the M-theory description
becomes more reliable.

As already discussed before, for arbitrary values of the deformation
parameters $\{\theta_i\}$, the configuration would break supersymmetry
completely, and its interpretation would be in terms of composites
configurations involving fundamental strings lying in the $x^1$
direction at $r=0$ and, generically, four different F7-branes
lying at $x^3=x^4=0$, $x^5=x^6=0$, $x^7=x^8=0$ and $x^9=x^\natural=0$,
respectively.  It is the presence of the F7-branes that breaks
supersymmetry completely.

\begin{table}[h!]
  \begin{center}
    \setlength{\extrarowheight}{3pt}
    \begin{tabular}{|>{$}c<{$}|c|>{$}c<{$}|}
      \hline
      \nu & Object & \multicolumn{1}{c|}{Subalgebra}\\
      \hline
      \hline
      \frac14 & FA$\parallel$F5 & \fsu(2)\\
      \frac18 & FA$\parallel$F3 & \fsu(3)\\
      \frac18 & FA$\parallel$F1 & \fsp(2)\\
      \frac{1}{16} & FA$\parallel$F1 & \fsu(4)\\[3pt]
      \hline
    \end{tabular}
    \vspace{8pt}
    \caption{Supersymmetric configurations of fundamental strings (FA)
      and fluxbranes}
    \label{tab:M2}
  \end{center}
\end{table}

On the other hand, there are five different types of supersymmetric
configurations
\begin{itemize}
\item[(1)] If $\theta_i=0$ for all $i$, this is the standard type IIA
  configuration describing fundamental strings streching 
  along the $x^1$ direction at $r=0$ and preserving $\nu=1/2$ of the
  spacetime supersymmetry.

\item[(2)] Setting $\theta_3=\theta_4=0$ and $\theta_1=\eta_2\theta_2$
  corresponds to fundamental strings streching along the
  $x^1$ direction at $r=0$ and lying on an F5-brane that sits
  on $x^3=x^4=x^5=x^6=0$.  This configuration preserves $\nu=1/4$ of
  the spacetime supersymmetry.
  
\item[(3)] Setting $\theta_4=0$ and $\theta_1=\eta_2\theta_2 +
  \eta_3\theta_3$ corresponds to fundamental strings
  streching along the $x^1$ direction at $r=0$ and lying on an
  F3-brane that sits on $x^3=x^4=x^5=x^6=x^7=x^8=0$.  This
  configuration preserves $\nu=1/8$ of the spacetime supersymmetry.
  
\item[(4)] Setting $\theta_1=\eta_2\theta_2$ and
  $\theta_3=\eta_4\theta_4$ corresponds to fundamental strings and
  $\tfrac14$-BPS fluxstrings lying on $x^1$ at $r=0$.  It preserves
  $\nu=1/8$.
  
\item[(5)] Setting $\theta_1=\eta_2\theta_2
  +\eta_3\theta_3+\eta_4\theta_4$, corresponds to fundamental strings
  and fluxstrings lying on $x^1$ at $r=0$.  It preserves $\nu=1/16$.
\end{itemize}

The allowed supersymmetric configurations are summarised in
Table~\ref{tab:M2}.

If one sets the charge of the original M2-brane to zero, one recovers
the corresponding fluxbrane configurations reviewed in \ref{sec:flat}.
These do have some notion of flux associated with the integral over
the transverse sections to the fluxbrane of $F_2=dA_1$, or wedge
products of it.  It is natural to compute this flux, when fundamental
strings are switched on.  We shall concentrate on F5-branes for
simplicity.  Notice that due to the presence of the fundamental
strings, the RR 1-form potential depends on the point
$\bx=(x^3,x^4,x^5,x^6)$, but it is still invariant along the $x^1$
direction.  Thus, the flux may depend on the point where we fix the
transverse section along which we compute it.  Fixing this point
$\bx$, and computing the integral
\begin{equation*}
  \frac{1}{4\pi}\int_{\RR^4(\bx)} F_2\wedge F_2~,
\end{equation*}
afterwards, one can check that the flux equals $\theta^{-2}$ (as in
flat case) everywhere except at $\bx=0$, where the string lies.  In
that point, the flux vanishes.  This absence of flux at $\bx=0$ seems
to be consistent with the fact that there is no moduli associated with
relative motions among fundamental strings and F5-branes.  Indeed, a
probe computation shows that fundamental strings do always feel a
force when they sit away from $r=0$.

We shall next discuss the Kaluza-Klein reductions involving
translations which are null.  From our general discussion on
freely-acting spacelike Killing vectors $\xi$, we already know the
only allowed possibilities are those in which
\begin{equation*}
  \xi= \partial_+ + \lambda~,
\end{equation*}
where $\lambda$ is a rotation acting on the transverse space to the
brane and without fixing any direction.  It is thus the same as the
one used previously, but this time no $\theta_i$ are allowed to
vanish.  By using our general formalism, the constant matrix B is
again given by \eqref{eq:Bmatrixm2a}.  This determines the
ten-dimensional metric to be
\begin{multline*}
  g = \tilde{\Lambda}^{1/2}\left(V^{-1/2}(dx^2)^2 +
  V^{1/2}ds^2(\EE^8)\right)\\
  -\tilde{\Lambda}^{-1/2}\,V^{1/2}\left( \theta_1 \omega^{34} + 
  \theta_2 \omega^{56} + \theta_3 \omega^{78} + \theta_4 \omega^{9\natural}
  + V^{-1}\,dx^- \right)^2~,
\end{multline*}
where $\omega^{ij} := x^i dx^j - x^j dx^i$.
On the other hand, there are non-trivial RR 1-form $A_{1}$, 
NS-NS 3-form field strength $H_{3}$ and dilaton $\Phi$ which are
listed below:
\begin{equation*}
  \begin{aligned}[m]
    A_1 &= \tilde{\Lambda}^{-1}\left\{\theta_1 \omega^{34} + 
    \theta_2 \omega^{56} + \theta_3 \omega^{78} + \theta_4 \omega^{9\natural}
    + V^{-1}\,dx^- \right\} \\
    H_3 &= dx^-\wedge dx^2 \wedge dV^{-1} \\
    \Phi &= \tfrac{3}{4}\log \left(\tilde{\Lambda}\cdot
      V^{1/3}\right)~.
  \end{aligned}
\end{equation*}
The above type IIA configuration depends on an scalar function
$\tilde{\Lambda}$ which was again defined in terms of the scalar
function $\Lambda$ appearing in the general discussion section, by
\begin{equation*}
\Lambda = V^{1/3}\cdot \tilde{\Lambda}~,
\end{equation*}
and equals
\begin{multline*}
  \tilde{\Lambda} = \left(\theta_1\right)^2\left[(x^3)^2 +
    (x^4)^2\right] + \left(\theta_2\right)^2\left[(x^5)^2
    +(x^6)^2\right] \\
  +\left(\theta_3\right)^2\left[(x^7)^2 +(x^8)^2\right] +
  \left(\theta_4\right)^2\left[(x^9)^2 +(x^\natural)^2\right]~.
\end{multline*}

As before, the string coupling constant blows up at large distances
but is bounded from above by $\left(\sum_i (\theta_i)^2
\right)^{3/4}\,Q^{1/4}$ at $r\to 0$.

For arbitrary values of the deformation parameters $\theta_i$
($\theta_i\neq 0$), the configuration would break supersymmetry
completely.  It is only when $\theta_1=\eta_2\theta_2 + \eta_3\theta_3
+ \eta_4\theta_4$ or $\theta_1=\eta_2\theta_2$ and $\theta_3
=-\eta_3\eta_4\theta_4$, that the above configuration preserves
$\nu=1/16$ or $\nu=1/8$, respectively.  We do not have a physical
interpretation for this set of configurations, and even though they
were obtained by Kaluza-Klein reduction along the orbits of some
Killing vectors which are spacelike everywhere, we were not able to
prove whether these spacetimes have no causal singularities.

Let us finally move to the third possibility, the one involving
timelike translations.  The Killing vector is given by
$\xi=a\partial_0 + \lambda$, $\lambda$ standing for the spacetime
rotation
\begin{equation*}
  \begin{aligned}[m]
    \lambda & = \theta_1\left(x^1\partial_2 -x^2\partial_1\right)
    + \theta_2\left(x^3\partial_4 - x^4\partial_3\right) \\
    & + \theta_3\left(x^5\partial_6 - x^6\partial_5\right) +
    \theta_4\left(x^7\partial_8 - x^8\partial_7\right) +
    \theta_5\left(x^9\partial_\natural - x^\natural\partial_9\right)~,
  \end{aligned}
\end{equation*}
where the timelike translation parameter is bound by $0<|a| < \mu$, 
with $\mu^2$ given by \eqref{eq:muM2}.

In this case, the constant matrix $B$ is a $10\times 10$ matrix, which
in the basis $\{x^1,x^2,\dots\,x^9,x^\natural\}$ can be written as
\begin{equation}
  \label{eq:Bmatrixm2b}
  B= 
  \begin{pmatrix}
    0 & -\theta_1 & 0 & 0 & 0 & 0 & 0 & 0 & 0 & 0 \\
    \theta_1 & 0 &  0 & 0 & 0 & 0 & 0 & 0 & 0 & 0 \\
    0 & 0 & 0 & -\theta_2 & 0 & 0 & 0 & 0 & 0 & 0 \\
    0 & 0 & \theta_2 & 0 & 0 & 0 & 0 & 0 & 0 & 0 \\
    0 & 0 & 0 & 0 & 0 & -\theta_3 & 0 & 0 & 0 & 0 \\
    0 & 0 & 0 & 0 & \theta_3 & 0 & 0 & 0 & 0 & 0 \\
    0 & 0 & 0 & 0 & 0 & 0 & 0 & -\theta_4 & 0 & 0 \\
    0 & 0 & 0 & 0 & 0 & 0 & \theta_4 & 0 & 0 & 0 \\
    0 & 0 & 0 & 0 & 0 & 0 & 0 & 0 & 0 & -\theta_5 \\
    0 & 0 & 0 & 0 & 0 & 0 & 0 & 0 & \theta_5 & 0 \\
  \end{pmatrix}~.
\end{equation}
The corresponding type IIA configurations have a ten-dimensional
metric given by\footnote{We are grateful to Hannu Rajaniemi for
  spotting a small error in a previous version of this formula.}
\begin{multline*}
  g = \tilde{\Lambda}^{1/2}\left\{V^{-1}ds^2(\EE^{2}) 
    + ds^2(\EE^8)\right\} \\
  - \tilde{\Lambda}^{-1/2}V^{-1} \left[\theta_1 \omega^{12} +
    V\left\{\theta_2 \omega^{34} + \theta_3 \omega^{56} + \theta_4
      \omega^{78} + \theta_5 \omega^{9\natural} \right\}\right]^2~,
\end{multline*}
where $\omega^{ij} := x^i dx^j - x^j dx^i$.  
The RR 1-form $A_{1}$, NS-NS 3-form field strength $H_{3}$ and
dilaton $\Phi$ are given by
\begin{equation*}
  \begin{aligned}[m]
    A_1 &= \tilde{\Lambda}^{-1}\left\{\theta_1 \omega^{12} +
    V\left\{\theta_2 \omega^{34} + \theta_3 \omega^{56} + \theta_4
    \omega^{78} + \theta_5 \omega^{9\natural} \right\}\right\} \\
    H_3 &= a\dvol\EE^{2}\wedge dV^{-1} \\
    \Phi &= \tfrac{1}{2}\log \left(\tilde{\Lambda}^{3/2}\cdot
    V^{-1}\right)~.
  \end{aligned}
\end{equation*}
The configuration depends on an scalar function $\tilde{\Lambda}$
which is defined in terms of the scalar function $\Lambda$ appearing
in the general discussion section, by
\begin{equation*}
\Lambda = V^{-2/3}\cdot \tilde{\Lambda}~,
\end{equation*}
and equals
\begin{footnotesize}
  \begin{multline*}
    \tilde{\Lambda} = -a^2 + \left(\theta_1\right)^2 \left[(x^1)^2 +
      (x^2)^2\right] + V\left\{\left(\theta_2\right)^2
      \left[(x^3)^2 + (x^4)^2\right]   \right. \\
    \left. + \left(\theta_3\right)^2\left[(x^5)^2 +(x^6)^2\right] +
      \left(\theta_4\right)^2\left[(x^7)^2 +(x^8)^2\right] +
      \left(\theta_5\right)^2\left[(x^9)^2
        +(x^\natural)^2\right]\right\}~.
  \end{multline*}
\end{footnotesize}

For generic values of the five parameters $\{\theta_i\}$,
$i=1,2,\dots,5$, the above configuration breaks supersymmetry.  There
are several loci in this five-dimensional parameter space where
supersymmetry is restored.  If the rotation along the M2-brane is
non-vanishing, $\theta_1\neq 0$, there are two possibilities to be
discussed:
\begin{itemize}
\item[(1)] $\sum_i \mu_i\theta_i = 0$ such that $\mu_i^2=1$.  The full
  rotation belongs to the $\fsu(5)$ spinor isotropy subalgebra.  This
  configuration preserves $\nu=1/32$ of the spacetime supersymmetry.
  \item[(2)] If the rotation belongs to the $\fsu(2)\times\fsu(3)$
  spinor isotropy subalgebra, there are two subcases to be considered
  due to the isometries of the starting M2-brane configuration.
  Indeed, 
  \begin{itemize}
  \item[(2.1)] If the rotation on the brane $(\theta_1)$ belongs to
    the $\fsu(2)$ subalgebra.  In this case,
    $\mu_1\theta_1+\mu_2\theta_2=0$ and $\sum_{i=3,4,5}
    \mu_i\theta_i=0$.
  \item[(2.2)] If the rotation on the brane $(\theta_1)$ belongs to
    the $\fsu(3)$ subalgebra.  In this case,
    $\mu_4\theta_4+\mu_5\theta_5=0$ and $\sum_{i=1,2,3}
    \mu_i\theta_i=0$.
  \end{itemize}
  Notice that all other possibilities are conjugate to the ones
  selected above.  Both of them preserve $\nu=1/16$.
\end{itemize}

On the other hand, if $\theta_1=0$, one is just left with a transverse
rotation $\lambda=\rho_\perp$.  Due to the isometries of the
background configuration, we are just left to consider two
possibilities:
\begin{itemize}
  \item[(1)] If $\rho_\perp$ belongs to the $\fsu(4)$ spinor isotropy
  subalgebra.  In that case, $\sum_{i=2,\dots,5}\mu_i\theta_i = 0$.
  The resulting configuration preserves $\nu=1/16$.
\item[(2)] If $\rho_\perp$ belongs to the $\fsp(1)\times\fsp(1)$
  subalgebra.  In that case, $\mu_2\theta_2 + \mu_3\theta_3=0$ and
  $\mu_4\theta_4 + \mu_5\theta_5=0$.  The resulting configuration
  preserves $\nu=1/8$.
\end{itemize}

As in the former family of reductions, the physical interpretation
and the causal structure of the above spacetimes is still missing.

\subsection{Supersymmetric reductions of the delocalised M2-brane}
\label{sec:M2d}

The standard Kaluza--Klein reduction of the M2-brane to obtain the
D2-brane, requires that the M2-brane be \emph{delocalised} along one
transverse direction; that is, that the M2-brane admit a Killing
vector which is a translation along a transverse direction.  The
metric of the spacetimes exterior to such a membrane is again of the
general form \eqref{eq:elemint} but now with three factors:
\begin{equation}
  \label{eq:mM2d}
  g = V^{-2/3} ds^2(\EE^{1,2}) + V^{1/3} dz^2 + V^{1/3} ds^2(\EE^7)~,
\end{equation}
where $z$ is the transverse coordinate along which the membrane is
delocalised and $V = 1 + |Q|/r^5$ is a harmonic function on $\EE^7$
depending only on the radial distance.  The symmetry group is now
\begin{equation}
  \label{eq:GM2d}
  G = \ISO(1,2) \times \RR \times \SO(7)~,
\end{equation}
with Lie algebra
\begin{equation}
  \label{eq:gM2d}
  \fg = \left( \RR^{1,2} \rtimes \fso(1,2) \right) \times \RR \times
  \fso(7)~.
\end{equation}
Therefore a Killing vector may be decomposed as
\begin{equation}
  \label{eq:KVM2d}
  \xi = \tau_\parallel + \tau_\perp + \lambda_\parallel +
  \rho_\perp~,
\end{equation}
with the same notation as above.

\subsubsection{Freely-acting spacelike isometries}

We proceed as before by using the freedom of acting by $G$ in order to
bring $\lambda$ to a normal form.  Either $\lambda=0$ or else it can
be brought into one of three normal forms: an infinitesimal boost,
rotation or null rotation.  We can again discard the boost since this
leads to a $\xi$ which is not spacelike.  The same reason forces
$\tau_\parallel = 0$ in the case where $\lambda$ is a null rotation.
Arguing as in the previous section, we are left with the following two
cases of freely-acting spacelike Killing vectors:
\begin{enumerate}
\item[(A)] $\xi = \tau_\perp + \nu_\parallel + \rho_\perp$, where
  $\tau \neq 0$ for otherwise $\xi$ does not act freely;
\item[(B)] $\xi = \tau_\parallel + \tau_\perp + \rho_\parallel +
  \rho_\perp$, where $\tau_\parallel + \tau_\perp$ is spacelike, but
  where
  \begin{enumerate}
  \item[(i)] if $\rho_\parallel = 0$ then $\tau_\parallel$ can be
    either spacelike, timelike or null; and
  \item[(ii)] if $\rho_\parallel \neq 0$, then $\tau_\parallel$
    is timelike.
  \end{enumerate}
\end{enumerate}
In all cases the decomposition of $\xi$ is orthogonal relative to the
brane metric.  We also remind the reader that $\tau$, $\rho$ and $\nu$
stand, respectively, for a translation, a rotation and a null
rotation, and that the subscripts $\parallel$ and $\perp$ denote
directions tangent to and perpendicular to the brane, respectively.
Notice that $\rho_\perp$ here always fixes at least one direction
since it defines a tangent vector field on an even-dimensional
sphere.

\subsubsection{Moduli space of smooth reductions}

We now describe the different strata of the moduli space of smooth
reductions.  In case (A), coordinates can be chosen so that the
Killing vector $\xi$ takes the form
\begin{equation*}
  \xi = a \d_z + N_{+2} + \theta_2 R_{34} + \theta_3 R_{56} +
  \theta_4 R_{78}~.
\end{equation*}
The moduli space is obtained by projectivising and quotienting by the
action of the Weyl group and hence this stratum of the moduli space is
three-dimensional.  Supersymmetry will then select a two-dimensional
locus.

In case (B) with $\rho_\parallel \neq 0$, $\tau_\parallel$ must be
timelike, whence
\begin{equation*}
  \xi = a \d_0 + b \d_9 + \theta_1 R_{12} + \theta_2 R_{34} + \theta_3
  R_{56} + \theta_4 R_{78}~,
\end{equation*}
where $|a|<|b|$.  There are six free parameters, whence this
stratum is five-dimensional after projectivisation.  Supersymmetry
will then select a four-dimensional locus.

Finally in case (B) with $\rho_\parallel = 0$, we have to distinguish
between three cases depending on whether $\tau_\parallel$ is timelike,
spacelike or null.  If $\tau_\parallel$ is timelike we can bring $\xi$
to the form
\begin{equation*}
  \xi = a \d_0 + b \d_9 + \theta_2 R_{34} + \theta_3 R_{56} + \theta_4
  R_{78}~,
\end{equation*}
with $|a| < |b|$.  As a result there are five free parameters
yielding a four-dimensional moduli space.  Supersymmetry will further
select a three-dimensional locus.  Similarly if $\tau_\parallel$ is
spacelike, we can bring $\xi$ to the form
\begin{equation*}
  \xi = a \d_1 + b \d_9 + \theta_2 R_{34} + \theta_3 R_{56} + \theta_4
  R_{78}~,
\end{equation*}
where now $a$ and $b$ cannot both be zero.  Again we have a
four-dimensional moduli space of smooth reductions with a
codimension-one locus of supersymmetric reductions.  Finally if
$\tau_\parallel$ is null, $\xi$ takes the form
\begin{equation*}
  \xi = \d_+ + b \d_9 + \theta_2 R_{34} + \theta_3 R_{56} + \theta_4
  R_{78}~,
\end{equation*}
with $b\neq 0$.  This gives rise to a three-dimensional moduli space
of smooth reductions with a two-dimensional locus of supersymmetric
reductions.

\subsubsection{Absence of closed causal curves}
\label{sec:M2dCCC}

The purpose of the present subsection is to analytically prove that
despite the intuition, the above spacetimes do not have closed causal
curves.  We shall concentrate on spacetimes reduced along the orbits
infinitesimally generated by Killing vectors $\xi$ which act
non-trivially either on a timelike or a lightlike direction.  For
simplicity, we will not allow the transverse rotation parameters to be
arbitrary, but set all of them to zero.

Let us start by analysing the problem of existence of closed causal
curves in an M2-brane background delocalised in one transverse
direction $z$ reduced along the orbits of the Killing vector
\begin{equation*}
  \xi = a\partial_0 + b\partial_z~.
\end{equation*}
The only condition that such a Killing vector is required to satisfy
is to be spacelike everywhere.  This requirement provides us with the
constraint that, for all $r$,
\begin{equation}
  \label{eq:space1}
  \|\xi\|^2 = V^{-2/3}(r)\left(-a^2 + V(r)b^2\right) > 0~.
\end{equation}
 
To analyse this question, it is convenient to change coordinates to an
adapted coordinate system, in which the Killing vector becomes a
single spacelike translation $\xi=\partial_{z'}$.  In this case, this
is easily achieved by a linear transformation in the original
$\{x^0,z\}$ space.  In the new coordinate system $\{t',z'\}$, the
eleven-dimensional takes the form
\begin{multline*}
  g = -V^{-2/3}\left[a^2 (dz')^2 + b^{-2}(dt')^2 + 2ab^{-1}
  dz' dt'\right] + b^2 V^{1/3}(dz')^2 \\
  + V^{1/3}ds^2(\EE^7)~.
\end{multline*}

What we would like to know is whether there exist closed causal curves
$x(\lambda)$, i.e., $\|\frac{dx}{d\lambda}\|^2 \leq 0$ joining the
points $(t'_0, x^i_0, z'_0)$ and $(t'_0, x^i_0, z'_0 + \Delta)$, since
they become identified in the quotient.  Let us assume that such a
curve exists.  If so, there must exist at least one value $\lambda^*$
of the affine parameter $\lambda$ where the timelike component of the
tangent vector to the curve vanishes:
\begin{equation*}
  \exists \lambda^\star \quad \text{such that} \quad
  \left.\frac{dt'}{d\lambda}\right|_{\lambda^*}=0~.
\end{equation*}
If one computes the norm of such a tangent vector at $\lambda^*$,
one derives the inequality
\begin{equation*}
  \|\xi\|^2(\lambda^*)\left.\frac{dz'}{d\lambda}
  \right|^2_{\lambda^*} + V^{1/3}\sum_i \left.\frac{dx^i}{d\lambda}
  \right|^2_{\lambda^*} \leq 0~.
\end{equation*}
Due to the constraint \eqref{eq:space1}, it is clear that the left
hand side of the above norm is the sum of positively defined terms, so
that the inequality can never be satisfied.  This already shows the
non-existence of closed timelike curves.  Furthermore, the only
possibility for the equality to be satisfied is whenever, for all $i$,
\begin{equation*}
  \left.\frac{dz'}{d\lambda}\right|_{\lambda^*} =
  \left.\frac{dx^i}{d\lambda}\right|_{\lambda^*}= 0~.
\end{equation*}
But the existence of one point where the tangent vector to the causal
curve vanishes identically violates the fact that $\lambda$ is an
affine parameter.  We thus conclude that no closed lightlike curves
are allowed in this spacetime.

The corresponding proof for the action generated by $\xi = \partial_+
+ b\partial_z$ involves similar ideas and techniques.  In this case,
the requirement of having an everywhere spacelike Killing vector gives
rise to the condition
\begin{equation*}
  V(r)^{1/3}\,b^2 > 0 \quad \Rightarrow \quad |b| > 0~.
\end{equation*}
By a linear transformation, we can move to an adapted coordinate
system $\{z',x^{+\prime}\}$ in which the Killing vector becomes
a single translation $\xi=\partial_{z'}$ and the eleven-dimensional
metric takes the form
\begin{multline*}
  g = 2V^{-2/3}dx^-\left(dz' + b^{-1}dx^{+\prime}\right) +
  V^{1/3}b^2 (dz')^2 \\
  + V^{1/3}ds^2(\EE^7) + V^{-2/3}(dx^2)^2~.
\end{multline*}

If we again assume the existence of a closed causal curve of affine
parameter $\lambda$ joining the points $(x^{+\prime}_0, x^-_0,x^i_0,
z'_0)$ and $(x^{+\prime}_0, x^-_0,x^i_0, z'_0 + \Delta)$, there must
necessarily exist at least one value for this affine parameter
$\lambda^*$ where
\begin{equation*}
  \exists \lambda^* \quad \text{such that} \quad
  \left.\frac{dx^-}{d\lambda}\right|_{\lambda^*}=0~.
\end{equation*}
By computing the norm of the tangent vector to the causal curve at the
point $\lambda^*$, and using the fact that $|b|>0$, it is immediate to
show the non-existence of such closed causal curves by the same
argument used before.

\subsubsection{Supersymmetry}

Now we determine the locus of supersymmetric reductions.  In all the
cases of smooth reductions, the rotation component of the Killing
vector takes the general form
\begin{equation*}
  \rho = \theta_1 R_{12} + \theta_2 R_{34} + \theta_3 R_{56} +
  \theta_4 R_{78}~,
\end{equation*}
which is the special case of \eqref{eq:so10rotation} corresponding to
$\theta_5 =0$.  This allows us to reduce the determination of the
supersymmetric locus to the case of the M2-brane, with the added
feature that when in addition $\theta_1=0$ we have the option of
adding a null rotation to $\xi$ with the effect of halving the
fraction of supersymmetry, as described in Section~\ref{sec:method}.
We will not repeat the arguments here and simply state the results,
which are illustrated in Table~\ref{tab:M2dsusy}.

Before moving into the explicit Kaluza--Klein reductions, let us
stress that the previous classification gives rise to a wealth of
smooth supersymmetric M-theory backgrounds, $\eM_{\text{M2}}/\Gamma_0$
by considering discrete subgroups $\Gamma_0\subset\Gamma$.  These
include a stack of delocalised M2-branes and eleven-dimensional
fluxbranes $(\xi= \partial_1 + \rho_\perp)$ or eleven-dimensional
nullbranes $(\xi= \partial_1 + \nu_\parallel)$.  The latter is an
example of an eleven-dimensional time-dependent background in which a
compact spacelike worldvolume dimension shrinks as time evolves down
to a minimum size and then re-expands.  It would be very interesting
to understand the physics on the throat of the brane in such an
scenario.

\begin{table}[h!]
  \begin{center}
    \setlength{\extrarowheight}{5pt}
    \begin{tabular}{|>{$}c<{$}|>{$}c<{$}|>{$}c<{$}|>{$}c<{$}|}
      \hline
      \multicolumn{1}{|c|}{Translation} & \text{Subalgebra} & \nu &
      \dim\\
      \hline
      \hline
      a\d_1 + b \d_z & \fsu(3) & \frac18~\left(\frac1{16}\right)&
      3~(2)\\
      a,b~\text{not both $0$}& \fsu(2) & \frac14~\left(\frac18\right)&
      2~(1)\\
      (a=0~,b\neq0)& \{0\} & \frac12~\left(\frac14\right)&
      1~(0)\\[3pt]
      \hline
      \d_+ + b\d_z & \fsu(3) & \frac18 & 2\\
      b\neq 0& \fsu(2) & \frac14 & 1\\
      & \{0\} & \frac12 & 0\\[3pt]
      \hline
      & \fsu(4) & \frac1{16} & 4\\
      a\d_0 + b\d_z & \fsu(3) & \frac18 & 3\\
      & \fsp(1) \times \fsp(1) & \frac18 & 3\\
      |b|>|a|>0& \fsu(2) & \frac14 & 2\\
      & \{0\} & \frac12 & 1 \\[3pt]
      \hline
    \end{tabular}
    \vspace{8pt}
    \caption{Supersymmetric reductions of the delocalised M2-brane.
      We indicate the form of the translation, the spinor isotropy
      subalgebra to which the rotation belongs, the fraction $\nu$ of
      the supersymmetry preserved and the dimension of the
      corresponding stratum of the moduli space $\eM$ of
      supersymmetric reductions.  The numbers in parentheses indicate
      the values in the presence of a null rotation, which can only
      occur when the translation is trasverse.}
    \label{tab:M2dsusy}
  \end{center}
\end{table}

\subsubsection{Explicit reductions}

In the following, we shall explicitly write down the different type
IIA configurations obtained by the inequivalent Kaluza-Klein
reductions identified and classified in previous subsections.  Let us
start by the subspace of the moduli space in which the parameter $a$
associated with the spacelike translation $a\partial_1$ is set to
zero.  In order to discuss both null and flux branes at the same time,
we shall present the Kaluza-Klein reduction along the orbits of the
Killing vector $\xi= \partial_z + \lambda$, where the infinitesimal
transformation $\lambda$ is given by
\begin{equation*}
  \lambda = \beta B_{02} + \theta_1 R_{12} + \theta_2 R_{34} +
  \theta_3 R_{56} + \theta_4 R_{78}~.
\end{equation*}
Thus, whenever $\beta=0$, we will be discussing composite
configurations of D2-branes and flux branes; whenever
$|\beta|=|\theta_1|$, we will be discussing composite configurations
of D2-branes, null branes and flux branes.

The constant matrix $B$ is a $9\times 9$ matrix which does not act
both on the $x^9$ and $z$ directions.  Relative to the basis
$\{x^0,x^1,\dots,x^8\}$, it is given explicitly by
\begin{equation}
  \label{eq:Bmatrixm2c}
  B= 
  \begin{pmatrix}
    0 & 0 &  \beta & 0 & 0 & 0 & 0 & 0 & 0 \\
    0 & 0 & -\theta_1 & 0 & 0 & 0 & 0 & 0 & 0 \\
    \beta & \theta_1 & 0 & 0 & 0 & 0 & 0 & 0 & 0 \\
    0 & 0 & 0 & 0 & -\theta_2 & 0 & 0 & 0 & 0 \\
    0 & 0 & 0 & \theta_2 & 0 & 0 & 0 & 0 & 0 \\
    0 & 0 & 0 & 0 & 0 & 0 & -\theta_3 & 0 & 0 \\
    0 & 0 & 0 & 0 & 0 & \theta_3 & 0 & 0 & 0 \\
    0 & 0 & 0 & 0 & 0 & 0 & 0 & 0 & -\theta_4 \\
    0 & 0 & 0 & 0 & 0 & 0 & 0 & \theta_4 & 0 \\
  \end{pmatrix}~.
\end{equation}
The corresponding type IIA configurations have a ten-dimensional
metric given by
\begin{multline}
  \label{met:m2a}
  g = \tilde{\Lambda}^{1/2}\left\{V^{-1/2}ds^2(\EE^{1,2}) 
    + V^{1/2} ds^2(\EE^7)\right\}\\
  - \tilde{\Lambda}^{-1/2}V^{1/2}\left\{V^{-1}
    \left[\beta \omega^{02} + \theta_1 \omega^{12}\right] +
    \theta_2 \omega^{34} + \theta_3 \omega^{56} + \theta_4 \omega^{78}
  \right\}^2~,
\end{multline}
where again $\omega^{ij}=x^i dx^j-x^j dx^i$.  In addition, the RR
1-form $A_{1}$, NS-NS 3-form field strength $H_{3}$, RR 4-form $H_4$
and dilaton $\Phi$ are given by
\begin{footnotesize}
  \begin{equation}
    \label{oth:m2a}
    \begin{aligned}[m]
      A_1 &= \tilde{\Lambda}^{-1}\left\{V^{-1}\left[\beta \omega^{02}
          + \theta_1 \omega^{12}\right] + \theta_2 \omega^{34} +
        \theta_3
        \omega^{56} + \theta_4 \omega^{78} \right\} \\
      H_3 &= -\left(\beta x^2 dx^1\wedge dx^2 + \theta_1 x^2
        dx^0\wedge dx^2 +
        (\beta x^0 + \theta_1 x^1) dx^0\wedge dx^1\right)\wedge
      dV^{-1} \\
      H_4 &= \dvol\left(\EE^{1,2}\right) \wedge dV^{-1} \\
      \Phi &= \tfrac{3}{4}\log \left(\tilde{\Lambda}\cdot
        V^{1/3}\right)~.
    \end{aligned}
  \end{equation}
\end{footnotesize}
The configuration depends on an scalar function $\tilde{\Lambda}$
which is defined in terms of the scalar function $\Lambda$ appearing
in the general discussion section, by
\begin{equation*}
\Lambda = V^{1/3}\cdot \tilde{\Lambda}~,
\end{equation*}
and equals
\begin{footnotesize}
  \begin{multline*}
    \tilde{\Lambda} = 1 + V^{-1}\left\{(x^2)^2[\theta_1^2 - \beta^2] +
      (\beta x^0 + \theta_1 x^1)^2\right\} +
    \left(\theta_2\right)^2\left[(x^3)^2 + (x^4)^2\right]\\
    + \left(\theta_3\right)^2\left[(x^5)^2 +(x^6)^2\right]
    +\left(\theta_4\right)^2\left[(x^7)^2 +(x^8)^2\right]~.
  \end{multline*}
\end{footnotesize}
Whenever $|\theta_1|<|\beta|$, there always exists a Lorentz
transformation such that $\theta_1 R_{12} + \beta B_{02}$ becomes a
pure boost.  Therefore, such a configuration always breaks
supersymmetry.  Furthermore, the corresponding Killing vector is no
longer spacelike everywhere, pointing out to the existence of regions
in spacetime where there exist closed timelike curves.  These
configurations would correspond, whenever we restrict ourselves to
regions of spacetime with no causal sickness, to similar cosmological
scenarios to the ones discussed in \cite{KOSST, SeibergBC, CorCos, 
 Nekrasov}, but this time taking place on the worldvolume of
a 2+1 brane.  By switching on the moduli associated with transverse
rotations, one is just adding F7-branes into the discussion.  On the
other hand, if $|\theta_1|>|\beta|$, there always exists a Lorentz
transformation mapping $\theta_1 R_{12} + \beta B_{02}$ to a pure
rotation, whose physical interpretation has already been given.

Let us concentrate on the interpretation of the different
supersymmetric loci summarised in \ref{tab:M2dsusy}.  If we set
$\beta=0$, there are four different possibilities to be considered:
\begin{enumerate}
\item[(1)] The case $\theta_i=0$ for all $i$ corresponds to the
  well-known D2-brane in type IIA preserving $\nu=1/2$ of the
  spacetime supersymmetry.
  
\item[(2)] We must distinguish between two case of two non-vanishing
  $\theta$s:
  \begin{enumerate}
  \item[(i)] If $\theta_1=\eta_2\theta_2$, the configuration describes
    a D2-brane in the (12)-plane and an F5-brane along the
    (56789)-plane sitting at $x^1=x^2=x^3=x^4=0$.  It preserves
    $\nu=1/4$ of the spacetime supersymmetry, with Killing spinors
    being preserved by an $\fsu(2)$ subalgebra.

  \item[(ii)] If $\theta_2=\eta_3\theta_3$, the configuration
    describes a D2-brane in the (12)-plane and an F5-brane along the
    (12789)-plane sitting at $x^3=x^4=x^5=x^6=0$.  It preserves
    $\nu=1/4$ of the spacetime supersymmetry, with Killing spinors
    again preserved by an $\fsu(2)$ subalgebra.
  \end{enumerate}
\item[(3)] There are again two distinct case of three non-vanishing
  $\theta$s:
  \begin{enumerate}
  \item[(i)] If $\theta_1=\eta_2\theta_2+\eta_3\theta_3$, the
    configuration describes a D2-brane in the (12)-plane and
    an F3-brane along the (789)-plane sitting at
    $x^1=x^2=x^3=x^4=x^5=x^6=0$.  It preserves $\nu=1/8$ of the
    spacetime supersymmetry, with Killing spinors being preserved by
    an $\fsu(3)$ subalgebra.
    
  \item[(ii)] If $\theta_2=\eta_3\theta_3+\eta_4\theta_4$, the
    F3-brane extends along the (129)-plane sitting at
    $x^3=x^4=x^5=x^6=x^7=x^8=0$.  It preserves the same amount of
    supersymmetry as the previous case due to the existence of an 
    $\fsu(3)$ subalgebra preserving some Killing spinors.
  \end{enumerate}
\item[(4)] As explained in the general discussion about preservation
  of supersymmetry, there are two inequivalent ways of preserving
  Killing spinors when four of the $\theta$s are non-vanishing:
  \begin{enumerate}
  \item[(i)] $\theta_1=\eta_2\theta_2 + \eta_3\theta_3 +
    \eta_4\theta_4$.  This configuration describes a D2-brane in the
    (12)-plane and a flux string along the $x^9$ direction, preserving
    $\nu=1/16$ of the spacetime supersymmetry.  This is the one
    associated with the $\fsu(4)$ isotropy algebra discussed before.
    
  \item[(ii)] $\theta_1=\eta_2\theta_2$ and $\theta_3=\eta_4\theta_4$.
    This second possibility involves a D2-brane and a maximally
    supersymmetric flux string in the $x^9$ direction.  It preserves
    $\nu=1/8$ and it is associated with the $\fsp(1)\times\fsp(1)$
    isotropy algebra.
  \end{enumerate}
\end{enumerate}

If $\beta\neq 0$, the only allowed possibility preserving
supersymmetry requires $|\theta_1|=|\beta|$.  Let us thus concentrate
on this case.  Depending on whether the remaining parameters vanish or
satisfy certain linear relations, we distinguish between the following
configurations
\begin{enumerate}
  \item[(1)] If $\theta_i=0$ $i=2,3,4$, it describes a D2-brane in the
  (12)-plane and a null brane.  This composite configuration preserves
  $\nu=1/4$ of the spacetime supersymmetry.
  
\item[(2)] If one of the $\theta$s is non-vanishing, the corresponding
  configuration breaks supersymmetry completely, due to the presence
  of a F7-brane besides the previous D2-brane and null brane.
  
\item[(3)] If two of the $\theta$s are non-vanishing, we shall
  distinguish among two cases:
  \begin{enumerate}
  \item[(i)] If $\theta_2\neq \eta_3\theta_3$, the configuration
    describes a system of two intersecting F7-branes (besides the
    composite system of a D2-brane and a null brane), such that all
    supersymmetry is broken.
   
  \item[(ii)] If $\theta_2=\eta_3\theta_3$, the intersection among the
    F7-branes gives rise to the so-called F5-brane, extending along
    the (12789)-plane and sitting at $x^3=x^4=x^5=x^6=0$.  The full
    configuration includes the previous pair D2/null-brane system,
    thus preserving $\nu=1/8$.
  \end{enumerate}
\item[(4)] If all $\theta$s are non-vanishing, we need to distinguish
  between two cases:
  \begin{enumerate}
    \item[(i)] If $\theta_2\neq \eta_3\theta_3 + \eta_4\theta_4$, the
    configuration describes the intersection of three intersecting
    F7-branes, besides the D2-null-brane system, thus breaking all
    supersymmetry.
  \item[(ii)] If $\theta_2=\eta_3\theta_3 + \eta_4\theta_4$, the
    intersection of the F7-branes gives rise to a F3-brane extending
    along the (129)-plane and sitting at $x^k=0$ k=$3,\dots ,8$.
    Thus, there exists a composite configuration involving a D2-brane,
    null brane and F3-brane preserving $\nu=1/16$.
  \end{enumerate}
\end{enumerate}

The supersymmetric configurations discussed above are summarised
in table~\ref{tab:M2d2}.

\begin{table}[h!]
  \begin{center}
    \setlength{\extrarowheight}{3pt}
    \begin{tabular}{|>{$}c<{$}|c|>{$}l<{$}|}
      \hline
      \nu & Object & \multicolumn{1}{c|}{Subalgebra}\\
      \hline
      \hline
      \frac14 & D2$\perp$F5 & \fsu(2)\\
              & D2$\parallel$F5 & \\
      \frac14 & D2 + N & \RR \\
      \frac18 & D2$\perp$F3 & \fsu(3) \\
              & D2$\parallel$F3 & \\
      \frac18 & D2 + N + F5 & \fsu(2)\times \RR \\
      \frac1{16} & D2$\perp$F1 & \fsu(4)\\
      \frac1{16} & D2$\perp$F1 & \fsp(1)\times\fsp(1)\\
      \frac1{16} & D2 + N + F3 & \fsu(3)\times \RR \\[3pt]
      \hline
    \end{tabular}
    \vspace{8pt}
    \caption{Supersymmetric configurations of D2-branes (D2),
      fluxbranes and nullbranes.}
    \label{tab:M2d2}
  \end{center}
\end{table}

It is interesting to compute the fluxes associated with fluxbranes in
the presence of D2-branes.  We shall concentrate on the F5-brane, for
simplicity.  According to table \ref{tab:M2d2}, there are two
different cases to be discussed.  Let us start with $D2\parallel F5$.
This corresponds to setting $\theta_1=\theta_4=\beta=0$ and
$\theta\equiv\theta_2=\theta_3$ in \eqref{met:m2a} and
\eqref{oth:m2a}.  Notice, by inspection of \eqref{oth:m2a}, the
absence of NS-NS charge in this case and the fact that the RR 1-form
does not depend on the point of the F5-brane worldvolume, as it was
the case for the fundamental strings analysed in the previous section.
It is clear then that the flux ``carried'' by the F5-brane is exactly
the same as in flat space
\begin{equation*}
  \frac{1}{8\pi^2}\int_{\RR^4}F_2\wedge F_2 = \theta^{-2}~.
\end{equation*}

It is interesting to note that the solution we found is not expected
to be the most general one for this system, due to the existence of
moduli.  Indeed, if one probes the F5-brane background with a D2-brane
oriented as described above, the D2-brane does not feel any force for
an arbitrary transverse distance among both objects, whereas in the
solution described here the D2-brane lies on the F5-brane.  This is
certainly not necessary, as it was for fundamental strings.

When the D2-branes are transverse to the F5-brane, the RR 1-form does
depend on the radial distance along the (56789)-plane spanned by the
F5-brane, whereas there appears some NS-NS charge in the radial
direction on the (12)-plane, where the D2-brane lies.  Nevertheless,
proceeding as for the fundamental strings, that is, choosing a point
on the F5-brane and keeping it as a parameter, it can be shown that
\begin{equation*}
  F_2\wedge F_2 \propto \frac{\partial \left(r_2\cdot
      \tilde{\Lambda}^{-2} \right)}{\partial\,r_1}dr_1 \wedge
  d\varphi_1\wedge dr_2\wedge d\varphi_1~,
\end{equation*}
where $x^1+ix^2 = r_1\,e^{i\varphi_1}$ and $x^3+ix^4 =
r_1\,e^{i\varphi_2}$.  Therefore, the flux equals the one of the flat
spacetime, except at the origin where it vanishes.

Let us move to the region of the moduli space where the extra
spacelike translation parameter is non-vanishing.  In other words, let
us consider the Kaluza-Klein reduction along the orbits of the Killing
vector $\xi=\partial_z + \alpha$, where
\begin{equation*}
  \alpha = a\partial_2 + \theta_2 R_{34} +
  \theta_3 R_{56} + \theta_4 R_{78}~.
\end{equation*}
The constant matrix $B$ is a $7\times 7$ matrix which does not act on
the $\{x^0,x^1,x^9\}$ directions.  It is given explicitly, in the
basis $\{x^2,\dots,x^8\}$, by
\begin{equation}
  \label{eq:Bmatrixm2d}
  B= 
  \begin{pmatrix}
    0 & 0 & 0 & 0 & 0 & 0 & 0 \\
    0 & 0 & -\theta_2 & 0 & 0 & 0 & 0 \\
    0 & \theta_2 & 0 & 0 & 0 & 0 & 0 \\
    0 & 0 & 0 & 0 & -\theta_3 & 0 & 0 \\
    0 & 0 & 0 & \theta_3 & 0 & 0 & 0 \\
    0 & 0 & 0 & 0 & 0 & 0 & -\theta_4 \\
    0 & 0 & 0 & 0 & 0 & \theta_4 & 0 \\
  \end{pmatrix}~.
\end{equation}
Since $\alpha$ involves a translation, the constant vector $\bC$
defined in \eqref{eq:data} taking care of the inhomogeneous part of
the infinitesimal transformation is non-vanishing.  It is a given by a
7-vector
\begin{equation*}
  (\bC)^t = (a,\vec{0})~.
\end{equation*}
The corresponding type IIA configurations have a ten-dimensional
metric given by
\begin{multline*}
  g =
  \tilde{\Lambda}^{1/2}\left\{V^{-1/2}\left(ds^2(\EE^{1,1})+(dx^2)^2
    \right) + V^{1/2} \left(ds^2(\EE^6) + (dx^9)^2\right)\right\}\\
  - \tilde{\Lambda}^{-1/2}V^{1/2} \left\{aV^{-1} dx^2 + \theta_2
    \omega^{34} + \theta_3 \omega^{56} + \theta_4
    \omega^{78}\right\}^2~,
\end{multline*}
where we are still using the notation $\omega^{ij} = x^i dx^j - x^j
dx^i$.  In addition, the RR 1-form $A_{1}$, NS-NS 3-form field
strength $H_{3}$, RR 4-form $H_4$ and dilaton $\Phi$ are listed below:
\begin{equation*}
  \begin{aligned}[m]
    A_1 &= \tilde{\Lambda}^{-1}\left\{aV^{-1}dx^2 + \theta_2
      \omega^{34} + \theta_3 \omega^{56} + \theta_4 \omega^{78}
    \right\} \\
    H_3 &= -a \dvol\left(\EE^{1,1}\right)\wedge dV^{-1} \\
    H_4 &= \dvol\left(\EE^{1,1}\right)\wedge dx^2 \wedge dV^{-1} \\
    \Phi &= \tfrac{3}{4}\log \left(\tilde{\Lambda}\cdot
      V^{1/3}\right)~.
  \end{aligned}
\end{equation*}
The configuration depends on an scalar function $\tilde{\Lambda}$
which is defined in terms of the scalar function $\Lambda$ appearing
in the general discussion section, by
\begin{equation*}
  \Lambda = V^{1/3}\cdot \tilde{\Lambda}~,
\end{equation*}
and equals
\begin{multline*}
  \tilde{\Lambda} = 1 + a^2\,V^{-1}+
  \left(\theta_2\right)^2\left((x^3)^2 + (x^4)^2\right) \\
  + \left(\theta_3\right)^2\left((x^5)^2 +(x^6)^2\right)
  +\left(\theta_4\right)^2\left((x^7)^2 +(x^8)^2\right)~.
\end{multline*}

To begin with, we shall give an interpretation for the configuration
in which all transverse rotations are set to zero: $\theta_i=0$ for
$i=2,3,4$.  We would like to interpret it as the lift to ten
dimensions of a bound state of $(p,q)$-strings in $N{=}2$ $D{=}9$
supergravity, or as the T-dual of $(p,q)$-strings in type IIB, this
being the reason why fundamental strings are delocalised in the $x^2$
direction in the above supergravity solution.  This interpretation can
be inferred as follows.  The global symmetries of supergravity
theories compactified on torus have been studied extensively
\cite{CJLP,BHO}.  Let us concentrate on $N{=}2$ $D{=}9$ obtained by
reduction of $D{=}11$ on a 2-torus.  This nine-dimensional theory is
$\SL(2,\RR)$ invariant.  What this means, among other things, is that
configurations obtained from $D{=}11$ supergravity by reduction first
along the $z$ direction, and afterwards, along the $x$ direction, can
be mapped to those configurations in which one first reduces along the
$x$ direction, and then on the $z$ direction.  In this particular case
of transverse directions, there is indeed an $\SO(2)$ transformation
of angle $\pi/2$ relating both configurations.  There exist, of
course, more general transformations.  This is pretty close to what we
have been doing.  By reducing the M2-brane configuration along
$\partial_x$ and $\partial_z$, or the other way around, we are
describing $(p,0)$ or $(0,q)$-strings in $N{=}2$ $D{=}9$.  Under
$\SL(2,\RR)$ transformations, one can generate the full spectrum of
$(p,q)$-strings.  Notice that these transformations are nothing but
linear diffeomorphism transformations in $D{=}11$ supergravity which
map $\partial_z$ into $a\partial_z + b\partial_x$, which is the kind
of Killing vector we used to reduce the starting M2-brane
configuration.  Thus, the ten-dimensional configuration found above is
nothing but the lift to ten dimensions of one of these
$(p,q)$-strings, which is a bound state of D2-branes and delocalised
fundamental strings.

Once this background has been understood, it is easy to interpret the
effect of turning on the deformation parameters $\theta_i$.  Indeed,
whenever $\theta_i\neq 0$, the corresponding configurations are no
longer asymptotically flat.  They correspond to composite
configurations of the vacuum and fluxbranes.  The discussion of the
different supersymmetric and non-supersymmetric possibilities is
analogous to the ones already given before, so we refer the reader to
the Table~\ref{tab:M2(p,q)} summarising the results.

\begin{table}[h!]
  \begin{center}
    \setlength{\extrarowheight}{3pt}
    \begin{tabular}{|>{$}c<{$}|c|>{$}l<{$}|}
      \hline
      \nu & Object & \multicolumn{1}{c|}{Subalgebra}\\
      \hline
      \hline
      \frac12 & D2-FA & \{0\} \\
      \frac14 & D2-FA + F5 & \fsu(2) \\
      \frac18 & D2-FA + F3 & \fsu(3) \\[3pt]
      \hline
    \end{tabular}
    \vspace{8pt}
    \caption{Supersymmetric configurations of bound states made of
      D2-branes and delocalised fundamental strings (D2-FA) and
      fluxbranes.}
    \label{tab:M2(p,q)}
  \end{center}
\end{table}

Let us move to the region of the moduli space where there is
a non-vanishing null translation.  In other words, we shall discuss
the Kaluza-Klein reductions along the orbits of the Killing vectors
$\xi=\partial_z + \alpha$ where
\begin{equation*}
  \alpha = \partial_+ + \theta_2 R_{34} +
  \theta_3 R_{56} + \theta_4 R_{78}~.
\end{equation*}
The constant matrix $B$ is formally the same as in
\eqref{eq:Bmatrixm2d}, but this time being written in the basis
$\{x^+,\,x^3,\dots,x^7,\,x^8\}$.  Thus, besides the $z$ direction, it
leaves invariant the $\{x^-,\,x^2,\,x^9\}$ directions.  Since $\alpha$
involves a translation in the null direction $x^+$, there is a
non-trivial 7-vector $\bC$ describing the inhomogeneous part of the
isometry transformation
\begin{equation*}
  (\bC)^t = (1,\vec{0})~.
\end{equation*} 

The corresponding type IIA configurations have a ten-dimensional
metric given by
\begin{multline*}
  g = \tilde{\Lambda}^{1/2}\left\{V^{-1/2}\left(2dx^+dx^- +(dx^2)^2
    \right) + V^{1/2} \left(ds^2(\EE^6) + (dx^9)^2\right)\right\} \\
  - \tilde{\Lambda}^{-1/2}V^{1/2}\left\{V^{-1}dx^- +
    \theta_2\omega^{34} + \theta_3 \omega^{56} + \theta_4
    \omega^{78}\right\}^2~,
\end{multline*}
whereas the RR 1-form $A_{1}$, NS-NS 3-form field strength $H_{3}$, RR
4-form $H_4$ and dilaton $\Phi$ are listed below:
\begin{equation*}
  \begin{aligned}[m]
    A_1 &= \tilde{\Lambda}^{-1}\left\{V^{-1} dx^- + \theta_2
      \omega^{34} + \theta_3\omega^{56} + \theta_4 \omega^{78}
    \right\} \\
    H_3 &= -dx^-\wedge dx^2\wedge dV^{-1} \\
    H_4 &= dx^+\wedge dx^-\wedge dx^2 \wedge dV^{-1} \\
    \Phi &= \tfrac{3}{4}\log \left(\tilde{\Lambda}\cdot
      V^{1/3}\right)~.
  \end{aligned}
\end{equation*}
The configuration depends on an scalar function $\tilde{\Lambda}$
which is defined in terms of the scalar function $\Lambda$ appearing
in the general discussion section, by
\begin{equation*}
  \Lambda = V^{1/3}\cdot \tilde{\Lambda}~,
\end{equation*}
and equals
\begin{footnotesize}
  \begin{equation*}
    \tilde{\Lambda} = 1 + \left(\theta_2\right)^2  \left[(x^3)^2 +
      (x^4)^2\right] + \left(\theta_3\right)^2\left[(x^5)^2
      +(x^6)^2\right] + \left(\theta_4\right)^2\left[(x^7)^2
      +(x^8)^2\right]~.
  \end{equation*}
\end{footnotesize}
Even though we lack a good physical understanding of the above set of
configurations, it is instructive to look at the particular case in
which $\theta_i=0$ for all $i$.  We expect not to be describing any
fluxbrane, and manage to isolate the new effect associated with the
Kaluza-Klein reduction along an orbit involving lightlike
translations.  In such a case, the metric reduces to
\begin{equation*}
  g = V^{-1/2}ds^2(\EE^{1,2}) + V^{1/2}ds^2(\EE^7) -
  V^{-3/2}(dx^-)^2~.
\end{equation*}
There is the standard RR 4-form giving the expected charge carried by
a stack of D2-branes (and also the expected dilaton profile for a
D2-brane configuration), but also non-trivial RR 1-form and NS-NS
3-form field strength given respectively by
\begin{equation*}
  A_1 = V^{-1} dx^-  \qquad\text{and}\qquad
  H_3 = -dx^-\wedge dx^2 \wedge dV^{-1}~.
\end{equation*}
By inspection of the above expressions, it is clear that
asymptotically at spacelike infinity $(r\to\infty)$, the solution is
no longer Minkowski spacetime, but a wave background.  Thus, just as
fluxbranes induce some magnetic flux under Kaluza-Klein reduction and
also modify the spacelike asymptotics, the extra lightlike translation
induces the propagation of a lightlike perturbation at infinity.  In
the region $r\to 0$, one recovers the description close to a stack of
D2-branes, in the first approximation.  The stability and
supersymmetry of the configuration requires both $H_3$ and $F_2 =
dA_1$ to be null forms.

By switching on the $\theta_i$ parameters, one expects to add
fluxbranes to the above configuration.  Since the discussion of the
different possibilities does not give any new insight, we leave the
details to the interested reader.

Let us finally consider the region of the moduli space which involves
an extra timelike translation.  That is, let us discuss the
Kaluza-Klein reduction along the orbits of the Killing vector $\xi=
\partial_z + \alpha$, where
\begin{equation*}
  \alpha = a\partial_0 + \theta_2 R_{34} +
  \theta_3 R_{56} + \theta_4 R_{78}\quad , \quad |a|<1~.
\end{equation*}
The constant matrix $B$ is again a $7\times 7$ matrix given by 
\eqref{eq:Bmatrixm2d} in the basis $\{x^0,\,x^3,\dots,x^7,\,x^8\}$.
Thus, it leaves the $\{x^1,\,x^2,\,x^9\}$ directions invariant.
There is again a non-vanishing vector $\bC$ given by
\begin{equation*}
  (\bC)^t = (a,\vec{0})~.
\end{equation*}
The corresponding type IIA configurations have a ten-dimensional
metric given by
\begin{multline*}
  g = \tilde{\Lambda}^{1/2}\left\{V^{-1/2}ds^2(\EE^{1,2})
    + V^{1/2} ds^2(\EE^7)\right\}\\
  - \tilde{\Lambda}^{-1/2}V^{1/2} \left\{-aV^{-1} dx^0 + \theta_2
    \omega^{34} + \theta_3 \omega^{56} + \theta_4
    \omega^{78}\right\}^2~,
\end{multline*}
whereas the RR 1-form $A_{1}$, NS-NS 3-form field strength $H_{3}$,
RR 4-form $H_4$ and dilaton $\Phi$ are listed below:
\begin{equation*}
  \begin{aligned}[m]
    A_1 &= \tilde{\Lambda}^{-1}\left\{-V^{-1}a dx^0 + \theta_2
      \omega^{34} + \theta_3 \omega^{56} + \theta_4 \omega^{78}
    \right\} \\
    H_3 &= -a dx^1\wedge dx^2\wedge dV^{-1} \\
    H_4 &= \dvol\left(\EE^{1,2}\right)\wedge dV^{-1} \\
    \Phi &= \tfrac{3}{4}\log \left(\tilde{\Lambda}\cdot
      V^{1/3}\right)~.
  \end{aligned}
\end{equation*}
The configuration depends on an scalar function $\tilde{\Lambda}$
which is defined in terms of the scalar function $\Lambda$ appearing
in the general discussion section, by
\begin{equation*}
  \Lambda = V^{1/3}\cdot \tilde{\Lambda}~,
\end{equation*}
and equals
\begin{multline*}
  \tilde{\Lambda} = 1 -V^{-1}a^2 + \left(\theta_2\right)^2
  \left[(x^3)^2 + (x^4)^2\right]\\
  + \left(\theta_3\right)^2\left[(x^5)^2 +(x^6)^2\right] +
  \left(\theta_4\right)^2\left[(x^7)^2 +(x^8)^2\right]~.
\end{multline*}

Let us start by considering the particular configuration in which all
transverse rotation parameters vanish: $\theta_i=0$ for $i=2,3,4$.
The role played by the bound $|a|<1$ can be immediately appreciated by
inspection of the corresponding metric
\begin{equation*}
  g = -\left(V-a^2\right)^{-1/2}\,(dx^0)^2 +
  \left(V-a^2\right)^{1/2}\left\{ds^2(\EE^7) +
    V^{-1}ds^2(\EE^2)\right\}~.
\end{equation*}
It is now clear that the condition $|a|<1$ ensures the absence of
horizons in spacetime.  Furthermore, if $|a|\geq 1$ would have been
arbitrary, these horizons sitting at
\begin{equation*}
  r_{\text{H}}^5 = \frac{Q}{a^2-1}~,
\end{equation*}
would have divided spacetime into regions ($r> r_{\text{H}}$) having
closed timelike curves and regions ($r< r_{\text{H}}$) free of this
causal sickness.

Despite these features, the physical interpretation of these
configurations remains unclear.  Proceeding as in previous
configurations involving more than one spacelike translation, it would
be natural to interpret this configuration as a bound state of
D2-branes and delocalised E2-branes \cite{Chris, HullTimeLike,
  HullKhuriTimes}, the charge of the latter being constrained by the
bound $|a|<1$.  It is clear that the geometry in the region $r\to 0$
is the one describing the core of a stack of D2-branes, whereas in the
asymptotic spacelike infinity, this time we recover Minkowski
spacetime by a trivial rescaling of coordinates (notice that the
dilaton acquires a constant factor depending on $a$ in this asymptotic
limit).

Even though this configuration has no clear physical interpretation,
it is obvious that it allows the addition of fluxbranes by switching
on the $\theta_i$ parameters, while preserving some supersymmetry.
Since the discussion of these possibilities does not involve any new
features, we leave the details to the reader.

\section{Kaluza--Klein reductions of the M5-brane}
\label{sec:M5}

In this section we classify the set of M-theory backgrounds obtained
by modding out the M5-brane background by a one-parameter subgroup of
its isometry group and study the corresponding smooth supersymmetric
Kaluza--Klein reductions along the orbits of the Killing vectors
generating such subgroups.  We shall first describe the standard
M5-brane configuration in section~\ref{sec:M5l}.  Afterwards, we shall
discuss the M5-brane delocalised along one transverse direction in
section~\ref{sec:M5d}.

\subsection{Supersymmetric reductions of the M5-brane}
\label{sec:M5l}

The M-theory fivebrane \cite{Guven} is described by a metric of the
type \eqref{eq:elemint} with two factors,
\begin{equation}
  \label{eq:mM5}
  g = V^{-1/3} ds^2(\EE^{1,5}) + V^{2/3} ds^2(\EE^5)~,
\end{equation}
where $V = 1 + |Q|/r^3$ with $|Q|$ some positive constant and $r$ the
radial distance in the transverse $\EE^5$.  The $7$-form dual to the
$4$-form is given by
\begin{equation}
  \label{eq:FM5}
  * F_4 = \dvol(\EE^{1,5}) \wedge d V^{-1}~,
\end{equation}
up to a constant of proportionality.  The Killing spinors are of the
form
\begin{equation}
  \label{eq:SM5}
  \varepsilon = V^{-1/12} \varepsilon_\infty~,
\end{equation}
where $\varepsilon_\infty$ is a constant spinor satisfying
\begin{equation}
  \label{eq:PM5}
  \dvol(\EE^{1,5}) \cdot \varepsilon_\infty = \varepsilon_\infty~.
\end{equation}
The symmetry group is
\begin{equation}
  \label{eq:GM5}
  G = \ISO(1,5) \times \SO(5) \subset \ISO(1,10)~,
\end{equation}
with Lie algebra
\begin{equation}
  \label{eq:gG5}
  \fg = \left(\RR^{1,5} \rtimes \fso(1,5)\right) \times \fso(5)~,
\end{equation}
whence any Killing vector $\xi$ can be decomposed as
\begin{equation}
  \label{eq:KVM5}
  \xi = \tau_\parallel + \lambda_\parallel + \rho_\perp~,
\end{equation}
with the usual notation.

\subsubsection{Freely-acting spacelike isometries}

As before, to determine the freely-acting spacelike Killing vectors,
we exploit the freedom to conjugate by $G$ in order to bring $\xi$ to
a convenient normal form.  Conjugating first by the $\SO(1,5)$
subgroup, we can bring $\lambda$ to one of several normal forms.
Either $\lambda = 0$ or else it is conjugate to one of the following
three normal forms
\begin{enumerate}
\item $\lambda = \beta B_{01} + \theta_1 R_{23} + \theta_2 R_{45}$,
  with $\beta \neq 0$;
\item $\lambda = N_{+2} + \theta R_{45}$; or
\item $\lambda = \theta_1 R_{23} + \theta_2 R_{45}$,
\end{enumerate}
with the same notation introduced earlier.  The first case can be
easily discarded since for $\beta\neq 0$, $\xi$ is not everywhere
spacelike, regardless what $\rho_\perp$ and $\tau$ are.  Changing the
origin in the worldvolume of the brane, it is possible to set $\tau$
in the second case to $a\d_- + b \d_3$; but again unless $a=0$, $\xi$
will not be everywhere spacelike.  For a freely-acting $\xi$ one must
in addition have $\tau \neq 0$.  Similarly in the third and final
case, $\tau$ must be spacelike for $\xi$ to be everywhere spacelike,
hence we can conjugate $\tau$ to a spacelike direction orthogonal to
$\lambda$, say $\tau \propto \d_1$, where again for a free action
$\tau \neq 0$.  In this case the vector fields all integrate to a free
action because of the presence of the translation.  In summary, we
have three possible cases of freely-acting spacelike Killing vectors
in the M5-brane geometry:
\begin{enumerate}
\item[(A)] $\xi = \tau_\parallel + \rho_\parallel + \rho_\perp$, with
  $\tau \neq 0$ spacelike and where $\rho_\parallel$ can vanish; and
\item[(B)] $\xi = \tau_\parallel + \nu_\parallel + \rho_\parallel +
  \rho_\perp$, with $\nu_\parallel \neq 0$, $\tau \neq 0$ spacelike,
  and where $\rho_\parallel$ can vanish.
\end{enumerate}
We again remark that the above decompositions of $\xi$ are orthogonal
relative to the brane metric.

\subsubsection{Moduli space of smooth reductions}

In case (A) above, the Killing vector $\xi$ can be brought to the form
\begin{equation*}
  \xi = a \d_1 + \theta_1 R_{23} + \theta_2 R_{45} + \theta_3 R_{67} +
  \theta_4 R_{89}~,
\end{equation*}
with $a\neq 0$.  There are five free parameters, which after
projectivisation and modding out by the action of the Weyl group
yields a four-dimensional moduli space of smooth reductions and within
it a three-dimensional supersymmetric locus.

In case (B), the Killing vector $\xi$ can be brought to the form
\begin{equation*}
  \xi = a \d_1 + N_{+3} + \theta_2 R_{45} + \theta_3 R_{67} + \theta_4
  R_{89}~,
\end{equation*}
where $a \neq 0$.  There is now a three-dimensional moduli space of
smooth reductions and supersymmetry will select a two-dimensional
locus.

\subsubsection{Supersymmetry}

In both of the above cases the rotation component of the Killing
vector $\xi$ takes the general form
\begin{equation}
  \label{eq:so8rotation}
  \rho = \theta_1 R_{23} + \theta_2 R_{45} + \theta_3 R_{67} +
  \theta_4 R_{89}~.
\end{equation}
Relative to a basis dual to the $R_{ij}$ the weights of the subspace
$S_0$ of the half-spin representation of $\Spin(1,10)$ obeying
\eqref{eq:PM5} are given in equation \eqref{eq:weightsM5}.  The
supersymmetric locus is therefore the union of eight hyperplanes
\begin{equation}
  \label{eq:hyperplanesM5}
  \sum_{i=1}^4 \mu_i \theta_i = 0~,\qquad\text{where $\mu_i^2 = 1$.}
\end{equation}
(Again there are only eight hyperplanes, because the weights $\mu$ and
$-\mu$ determine the same hyperplane.)  A rotation $\rho$ belonging to
one and only one of these hyperplanes belongs to an $\fsu(4)$
subalgebra.  Two weights will annihilate such a rotation and hence the
associated reduction will preserve a fraction $\nu = \tfrac1{16}$ of
the supersymmetry.  Points which lie in the intersection of two
hyperplanes come in two flavours: those points where no $\theta$
vanish, which belong to an $\fsp(1) \times \fsp(1)$ subalgebra and
those for which one of the $\theta$s vanish, which belong to an
$\fsu(3)$ subalgebra.  In either case, such a rotation is annihilated
by four weights and hence the reduction will preserve a fraction $\nu
= \frac18$ of the supersymmetry.  Points which lie in the intersection
of three hyperplanes necessarily have two vanishing $\theta$s and
they belong to an $\fsu(2)$ subalgebra and their reductions preserve a
fraction $\nu = \frac14$ of the supersymmetry.  Finally the only point
which lies in the intersection of four hyperplanes (and hence in all
hyperplanes) is the origin.  This reduction preserves all the
supersymmetry of the M5-brane, hence a fraction $\nu=\half$.  This
concludes the analysis of case (A).  Case (B) corresponds to setting
$\theta_1 = 0$ and introducing a null rotation, whence the
supersymmetry is further halved.  There are now four hyperplanes in
the subspace $\theta_1=0$.  The generic points belong to an $\fsu(3)$
subalgebra and their reductions preserve a fraction $\nu=\frac1{16}$
of the supersymmetry.  Points in the intersection of two hyperplanes
belong to an $\fsu(2)$ subalgebra and preserve a fraction
$\nu=\frac18$.  Finally, the only point in the intersection of three
hyperplanes is the origin, which preserves a fraction $\nu=\frac14$ of
the supersymmetry.  This is summarised in Table~\ref{tab:M5susy}.

Notice that by constructing $\eM_{\text{M5}}/\Gamma_0$, where
$\Gamma_0\subset\Gamma$ is a discrete subgroup, the previous
classification gives rise to a whole set of smooth supersymmetric
eleven-dimensional configurations.  This set includes a stack of
M5-branes and eleven-dimensional fluxbranes $(\xi = \partial_1 +
\rho)$ or eleven-dimensional nullbranes $(\xi = \partial_1 +
\nu_\parallel)$.

\begin{table}[h!]
  \begin{center}
    \setlength{\extrarowheight}{5pt}
    \begin{tabular}{|c|>{$}c<{$}|>{$}c<{$}|>{$}c<{$}|}
      \hline
      \multicolumn{1}{|c|}{Null rotation?} & \text{Subalgebra} & \nu &
      \dim\\
      \hline
      \hline
      & \fsu(4) & \frac1{16} & 3\\
      & \fsu(3) & \frac18 & 2\\
      No & \fsp(1) \times \fsp(1) & \frac18 & 2\\
      & \fsu(2) & \frac14 & 1\\
      & \{0\} & \frac12 & 0\\[3pt]
      \hline
      & \fsu(3) & \frac1{16} & 2\\
      Yes & \fsu(2) & \frac18 & 1\\
      & \{0\} & \frac14 & 0 \\[3pt]
      \hline
    \end{tabular}
    \vspace{8pt}
    \caption{Supersymmetric reductions of the M5-brane.  All
      translations are spacelike and tangent to the M5-brane.  We
      indicate the spinor isotropy subalgebra to which the rotation
      belongs, the fraction $\nu$ of the supersymmetry preserved and
      the dimension of the corresponding stratum of the moduli space
      $\eM$ of supersymmetric reductions.}
    \label{tab:M5susy}
  \end{center}
\end{table}

\subsubsection{Explicit reductions}

It is possible to discuss the full set of inequivalent Kaluza-Klein
reductions of the M5-brane by a single computation, the one associated
with reductions along the orbits of the Killing vector $\xi =
\partial_z + \lambda$, where
\begin{equation*}
  \lambda = \beta B_{03} + \theta_1 R_{23} + \theta_2 R_{45}
  + \theta_3 R_{67} + \theta_4 R_{89}~,
\end{equation*}
and $z$ stands for the $x^1$ direction along the M5-brane.

The constant matrix $B$ is a $9\times 9$ matrix, which does not act on
the $x^\natural$ coordinate and which equals, formally, the one
appearing in \eqref{eq:Bmatrixm2c}, but this time in the basis
$\{x^0,x^2,\dots,x^9\}$.  The ten-dimensional metric obtained after
Kaluza-Klein reduction can be written as,
\begin{footnotesize}
\begin{multline}
\label{met:m5a}
g = \tilde{\Lambda}^{1/2}\left\{V^{-1/2}ds^2(\EE^{1,4})
  + V^{1/2} ds^2(\EE^5)\right\}\\
- \tilde{\Lambda}^{-1/2}V^{-1/2} \left\{\beta \omega^{03} + \theta_1
  \omega^{23} + \theta_2 \omega^{45} + V\left[\theta_3 \omega^{67} +
    \theta_4 \omega^{89}\right]\right\}^2~,
\end{multline}
\end{footnotesize}
whereas the RR 1-form $A_{1}$, RR 4-form $H_4$ and dilaton 
$\Phi$ are listed below:
\begin{equation}
\label{oth:m5a}
  \begin{aligned}[m]
    A_1 &= \tilde{\Lambda}^{-1}\left\{\beta \omega^{03} +
    \theta_1 \omega^{23} + \theta_2 \omega^{45} +
    V\left[\theta_3 \omega^{67} + \theta_4 \omega^{89} \right]
  \right\} \\
    H_4 &= -\star \dvol\left(\EE^{1,4}\right)\wedge dV^{-1} \\ 
    \Phi &= \tfrac34\log \left(\tilde{\Lambda}\cdot V^{-1/3}\right)~.
  \end{aligned}
\end{equation}
The configuration depends on an scalar function $\tilde{\Lambda}$
which is defined in terms of the scalar function $\Lambda$ appearing
in the general discussion section, by
\begin{equation*}
\Lambda = V^{-1/3}\cdot \tilde{\Lambda}~,
\end{equation*}
and equals
\begin{multline*}
      \tilde{\Lambda} = 1 + (x^3)^2\left[(\theta_1)^2 - \beta^2\right]
      + 
      \left(\beta x^0 + \theta_1 x^2\right)^2 +
      \left(\theta_2\right)^2\left[(x^4)^2 + (x^5)^2\right] \\
      + V\left\{\left(\theta_3\right)^2\left[(x^6)^2 +(x^7)^2\right]
      +\left(\theta_4\right)^2\left[(x^8)^2 +(x^9)^2\right]\right\}~.
\end{multline*}

As discussed for the M2-brane reductions, whenever
$|\theta_1|<|\beta|$, there is always a Lorentz observer who sees a
pure boost.  Such spacetime breaks supersymmetry and contains closed
timelike curves.  If we restrict to the regions of spacetime where
such closed causal curves do not exist, their interpretation would
give rise to similar cosmological models to the ones discussed in
\cite{KOSST, SeibergBC, CorCos, Nekrasov} but this time on the
worldvolume of a 1+5 brane.  On the other hand, whenever
$|\theta_1|>|\beta|$, there is always an observer who measures a pure
rotation, so that case would be related to fluxbranes.

Let us concentrate on the interpretation of the different regions of
the above reduction.  If we set $\beta=0$, there are five different
possibilities to be considered:
\begin{enumerate}
\item[(1)] The case $\theta_i=0$ for all $i$ corresponds to the
  well-known D4-brane in type IIA preserving $\nu=1/2$ of the
  spacetime supersymmetry.
  
\item[(2)] If one of the $\theta$s is non-vanishing, the configuration
  describes a composite state involving a D4-brane and an F7-brane.
  Depending on the chosen $\theta$, its location is a
  ($456789\natural$)-plane at $x^3=x^4=0$ (if $\theta_1\neq 0$) or a
  ($234589\natural$)-plane at $x^6=x^7=0$ (if $\theta_2\neq 0$).  In
  either case, supersymmetry is completely broken due to the presence
  of these F7-branes.
  
\item[(3)] If there are two non-vanishing $\theta$s we must
  distinguish between four cases:
  \begin{enumerate}
  \item[(i)] If $\theta_1=\eta_2\theta_2$, the configuration describes
    a D4-brane in the (2345)-plane and an F5-brane along the
    ($6789\natural$)-plane sitting at $x^2=x^3=x^4=x^5=0$.  It
    preserves $\nu=1/4$ of the spacetime supersymmetry, with Killing
    spinors being preserved by an $\fsu(2)$ subalgebra.
    
  \item[(ii)] If $\theta_3=\eta_4\theta_4$, the configuration
    describes a D4-brane in the (2345)-plane and an F5-brane along the
    ($2345\natural$)-plane sitting at $x^6=x^7=x^8=x^9=0$.  It
    preserves $\nu=1/4$ of the spacetime supersymmetry, with Killing
    spinors being preserved by an $\fsu(2)$ subalgebra.
    
  \item[(iii)] If $\theta_1=\eta_3\theta_3$, the configuration
    describes a D4-brane in the (2345)-plane and an F5-brane along the
    ($4589\natural$)-plane sitting at $x^2=x^3=x^6=x^7=0$.  It
    preserves $\nu=1/4$ of the spacetime supersymmetry, with Killing
    spinors again preserved by an $\fsu(2)$ subalgebra.
    
  \item[(iv)] If $\theta_i\neq \eta_j\theta_j$ $\text{i}\neq
    \text{j}$, the configuration describes a system of two
    intersecting F7-branes besides the aforementioned D4-brane.  It
    breaks supersymmetry completely.
  \end{enumerate}
  
\item[(4)] When there are three non-vanishing $\theta$s, we must
  distinguish between three cases:
  \begin{enumerate}
  \item[(i)] If $\theta_1=\eta_2\theta_2+\eta_3\theta_3$, the
    configuration describes a D4-brane in the (2345)-plane and an
    F3-brane along the ($89\natural$)-plane sitting at $x^k=0$
    $k=2,\dots ,7$.  It preserves $\nu=1/8$ of the spacetime
    supersymmetry, with Killing spinors being preserved by an
    $\fsu(3)$ subalgebra.
    
  \item[(ii)] If $\theta_2=\eta_3\theta_3+\eta_4\theta_4$, the
    F3-brane extends along the ($23\natural$)-plane sitting at 
    $x^k=0$ $k=4,\dots,9$.  It preserves the same amount of
    supersymmetry as the previous one due to the existence of an
    $\fsu(3)$ subalgebra preserving some Killing spinors.
    
  \item[(iii)] If $\theta_i\neq \eta_j\theta_j + \eta_k\theta_k$ 
    for all $i,j,k$ distinct, the configuration describes the
    intersection of three different F7-branes plus a D4-brane.
    Spacetime supersymmetry is completely broken.
  \end{enumerate}
  
\item[(5)] When four $\theta$s are nonvanishing, there are generically
  two inequivalent ways of preserving Killing spinors, but due to the
  symmetries of our configuration, these split into four:
  \begin{enumerate}
  \item[(i)] If $\theta_1=\eta_2\theta_2 + \eta_3\theta_3 +
    \eta_4\theta_4$, the configuration describes a D4-brane in the
    (2345)-plane and a flux string along the $x^\natural$ direction,
    preserving $\nu=1/16$ of the spacetime supersymmetry.  This is the
    one associated with the $\fsu(4)$ isotropy algebra discussed
    before.
    
  \item[(ii)] If $\theta_1=\eta_2\theta_2$ and
    $\theta_3=\eta_4\theta_4$, the configuration involves a D4-brane
    and a $\tfrac14$-BPS fluxstring in the $x^\natural$ direction.  It
    preserves $\nu=1/8$ and it is associated with the
    $\fsp(1)\times\fsp(1)$ isotropy algebra.
    
  \item[(iii)] The case $\theta_1=\eta_3\theta_3$ and
    $\theta_2=\eta_4\theta_4$, has the same interpretation as the
    previous one, but the $\fsp(1)\times\fsp(1)$ isotropy algebra is
    selected in a different way.
    
  \item[(iv)] If non of the three previous possibilities are
    satisfied, there are four intersecting F7-branes and a stack of
    coincident D4-branes breaking spacetime supersymmetry completely.
  \end{enumerate}
\end{enumerate}

If $\beta\neq 0$, the only allowed possibility preserving
supersymmetry requires $|\theta_1|=|\beta|$.  Let us thus concentrate
on this case.  Depending on whether the remaining parameters vanish or
satisfy certain linear relations, we distinguish between the following
configurations
\begin{enumerate}
\item[(1)] If $\theta_i=0$ for $i=2,3,4$, it describes a D4-brane in
  the (2345)-plane and a nullbrane.  This composite configuration
  preserves $\nu=1/4$ of the spacetime supersymmetry.
  
\item[(2)] If one of the $\theta$s is non-vanishing, the corresponding
  configuration breaks supersymmetry completely, due to the presence
  of an F7-brane besides the previous D4/nullbrane pair.
  
\item[(3)] If two of the $\theta$s are non-vanishing, we shall
  distinguish between three cases:
  \begin{enumerate}
  \item[(i)] If $\theta_i\neq \eta_j\theta_j$ $i\neq j$, the
    configuration describes a system of two intersecting F7-branes
    (besides the composite system of a D4-brane and a nullbrane), such
    that all supersymmetry is broken.
   
  \item[(ii)] If $\theta_2=\eta_3\theta_3$, the intersection among the
    F7-branes gives rise to the so-called F5-brane, extending along
    the ($2389\natural$)-plane and sitting at $x^4=x^5=x^6=x^7=0$.
    The full configuration includes the previous pair D4/nullbrane
    system, thus preserving $\nu=1/8$.
    
  \item[(iii)] If $\theta_3=\eta_4\theta_4$, one finds a second
    F5-brane, this time extending along the 2345$\natural$-plane and
    sitting at $x^6=x^7=x^8=x^9=0$.  As before, the configuration
    preserves $\nu=1/8$.
  \end{enumerate}
\item[(4)] If all $\theta$s are non-vanishing, we need to distinguish
  between two cases:
  \begin{enumerate}
  \item[(i)] If $\theta_2\neq \eta_3\theta_3 + \eta_4\theta_4$, the
    configuration describes the intersection of three intersecting
    F7-branes, besides the composite D4/nullbrane system, thus
    breaking all supersymmetry.
  \item[(ii)] If $\theta_2=\eta_3\theta_3 + \eta_4\theta_4$, the
    intersection of the F7-branes gives rise to an F3-brane extending
    along the ($23\natural$)-plane and sitting at $x^k=0$
    k=$4,\dots,9$.  Thus, there exists a composite configuration
    involving a D4-brane, a nullbrane and an F3-brane preserving
    $\nu=1/16$.
  \end{enumerate}
\end{enumerate}

The set of supersymmetric configurations described above is summarised
in Table~\ref{tab:M5susy1}, where the notation
$\text{Dp}\perp\text{Fq} (r)$ has been introduced.  The latter stands
for a D$p$-brane+F$q$-brane composite configuration sharing $r$
spacelike directions.

\begin{table}[h!]
  \begin{center}
    \setlength{\extrarowheight}{3pt}
    \begin{tabular}{|>{$}c<{$}|c|>{$}l<{$}|}
      \hline
      \nu & Object & \multicolumn{1}{c|}{Subalgebra}\\
      \hline
      \hline
              & D4$\perp$F5(0) &  \\
      \frac14 & D4$\perp$F5(4) & \fsu(2)\\
              & D4$\perp$F5(2) & \\
      \frac14 & D4 + N & \RR \\
      \frac18 & D4$\perp$F3(0) & \fsu(3) \\
              & D4$\perp$F3(2) & \\
      \frac18 & (D4$\perp$F5(4)) + N & \fsu(2)\times \RR \\
              & (D4$\perp$F5(2)) + N & \\
      \frac18 & D4$\perp$F1(0) & \fsp(1)\times\fsp(1) \\
      \frac1{16} & D4$\perp$F1(0) & \fsu(4)\\
      \frac1{16} & (D4$\perp$F3(2)) + N & \fsu(3)\times \RR \\[3pt]
      \hline
    \end{tabular}
    \vspace{8pt}
    \caption{Supersymmetric configurations of D4-branes, fluxbranes
      and nullbranes.}
    \label{tab:M5susy1}
  \end{center}
\end{table}

Before finishing this presentation, we would like to compute the
fluxes associated with F5-branes in the presence of D4-branes.  There
are three cases to be considered separately, as indicated in
Table~\ref{tab:M5susy1}, and discussed in the text above.  Whenever
the D4-branes are parallel to the F5-brane,
$\theta_1=\theta_2=\beta=0,\,\theta\equiv\theta_3= \theta_4$, the RR
1-form potential in \eqref{oth:m5a} depends on the direction
$x^\natural$ along the F5-brane but transverse to the D4-branes.  As
it happened for fundamental strings, it can be shown that the flux
carried by the F5-brane equals the one on flat spacetime, except at
$x^\natural =0$, where it vanishes.  It is precisely at
$x^\natural=0$, where the D4-branes lie.  A probe computation shows
that indeed D4-branes with the above orientation are only stable where
the flux vanishes.  On the other extreme, when the D4-branes are
completely tranverse to the F5-branes $\theta_3=\theta_4
=\beta=0,\,\theta\equiv\theta_1=\theta_2$, the RR 1-form potential in
\eqref{oth:m5a} is independent of the worldvolume F5-brane point.
Therefore the flux equals the one carried in flat spacetime
everywhere.  Finally, when they are relatively transverse,
$\theta_2=\theta_4 =\beta=0,\,\theta\equiv\theta_1=\theta_3$, the RR
1-form potential in \eqref{oth:m5a} depends on the radial distance in
the $89\natural$-plane spanned by the F5-brane.  By fixing a point in
this plane, it can be shown that
\begin{equation*}
  F_2\wedge F_2 \propto -2\frac{\partial\left(r_1
      \tilde{\Lambda}^{-2}\right)} {\partial\,r_2} dr_1\wedge
  d\theta_1\wedge dr_2\wedge d\theta_2~,
\end{equation*}
where we used the same parametrisation as for the corresponding
M2-brane discussion.  Its integral over $\RR^4$ equals the one in flat
spacetime everywhere except at the origin of the ($89\natural$)-plane,
where it vanishes.

\subsection{Supersymmetric reductions of the delocalised M5-brane}
\label{sec:M5d}

To obtain the NS5-brane by Kaluza--Klein reduction of the M5-brane, it
is necessary to delocalise the M5-brane along a transverse direction.
In this section we will classify the supersymmetric Kaluza--Klein
reductions of such a delocalised M5-brane.  The metric of the
spacetime exterior to such a fivebrane is again of the general form
\eqref{eq:elemint} but now with three factors:
\begin{equation}
  \label{eq:mM5d}
  g = V^{-1/3} ds^2(\EE^{1,5}) + V^{2/3} dz^2 + V^{2/3} ds^2(\EE^4)~,
\end{equation}
where $z$ is the transverse coordinate along which the fivebrane is
delocalised and $V = 1 + |Q|/r^2$ is a harmonic function on $\EE^4$
depending only on the radial distance.  The symmetry group is now
\begin{equation}
  \label{eq:GM5d}
  G = \ISO(1,5) \times \RR \times \SO(4)~,
\end{equation}
with Lie algebra
\begin{equation}
  \label{eq:gM5d}
  \fg = \left( \RR^{1,5} \rtimes \fso(1,5) \right) \times \RR \times
  \fso(4)~.
\end{equation}
Therefore a Killing vector may be decomposed as
\begin{equation}
  \label{eq:KVM5d}
  \xi = \tau_\parallel + \tau_\perp + \lambda_\parallel + \rho_\perp
\end{equation}
in the usual notation.

\subsubsection{Freely-acting spacelike isometries}

As for the localised fivebrane, either the Lorentz transformation
$\lambda=0$ or else it can be brought to one the following normal
forms:
\begin{enumerate}
\item $\lambda = \beta B_{01} + \theta_1 R_{23} + \theta_2 R_{45}$,
  with $\beta \neq 0$;
\item $\lambda = N_{+2} + \theta R_{45}$; or
\item $\lambda = \theta_1 R_{23} + \theta_2 R_{45}$.
\end{enumerate}
It is again easy to discard the first case, since $\beta\neq 0$ means
that $\xi$ is not everywhere spacelike.  In the second case,
$\tau_\parallel$ is spacelike and $\tau_\parallel + \tau_\perp$ cannot
be zero, because otherwise the action is not free: as in the M2 case,
there would be points outside the horizon with nontrivial stabilisers.
As in the M2-brane, the norm of the transverse rotation $\rho_\perp$
obeys a sharp bound
\begin{equation*}
  r^2 M^2 \geq \|\rho_\perp\|_\infty^2 \geq r^2 m^2~,
\end{equation*}
where $M \geq m \geq 0$ and $m$ can be nonzero since the transverse
sphere is three-dimensional and possesses infinitesimal isometries
without zeros.  This means that in the third case, the norm of the
Killing vector is bounded below by
\begin{equation*}
  \|\xi\|^2 \geq V^{-1/3} \left( \|\tau_\parallel\|^2_\infty +
  \|\rho_\parallel\|^2_\infty \right) + V^{2/3} \left(
  \|\tau_\perp\|^2_\infty + r^2 m^2 \right)~,
\end{equation*}
which is again sharp.  Because there are points where
$\rho_\parallel=0$, the bound can be improved to
\begin{equation*}
  \|\xi\|^2 \geq V^{-1/3} \|\tau_\parallel\|^2_\infty + V^{2/3} \left(
  \|\tau_\perp\|^2_\infty + r^2 m^2 \right)~,
\end{equation*}
which is still sharp.  The right-hand side of the above bound defines
a function $f$ of $r$ with the following asymptotic properties.  As
$r \to 0$,
\begin{equation*}
  f(r) = |Q|^{2/3} \|\tau_\perp\|^2_\infty r^{-4/3} + O(r^{2/3})~,  
\end{equation*}
whence it blows up if $\tau_\perp \neq 0$ and goes to zero otherwise.
In the asymptotic regime where $r\to\infty$,
\begin{equation*}
  f(r) = m^2 r^2 + \|\tau_\parallel\|^2_\infty +
  \|\tau_\perp\|^2_\infty + O(r^{-1})~,
\end{equation*}
which blows up for $m > 0$, and approaches (the square of) the flat
norm of $\tau_\parallel + \tau_\perp$ otherwise.  Therefore,
generically $f$ has a minimum at some critical value $r_0>0$; although
if either $\tau_\perp$ or $m$ vanish this may be either zero or not
exist, respectively.  As in the case of the M2-brane, there is a
positive number $\mu$ such that $\xi$ is everywhere spacelike if and
only if $\|\tau_\parallel\|^2_\infty > - \mu^2$.  This number is
obtained by solving $f'(r_0)=0$, which gives $\mu$ as a function of
$r_0$, and substituting this into $f(r_0)=0$ which can be solved for
$r_0$ and hence for $\mu$.  One finds that the critical radius obeys
\begin{equation}
  \label{eq:rcritM5d}
  r_0^4 m^2 = |Q| \|\tau_\perp\|^2_\infty~,
\end{equation}
should one exist (it does if $m>0$) and hence that
\begin{equation}
  \label{eq:muM5d}
  \mu = \|\tau_\perp\|_\infty + m |Q|^{1/2}~.
\end{equation}
If $m=0$ we see that it is enough that $\tau_\parallel + \tau_\perp$
be asymptotically spacelike.

In summary, the freely-acting spacelike Killing vectors of the
delocalised fivebrane geometry fall into two cases:
\begin{enumerate}
\item[(A)] $\xi = \tau_\parallel + \tau_\perp + \nu_\parallel +
  \rho_\parallel + \rho_\perp$, with $\nu_\parallel \neq 0$,
  $\tau_\parallel + \tau_\perp$ spacelike and where $\tau_\parallel$
  if nonzero must also be spacelike; and
\item[(B)] $\xi = \tau_\parallel + \tau_\perp + \rho_\parallel +
  \rho_\perp$, with $\tau_\parallel$ satisfying a norm constraint of
  the form $\|\tau_\parallel\|_\infty^2 > -\mu^2$, where $\mu$ is
  given by equation \eqref{eq:muM5d}.  In particular, if $\rho_\perp$
  has zeros, then $\tau_\parallel + \tau_\perp$ must be asymptotically
  spacelike.
\end{enumerate}
The rotation $\rho_\parallel$ is allowed to vanish in both cases and,
once again, the decompositions of $\xi$ are orthogonal with respect to
the brane metric.  In both cases, the translation $\tau_\parallel +
\tau_\perp$ must be nonzero for the action of $\xi$ to be free.

\subsubsection{Moduli space of smooth reductions}

In case (A) we can always choose coordinates so that the Killing
vector $\xi$ takes the form
\begin{equation*}
  \xi = a \d_1 + b \d_z + N_{+2} + \theta_2 R_{45} + \theta_3 R_{67} +
  \theta_4 R_{89}~,
\end{equation*}
where $a$ and $b$ cannot both be zero.  There are five free parameters
which yield a four-dimensional moduli space of smooth reductions after
we projectivise and quotient by the action of the (discrete) Weyl
group.  Supersymmetry will then select a three-dimensional locus.

Case (B) breaks up into three cases depending on the nature of
$\tau_\parallel$: whether it is timelike, spacelike or null.  In the
first case we can bring $\xi$ to the form
\begin{equation*}
  \xi = a \d_0 + b \d_z + \theta_1 R_{23} + \theta_2 R_{45} + \theta_3
  R_{67} + \theta_4 R_{89}~,
\end{equation*}
which gives rise to a five-dimensional moduli space of smooth
reductions with a four-dimensional supersymmetric locus.  The results
are similar for $\tau_\parallel$ spacelike.  Finally, if $\xi$ is null
then we can always bring it to the form
\begin{equation*}
  \xi = \d_+ + b \d_z + \theta_1 R_{23} + \theta_2 R_{45} + \theta_3
  R_{67} + \theta_4 R_{89}~,
\end{equation*}
whence we have one less parameter.  Therefore the moduli space of
smooth reductions will be four-dimensional with a codimension-one
locus of supersymmetric reductions.

\subsubsection{Absence of closed causal curves}
\label{sec:M5dCCC}

It is rather straightforward to extend the proofs of absence of closed
causal curves given for the quotients of the M2-brane background
to the M5-brane background discussed here.  As in that case, we shall
set all $\theta_i$ and $\beta$ to zero.

Let us start by analysing the problem of existence of closed causal
curves in an M5-brane background delocalised in one transverse
direction ($z$) reduced along the orbits of the Killing vector
\begin{equation*}
  \xi = a\partial_0 + b\partial_z~.
\end{equation*}
The only condition that such a Killing vector is required to satisfy
is to be spacelike everywhere.  This requirement provides us with
the constraint
\begin{equation}
  \|\xi\|^2 = V(r)^{-1/3}\left(-a^2 + V(r)b^2\right) > 0~.
 \label{space3}
\end{equation}
Writing the metric in an adapted coordinate system, in which 
$\xi=\partial_{z'}$, one finds
\begin{multline*}
  g = -V^{-1/3}\left[a^2 (dz')^2 + b^{-2}(dt')^2 + 2ab^{-1}
  dz' dt'\right] + b^2 V^{2/3}(dz')^2 \\
  + V^{-1/3}ds^2(\EE^5) + V^{2/3}ds^2(\EE^4)~.
\end{multline*}

Let us assume the existence of causal curves $x(\lambda)$, i.e.
$\|\frac{dx}{d\lambda}\|^2 \leq 0$, joining the points $(t'_0, x^i_0,
z'_0)$ and $(t'_0, x^i_0, z'_0 + \Delta)$.  As argued for the M2-brane
background, there must exist at least one value of the affine
parameter $\lambda$ where the timelike component of the tangent vector
to the curve vanishes:
\begin{equation*}
  \exists \lambda^* \quad \text{such that} \quad
  \left.\frac{dt'}{d\lambda}\right|_{\lambda^*}=0~.
\end{equation*}

If one computes the norm of such a tangent vector at $\lambda^*$, one
derives the inequality
\begin{equation*}
  \|\xi\|^2(\lambda^*)\left.\frac{dz'}{d\lambda}
  \right|^2_{\lambda^*} + V^{-1/3}\sum_{i=1}^5
  \left.\frac{dx^i}{d\lambda}\right|^2_{\lambda^*} +
  V^{2/3}\sum_{i=6}^9 \left.\frac{dx^i}{d\lambda}\right|^2_{\lambda^*}
  \leq 0~.
\end{equation*}
Due to the constraint \eqref{space3}, it is clear that the left hand
side of the above norm is the sum of positive-definite terms, so that
the inequality can never be satisfied.  This already shows the
non-existence of closed timelike curves.  Furthermore, the only
possibility for the equality to be satisfied is whenever for all $i$, 
\begin{equation*}
  \left.\frac{dz'}{d\lambda}\right|_{\lambda^*} =
  \left.\frac{dx^i}{d\lambda}\right|_{\lambda^*} = 0~,
\end{equation*}
which violates the definition of $\lambda$ being an affine parameter.
We thus conclude that no closed lightlike curves are allowed in this
spacetime.

The corresponding proof for the action generated by $\xi = \partial_+
+ b\partial_z$ involves similar ideas and techniques.  In this case,
the requirement of having an everywhere spacelike Killing vector gives
rise to the condition
\begin{equation*}
  V(r)^{2/3}\,b^2 > 0 \quad \Rightarrow \quad |b| > 0~.
\end{equation*}
By a linear transformation, we can move to an adapted coordinate
system $\{z',x^{+\prime}\}$ in which the Killing vector becomes
a single translation $\xi=\partial_{z'}$ and the eleven-dimensional
metric takes the form
\begin{multline*}
  g = 2V^{-1/3}dx^-\left(dz' + b^{-1}dx^{+\prime}\right) +
  V^{2/3}b^2 (dz')^2 \\
  + V^{2/3}ds^2(\EE^4) + V^{-1/3}ds^2(\EE^4)^2~.
\end{multline*}

If we again assume the existence of a closed causal curve of affine
parameter $\lambda$ joining the points $(x^{+\prime}_0, x^-_0,x^i_0,
z'_0)$ and $(x^{+\prime}_0, x^-_0,x^i_0, z'_0 + \Delta)$, there must
necessarily exist at least one value for this affine parameter
$\lambda^*$ where
\begin{equation*}
  \exists \lambda^* \quad \text{such that} \quad
  \left.\frac{dx^-}{d\lambda}\right|_{\lambda^*}=0~.
\end{equation*}
By computing the norm of the tangent vector to the causal curve at the
point $\lambda^*$, and using the fact that $|b|>0$, it is immediate to
show the non-existence of such closed causal curves by the same
argument used before.

\subsubsection{Supersymmetry}

The determination of the supersymmetric locus can be read off from the
results of the M5-brane.  We will not repeat the arguments simply
state the results, which are contained in Table~\ref{tab:M5dsusy}.

\begin{table}[h!]
  \begin{center}
    \setlength{\extrarowheight}{5pt}
    \begin{tabular}{|>{$}c<{$}|>{$}c<{$}|>{$}c<{$}|>{$}c<{$}|}
      \hline
      \multicolumn{1}{|c|}{Translation} & \text{Subalgebra} & \nu &
      \dim\\
      \hline
      \hline
      & \fsu(4) & \frac1{16}& 4\\
      a\d_1 + b \d_z & \fsp(1) \times \fsp(1) & \frac18& 3\\
      & \fsu(3) & \frac18~\left(\frac1{16}\right)& 3~(3)\\
      a,b~\text{not both $0$}& \fsu(2) & \frac14~\left(\frac18\right)&
      2~(2)\\
      & \{0\} & \frac12~\left(\frac14\right)& 1~(1)\\[3pt]
      \hline
      & \fsu(4) & \frac1{16}& 3\\
      \d_+ + b \d_z & \fsp(1) \times \fsp(1) & \frac18& 2\\
      & \fsu(3) & \frac18 & 2\\
      b \neq 0 & \fsu(2) & \frac14 & 1\\
      & \{0\} & \frac12& 0\\[3pt]
      \hline
      & \fsu(4) & \frac1{16} & 4\\
      a\d_0 + b\d_z & \fsp(1) \times \fsp(1) & \frac18 & 3\\
      a,b~\text{not both $0$}& \fsu(3) & \frac18 & 3\\
      |a|<\mu& \fsu(2) & \frac14 & 2\\
      & \{0\} & \frac12 & 1 \\[3pt]
      \hline
    \end{tabular}
    \vspace{8pt}
    \caption{Supersymmetric reductions of the delocalised M5-brane.
      We indicate the form of the translation, the spinor isotropy
      subalgebra to which the rotation belongs, the fraction $\nu$ of
      the supersymmetry preserved and the dimension of the
      corresponding stratum of the moduli space $\eM$ of
      supersymmetric reductions.  The numbers in parentheses indicate
      the values in the presence of a null rotation.}
    \label{tab:M5dsusy}
  \end{center}
\end{table}

\subsubsection{Explicit reductions}

Let us start the discussion on the explicit configurations by
concentrating on the region of the moduli space involving an extra
spacelike translation in addition to the one on the delocalised
transverse direction $z$.  In other words, we shall start by reducing
along the orbits of the Killing vector $\xi=\partial_z + \alpha$,
where
\begin{equation*}
  \alpha = a\partial_1 + \beta B_{02} + \theta_1 R_{23} + \theta_2
  R_{45} + \theta_3 R_{67} + \theta_4 R_{89}~.
\end{equation*}
Notice that by parametrising the reduction in this way, we will be
able to discuss both the possibility of fluxbranes and nullbranes at
the same time.

The constant matrix $B$ is now $10\times 10$.  In the basis
$\{x^0,\,x^1,\dots,x^9\}$, it can be written as
\begin{equation}
  \label{eq:Bmatrixm5a}
  B= 
  \begin{pmatrix}
    0 & 0 & \beta & 0 & 0 & 0 & 0 & 0 & 0 & 0 \\
    0 & 0 &  0 & 0 & 0 & 0 & 0 & 0 & 0 & 0 \\
    \beta  & 0 & 0 & -\theta_1 & 0 & 0 & 0 & 0 & 0 & 0 \\
    0 & 0 & \theta_1 & 0 & 0 & 0 & 0 & 0 & 0 & 0 \\
    0 & 0 & 0 & 0 & 0 & -\theta_2 & 0 & 0 & 0 & 0 \\
    0 & 0 & 0 & 0 & \theta_2 & 0 & 0 & 0 & 0 & 0 \\
    0 & 0 & 0 & 0 & 0 & 0 & 0 & -\theta_3 & 0 & 0 \\
    0 & 0 & 0 & 0 & 0 & 0 & \theta_3 & 0 & 0 & 0 \\
    0 & 0 & 0 & 0 & 0 & 0 & 0 & 0 & 0 & -\theta_4 \\
    0 & 0 & 0 & 0 & 0 & 0 & 0 & 0 & \theta_4 & 0 \\
  \end{pmatrix}~.
\end{equation}
Whenever the extra spacelike translation is non-vanishing $(a\neq 0)$,
there will be a non-vanishing 10-vector $\bC$ taking care of the
inhomogeneous part of the symmetry transformation.  In the same basis
as the one used in the matrix \eqref{eq:Bmatrixm5a}, this vector is
\begin{equation*}
  (\bC)^t = (0,\,a,\vec{0})~.
\end{equation*}

The corresponding type IIA configurations have a ten-dimensional
metric given by
\begin{multline}
\label{met:m5b}
  g = \tilde{\Lambda}^{1/2}\left\{ds^2(\EE^{1,5}) 
    + V\,ds^2(\EE^4)\right\}\\
  - \tilde{\Lambda}^{-1/2}V \left\{V^{-1}\left[adx^1 +
      \beta\omega^{02}  + \theta_1 \omega^{23} + \theta_2
      \omega^{45}\right] + \theta_3 \omega^{67} + \theta_4 \omega^{89}
  \right\}^2~,
\end{multline}
whereas the RR 1-form $A_{1}$, NS-NS 7-form field strength $H_{7}$,
RR 6-form $H_6$ and dilaton $\Phi$ are listed below,
\begin{footnotesize}
\begin{equation}
\label{oth:m5b}
  \begin{aligned}[m]
    A_1 &= \tilde{\Lambda}^{-1}\left\{V^{-1}\left[adx^1 + 
        \beta \omega^{02} + \theta_1 \omega^{23} + \theta_2
        \omega^{45}\right] + \theta_3 \omega^{67} + \theta_4
      \omega^{89} \right\} \\
    H_7 &= \dvol\left(\EE^{1,5}\right)\wedge dV^{-1} \\
    H_6 &= -\beta (x^2dx^2-x^0dx^0)\wedge dx^1\wedge dx^3\wedge
    dx^4\wedge dx^5 \wedge dV^{-1} \\
    & - a dx^0\wedge dx^2\wedge dx^3 \wedge dx^4 \wedge dx^5 \wedge
    dV^{-1} \\
    & - \theta_1 dx^0\wedge dx^1\wedge (x^2 dx^2+ x^3 dx^3)\wedge
    dx^4\wedge dx^5 \wedge dV^{-1} \\
    & - \theta_2 dx^0\wedge dx^1\wedge dx^2 \wedge dx^3
    \wedge (x^4dx^4 + x^5 dx^5)\wedge dV^{-1} \\ 
    \Phi &= \tfrac34\log \left(\tilde{\Lambda}\cdot V^{2/3}\right)~.
  \end{aligned}
\end{equation}
\end{footnotesize}

The configuration depends on an scalar function $\tilde{\Lambda}$
which is defined in terms of the scalar function $\Lambda$ appearing
in the general discussion section, by
\begin{equation*}
\Lambda = V^{2/3}\cdot \tilde{\Lambda}~,
\end{equation*}
and equals
\begin{footnotesize}
  \begin{multline*}
    \tilde{\Lambda} = 1 + V^{-1}\left\{a^2 + (x^2)^2(\theta_1^2 -
      \beta^2) + (\beta x^0 - \theta_1 x^3)^2\right\} +
    \left(\theta_2\right)^2\left[(x^3)^2 + (x^4)^2\right] \\
    + \left(\theta_3\right)^2\left[(x^6)^2 +(x^7)^2\right]
    +\left(\theta_4\right)^2\left[(x^8)^2 +(x^9)^2\right]~.
  \end{multline*}
\end{footnotesize}

We shall discuss the interpretation of the different solutions
proceeding in an analogous way to the one followed for the M2-brane
Kaluza--Klein reductions.  Thus, let us start by examining the
subspace of the moduli space of reductions defined by $a=0$.  We
already know that whenever $|\theta_1|<|\beta|$, there is always a
Lorentz observer who sees a pure boost.  Such spacetime breaks
supersymmetry and contains closed timelike curves.  Restricting
ourselves to regions of spacetime where such closed causal curves do
not exist, their interpretation would give rise to similar
cosmological models to the ones discussed in \cite{KOSST,
  SeibergBC, CorCos, Nekrasov}.  On the other hand, whenever
$|\theta_1|>|\beta|$, there is always an observer who measures a pure
rotation, so that case would be related to fluxbranes.

After this brief comment, let us study the subset defined by
$\beta=0$, that is, the one involving no nullbranes.  There are five
different possibilities to be considered:
\begin{enumerate}
\item[(1)] If all $\theta_i=0$ we have the well-known NS5-brane in
  type IIA preserving $\nu=1/2$ of the spacetime supersymmetry.
  
\item[(2)] If one of the $\theta$s is non-vanishing, the configuration
  describes a composite state involving an NS5-brane and an F7-brane.
  Depending on the chosen $\theta$, its location is a
  ($456789\natural$)-plane at $x^3=x^4=0$ (if $\theta_1\neq 0$) or a
  ($234589\natural$)-plane at $x^6=x^7=0$ (if $\theta_2\neq 0)$.  In
  either case, supersymmetry is completely broken due to the presence
  of these F7-branes.
  
\item[(3)] If there are two non-vanishing $\theta$s, we must
  distinguish between four cases:
  \begin{enumerate}
  \item[(i)] If $\theta_1=\eta_2\theta_2$, the configuration describes
    a NS5-brane in the (12345)-plane and an F5-brane along the
    (16789)-plane sitting at $x^2=x^3=x^4=x^5=0$.  It preserves
    $\nu=1/4$ of the spacetime supersymmetry, with Killing spinors
    being preserved by an $\fsu(2)$ subalgebra.
    
  \item[(ii)] If $\theta_3=\eta_4\theta_4$, the configuration
    describes an NS5-brane in the (12345)-plane and an F5-brane along
    the (12345)-plane sitting at $x^6=x^7=x^8=x^9=0$.  It preserves
    $\nu=1/4$ of the spacetime supersymmetry, with Killing spinors
    being preserved by an $\fsu(2)$ subalgebra.
      
  \item[(iii)] If $\theta_1=\eta_3\theta_3$, the configuration
    describes an NS5-brane in the (12345)-plane and an F5-brane along
    the (14589)-plane sitting at $x^2=x^3=x^6=x^7=0$.  It preserves
    $\nu=1/4$ of the spacetime supersymmetry, with Killing spinors
    again preserved by an $\fsu(2)$ subalgebra.
    
  \item[(iv)] If $\theta_i\neq \eta_j\theta_j$ for distinct $i,j$, the
    configuration describes a system of two intersecting F7-branes
    besides the aforementioned NS5-brane.  It breaks supersymmetry
    completely.
  \end{enumerate}
  
\item[(4)] If there are three non-vanishing $\theta$s, we must
  distinguish among three cases:
  \begin{enumerate}
  \item[(i)] If $\theta_1=\eta_2\theta_2+\eta_3\theta_3$, the
    configuration describes an NS5-brane in the (12345)-plane and an
    F3-brane along the (189)-plane sitting at $x^k=0$ $k=2,\dots ,7$.
    It preserves $\nu=1/8$ of the spacetime supersymmetry, with
    Killing spinors being preserved by an $\fsu(3)$ subalgebra.
    
  \item[(ii)] $\theta_2=\eta_3\theta_3+\eta_4\theta_4$.  In this case,
    the F3-brane extends along the (123)-plane sitting at $x^k=0$
    $k=4,\dots ,9$.  It preserves the same amount of supersymmetry as
    the previous one due to the existence of an $\fsu(3)$ subalgebra
    preserving some Killing spinors.
    
  \item[(iii)] If $\theta_i\neq \eta_j\theta_j + \eta_k\theta_k$ for
    distinct $i,j,k$, the configuration describes the intersection of
    three different F7-branes plus a NS5-brane.  Spacetime
    supersymmetry is completely broken.
  \end{enumerate}
  
\item[(5)] Finally, in the case of four non-vanishing $\theta$s, there
  are generically two inequivalent ways of preserving Killing spinors
  when four of our deformation parameters are non-vanishing, but due
  to the symmetries of our configuration, these split into four:
  \begin{enumerate}
  \item[(i)] If $\theta_1=\eta_2\theta_2 + \eta_3\theta_3 +
    \eta_4\theta_4$, the configuration describes an NS5-brane in the 
    (12345)-plane and a fluxstring along the $x^1$ direction,
    preserving $\nu=1/16$ of the spacetime supersymmetry.  This is the
    one associated with the $\fsu(4)$ isotropy algebra discussed
    before.
    
  \item[(ii)] If $\theta_1=\eta_2\theta_2$ and
    $\theta_3=\eta_4\theta_4$, the configuration involves an NS5-brane
    and a $\tfrac14$-BPS fluxstring in the $x^1$ direction.  It
    preserves $\nu=1/8$ and it is associated with the
    $\fsp(1)\times\fsp(1)$ isotropy algebra.
    
  \item[(iii)] If $\theta_1=\eta_3\theta_3$ and
    $\theta_2=\eta_4\theta_4$, we have the same interpretation as the
    previous one, but the $\fsp(1)\times\fsp(1)$ isotropy algebra is
    selected in a different way.
    
  \item[(iv)] If all other cases, there are four intersecting
    F7-branes and a stack of coincident NS5-branes breaking spacetime
    supersymmetry completely.
  \end{enumerate}
\end{enumerate}

If $\beta\neq 0$, the only allowed possibility preserving
supersymmetry requires $|\theta_1|=|\beta|$, which is the nullbrane
sector.  Depending on whether the remaining parameters vanish or
satisfy certain linear relations, we distinguish between the following
configurations:
\begin{enumerate}
\item[(1)] If $\theta_i=0$ for $i=2,3,4$, the configuration describes
  an NS5-brane in the (12345)-plane and a nullbrane.  This composite
  configuration preserves $\nu=1/4$ of the spacetime supersymmetry.
  
\item[(2)] If one of the $\theta$s is non-vanishing, the corresponding
  configuration breaks supersymmetry completely, due to the presence
  of an F7-brane besides the previous NS5/nullbrane pair.
  
\item[(3)] If two of the $\theta$s are non-vanishing, we shall
  distinguish between three cases:
  \begin{enumerate}
  \item[(i)] If $\theta_i\neq \eta_j\theta_j$, for $i\neq j$, the
    configuration describes a system of two intersecting F7-branes
    (besides the composite system of a NS5-brane and a nullbrane),
    such that all supersymmetry is broken.
   
  \item[(ii)] If $\theta_2=\eta_3\theta_3$, the intersection among the
    F7-branes gives rise to the so-called F5-brane, extending along
    the (12389)-plane and sitting at $x^4=x^5=x^6=x^7=0$.  The full
    configuration includes the previous pair NS5/nullbrane system,
    thus preserving $\nu=1/8$.
    
  \item[(iii)] If $\theta_3=\eta_4\theta_4$, one finds a second
    F5-brane, this time extending along the (12345)-plane and sitting
    at $x^6=x^7=x^8=x^9=0$.  As before, the configuration preserves
    $\nu=1/8$.
  \end{enumerate}
\item[(4)] If all $\theta$s are non-vanishing, we need to distinguish
  between two cases:
  \begin{enumerate}
  \item[(i)] If $\theta_2\neq \eta_3\theta_3 + \eta_4\theta_4$, the
    configuration describes the intersection of three F7-branes,
    in addition to the NS5/nullbrane system, thus breaking all 
    supersymmetry.
  \item[(ii)] If $\theta_2=\eta_3\theta_3 + \eta_4\theta_4$, the
    intersection of the F7-branes gives rise to an F3-brane extending
    along the (123)-plane and sitting at $x^k=0$ k=$4,\dots ,9$.
    Thus, there exists a composite configuration involving a
    NS5-brane, nullbrane and F3-brane preserving $\nu=1/16$.
  \end{enumerate}
\end{enumerate}

The set of supersymmetric configurations described above is summarised
in Table~\ref{tab:M5susy2}, where a similar notation to the one
introduced for the M5-brane Kaluza--Klein reductions has been used.

\begin{table}[h!]
  \begin{center}
    \setlength{\extrarowheight}{3pt}
    \begin{tabular}{|>{$}c<{$}|c|>{$}l<{$}|}
      \hline
      \nu & Object & \multicolumn{1}{c|}{Subalgebra}\\
      \hline
      \hline
              & NS5$\perp$F5(1) &  \\
      \frac14 & NS5$\perp$F5(3) & \fsu(2)\\
              & NS5$\perp$F5(5) & \\
      \frac14 & NS5 + N & \RR \\
      \frac18 & NS5$\perp$F3(1) & \fsu(3) \\
              & NS5$\perp$F3(3) & \\
      \frac18 & (NS5$\perp$F5(3)) + N & \fsu(2)\times \RR \\
              & (NS5$\perp$F5(5)) + N & \\
      \frac18 & NS5$\perp$F1(1) & \fsp(1)\times\fsp(1) \\
      \frac1{16} & NS5$\perp$F1(1) & \fsu(4)\\
      \frac1{16} & (NS5$\perp$F3(3)) + N & \fsu(3)\times \RR \\[3pt]
      \hline
    \end{tabular}
    \vspace{8pt}
    \caption{Supersymmetric configurations of NS5-branes (NS5),
      fluxbranes and nullbranes.}
    \label{tab:M5susy2}
  \end{center}
\end{table}

Before finishing the discussion of the $a=0$ sector, we would like to
compute the fluxes associated with F5-branes in the presence of
NS5-branes.  There are three cases to be considered separately, as
indicated in Table~\ref{tab:M5susy2}, and discussed in the text above.
Whenever the NS5-branes share a single direction with the F5-brane,
$\theta_3=\theta_4=\beta=0,\, \theta\equiv\theta_1=\theta_2$, the RR
1-form potential in \eqref{oth:m5b} depends on the distance to the
origin of the F5-brane plane through the harmonic function $V(r)$.  It
can be shown that the flux carried by the F5-brane equals the one on
flat spacetime, except at $r=0$, where it vanishes.  On the other
extreme, when the NS5-branes are parallel to the F5-branes
$\theta_1=\theta_2=\beta=0,\,\theta\equiv \theta_3=\theta_4$, the RR
1-form potential in \eqref{oth:m5b} is independent of the worldvolume
F5-brane point.  Therefore the flux equals the one carried in flat
spacetime everywhere.  Finally, when they are two relatively
transverse dimensions, $\theta_2=\theta_4=\beta=0,\,\theta\equiv
\theta_1=\theta_3$, the RR 1-form potential in \eqref{oth:m5b} depends
on the relative radial distance in the (89)-plane spanned by the
F5-brane.  By fixing a point on this plane, it can be shown that
\begin{equation*}
  F_2\wedge F_2 \propto -2\frac{\partial\left(r_2
      \tilde{\Lambda}^{-2}\right)} {\partial\,r_1} dr_1\wedge
  d\theta_1\wedge dr_2\wedge d\theta_2~.
\end{equation*}
Its integral over $\RR^4$ equals the one in flat spacetime everywhere
except at the origin of the (89)-plane, where it vanishes.

All previous considerations were restricted to the $a=0$ subspace.
Let us move to the subspace where $a\neq 0$.  It is useful to set all
the rotation parameters $\theta_i$ and $\beta$ to zero.  In this case,
a similar discussion to the one giving rise to a bound state of
D2-branes and delocalised fundamental strings applies here.  Indeed,
the construction is entirely analogous just differing in the starting
eleven-dimensional background, which now is that of a delocalised
M5-brane.  Thus, following the same arguments, we will interpret this
system as a bound state of NS5-branes and delocalised D4-branes, which
still preserves one half of the spacetime supersymmetries.  This
vacuum allows further supersymmetric configurations both in the
fluxbrane and nullbrane sectors.  Since the detailed discussion of all
these possibilities does not give any new insight and follows closely
previous classifications, we simply summarise the results in
Table~\ref{tab:M5susy3}.

\begin{table}[h!]
  \begin{center}
    \setlength{\extrarowheight}{3pt}
    \begin{tabular}{|>{$}c<{$}|c|>{$}l<{$}|}
      \hline
      \nu & Object & \multicolumn{1}{c|}{Subalgebra}\\
      \hline
      \hline
      \frac12 & NS5-D4 & \{0\} \\
              & (NS5-D4)$\perp$F5(1) &  \\
      \frac14 & (NS5-D4)$\perp$F5(3) & \fsu(2)\\
              & (NS5-D4)$\perp$F5(5) & \\
      \frac14 & (NS5-D4) + N & \RR \\
      \frac18 & (NS5-D4)$\perp$F3(1) & \fsu(3) \\
              & (NS5-D4)$\perp$F3(3) & \\
      \frac18 & ((NS5-D4)$\perp$F5(3)) + N & \fsu(2)\times \RR \\
              & ((NS5-D4)$\perp$F5(5)) + N & \\
      \frac18 & (NS5-D4)$\perp$F1(1) & \fsp(1)\times\fsp(1) \\
      \frac1{16} & (NS5-D4)$\perp$F1(1) & \fsu(4)\\
      \frac1{16} & ((NS5-D4)$\perp$F3(3)) + N & \fsu(3)\times \RR
      \\[3pt]
      \hline
    \end{tabular}
    \vspace{8pt}
    \caption{Supersymmetric configurations of bound states made of
    NS5-branes and delocalised D4-branes (NS5-D4), fluxbranes and
    nullbranes.}
    \label{tab:M5susy3}
  \end{center}
\end{table}

Notice that the previous discussion does not cover the particular case
$b=0$ and $a\neq 0$ in \ref{tab:M5dsusy}. That would give rise to delocalised
D4-branes in the presence of fluxbranes, whenever $\theta_i\neq 0$.
Since we are not particularly interested in the study of delocalised branes
in the presence of fluxbranes, we shall not present the details for these
configurations.

Let us move to the region of the moduli space where the extra
translation is along a null direction.  That is, the starting Killing
vector is decomposed as $\xi= \partial_z + \alpha$, where
\begin{equation*}
  \alpha = \partial_+ + \theta_1 R_{23} + \theta_2 R_{45}
  + \theta_3 R_{67} + \theta_4 R_{89}~.
\end{equation*}
The constant matrix $B$ is a $9\times 9$ one, leaving the second null
direction $x^-$ invariant.  In the basis $\{x^+,x^2,\dots ,x^9\}$, it
can be written as
\begin{equation}
  \label{eq:Bmatrixm5b}
  B= 
  \begin{pmatrix}
    0 & 0 &  0 & 0 & 0 & 0 & 0 & 0 & 0  \\
    0 & 0 & -\theta_1 & 0 & 0 & 0 & 0 & 0 & 0 \\
    0 & \theta_1 & 0 & 0 & 0 & 0 & 0 & 0 & 0 \\
    0 & 0 & 0 & 0 & -\theta_2 & 0 & 0 & 0 & 0 \\
    0 & 0 & 0 & \theta_2 & 0 & 0 & 0 & 0 & 0 \\
    0 & 0 & 0 & 0 & 0 & 0 & -\theta_3 & 0 & 0 \\
    0 & 0 & 0 & 0 & 0 & \theta_3 & 0 & 0 & 0 \\
    0 & 0 & 0 & 0 & 0 & 0 & 0 & 0 & -\theta_4 \\
    0 & 0 & 0 & 0 & 0 & 0 & 0 & \theta_4 & 0 \\
  \end{pmatrix}~.
\end{equation}
Since there is an extra null translation (its parameter can always be
set to one), there is a non-vanishing 9-vector $\bC$ taking care of
the inhomogeneous part of the isometry transformation.  In the same
basis as the one used in the matrix \eqref{eq:Bmatrixm5b}, this vector
is
\begin{equation*}
  (\bC)^t = (1,\vec{0})~.
\end{equation*}

The corresponding type IIA configurations have a ten-dimensional
metric given by
\begin{multline*}
  g = \tilde{\Lambda}^{1/2}\left\{ds^2(\EE^{1,5}) 
    + Vds^2(\EE^4)\right\}\\
  - \tilde{\Lambda}^{-1/2}V \left\{\theta_3 \omega^{67} + \theta_4
    \omega^{89} + V^{-1}\left[dx^- + \theta_1 \omega^{23} + \theta_2
      \omega^{45}\right]\right\}^2~,
\end{multline*}
whereas the RR 1-form $A_{1}$, NS-NS 7-form field strength $H_{7}$,
RR 6-form $H_6$ and dilaton $\Phi$ are listed below
\begin{footnotesize}
\begin{equation*}
  \begin{aligned}[m]
    A_1 &= \tilde{\Lambda}^{-1}\left\{V^{-1}\left[dx^- + 
    \theta_1 \omega^{23} + \theta_2 \omega^{45}\right]
    + \theta_3\omega^{67} + \theta_4 \omega^{89} \right\} \\
    H_7 &= \dvol\left(\EE^{1,5}\right)\wedge dV^{-1} \\
    H_6 &= dx^-\wedge \dvol\left(\EE^4\right)\wedge dV^{-1} \\
    & - \theta_1 dx^+\wedge dx^-\wedge (x^2 dx^2+ x^3 dx^3)\wedge
    dx^4\wedge dx^5 \wedge dV^{-1} \\
    & - \theta_2 dx^+\wedge dx^-\wedge dx^2 \wedge dx^3
    \wedge (x^4dx^4 + x^5 dx^5)\wedge dV^{-1} \\ 
    \Phi &= \tfrac34\log \left(\tilde{\Lambda}\cdot V^{2/3}\right)~.
  \end{aligned}
\end{equation*}
\end{footnotesize}
The configuration depends on an scalar function $\tilde{\Lambda}$
which is defined in terms of the scalar function $\Lambda$ appearing
in the general discussion section, by
\begin{equation*}
  \Lambda = V^{2/3}\cdot \tilde{\Lambda}~,
\end{equation*}
and equals
\begin{footnotesize}
  \begin{multline*}
      \tilde{\Lambda} = 1 + V^{-1}\left\{(\theta_1)^2\left[(x^2)^2 
      + (x^3)^2\right] + \left(\theta_2\right)^2
      \left[(x^4)^2 + (x^5)^2\right]\right\} \\
      + \left(\theta_3\right)^2\left[(x^6)^2 +(x^7)^2\right]
      +\left(\theta_4\right)^2\left[(x^8)^2 +(x^9)^2\right]~.
  \end{multline*}
\end{footnotesize}

Whenever all $\theta$s vanish, and following a similar discussion to
the one presented when dealing with a delocalised M2-brane, one should
expect to get a configuration which interpolates among a wave
background at asymptotic infinity and a linear dilaton background
\cite{ABKS}, which is the corresponding geometry close to a stack of
NS5-branes.  This is indeed straightforward to check.  The stability
and supersymmetry of the configuration require both $H_6$ and $F_2
=dA_1$ to be null forms, but its role is not clear to us.  Switching
on the $\theta_i$ parameters, one is adding fluxbranes to the previous
configuration.

Let us finally move to the region of the moduli space where the extra
translation is along a time direction.  That is, the starting Killing
vector is decomposed as $\xi= \partial_z + \alpha$, where
\begin{equation*}
  \alpha = a\partial_0 + \theta_1 R_{23} + \theta_2 R_{45}
  + \theta_3 R_{67} + \theta_4 R_{89}~.
\end{equation*}
The constant matrix $B$ is now $9\times 9$, leaving the spacelike
direction $x^1$ invariant.  Formally, it is given by the matrix
\eqref{eq:Bmatrixm5b}, but this time written in the basis
$\{x^0,x^2,\dots,x^9\}$.  The inhomogeneous part of the isometry
transformation defines a non-trivial 9-vector $\bC$
\begin{equation*}
  (\bC)^t = (a,\vec{0})~.
\end{equation*}

The corresponding type IIA configurations have a ten-dimensional
metric given by,
\begin{multline*}
  g = \tilde{\Lambda}^{1/2}\left\{ds^2(\EE^{1,5}) 
    + Vds^2(\EE^4)\right\}\\
  - \tilde{\Lambda}^{-1/2}V \left\{\theta_3 \omega^{67} +
    \theta_4 \omega^{89} + V^{-1}\left[-adx^0 + \theta_1
      \omega^{23} + \theta_2 \omega^{45} \right]\right\}^2~,
\end{multline*}
where we remind the reader that $\omega^{ij} := x^i dx^j - x^j dx^i$.
In addition the RR 1-form $A_{1}$, NS-NS 7-form field strength
$H_{7}$, RR 6-form $H_6$ and dilaton $\Phi$ are listed below
\begin{footnotesize}
\begin{equation*}
  \begin{aligned}[m]
    A_1 &= \tilde{\Lambda}^{-1}\left\{V^{-1}\left[-adx^0 + 
    \theta_1 \omega^{23} + \theta_2 \omega^{45}\right]
    + \theta_3 \omega^{67} + \theta_4 \omega^{89} \right\} \\
    H_7 &= \dvol\left(\EE^{1,5}\right)\wedge dV^{-1} \\
    H_6 &= adx^0\wedge \dvol\left(\EE^5\right)\wedge dV^{-1} \\
    & - \theta_1 dx^0\wedge dx^1\wedge (x^2 dx^2+ x^3 dx^3)\wedge
    dx^4\wedge dx^5 \wedge dV^{-1} \\
    & - \theta_2 dx^0\wedge dx^1\wedge dx^2 \wedge dx^3
    \wedge (x^4dx^4 + x^5 dx^5)\wedge dV^{-1} \\ 
    \Phi &= \tfrac34 \log \left(\tilde{\Lambda}\cdot V^{2/3}\right)~.
  \end{aligned}
\end{equation*}
\end{footnotesize}

The configuration depends on an scalar function $\tilde{\Lambda}$
which is defined in terms of the scalar function $\Lambda$ appearing
in the general discussion section, by
\begin{equation*}
\Lambda = V^{2/3}\cdot \tilde{\Lambda}~,
\end{equation*}
and equals
\begin{footnotesize}
  \begin{multline*}
    \tilde{\Lambda} = 1 + V^{-1}\left\{-a^2 +
      (\theta_1)^2\left[(x^2)^2 + (x^3)^2\right] +
      \left(\theta_2\right)^2\left[(x^4)^2 + (x^5)^2\right]\right\} \\
    + \left(\theta_3\right)^2\left[(x^6)^2 +(x^7)^2\right]
    +\left(\theta_4\right)^2\left[(x^8)^2 +(x^9)^2\right]~.
  \end{multline*}
\end{footnotesize}

We shall follow the same strategy as for the similar construction
regarding the M2-brane.  If we set $\theta_i=0$ $\forall$ i, we are
left with a ten-dimensional configuration whose geometry is given by
\begin{multline*}
  g = -\left(V-a^2\right)^{-1/2}V^{1/2}(dx^0)^2\\
  + \left(V-a^2\right)^{1/2}\left\{V^{-1/2}ds^2(\EE^5) + V^{1/2}
    ds^2(\EE^4)\right\}~.
\end{multline*}
As before, the condition $|a|<1$ ensures the absence of horizons
in spacetime.  If we would have allowed $|a|\geq 1$, there would have
been horizons at
\begin{equation*}
  r_{\text{H}}^2 = \frac{Q}{a^2-1}~,
\end{equation*}
dividing spacetime into regions ($r> r_{\text{H}}$) having closed
timelike curves and regions ($r< r_{\text{H}}$) free of these causal
singularities.

We shall not add any further comments regarding the physical
interpretation of these configurations, besides the possibility of
looking at them as bound states of NS5-branes and delocalised
E4-branes \cite{Chris, HullTimeLike, HullKhuriTimes}.  They are again
interpolating among flat spacetime and the linear dilaton background
\cite{ABKS}.  It is straightforward to add fluxbranes by switching on
the $\theta_i$ parameters, while still preserving some supersymmetry.

Looking at table \ref{tab:M5dsusy}, we learn that the previous Kaluza--Klein
reduction does not cover the case $b=0$ and $a\neq 0$, which is certainly
allowed if both $\theta_3$ and $\theta_4$ are non-vanishing. We include
the corresponding ten dimensional configuration below for completeness, even 
though its physical interpretation is unclear to us and the final background
is delocalised in the $z$ direction. The type IIA metric is given by
\begin{multline*}
  g = V^{-1/2}\tilde{\Lambda}^{1/2}ds^2(\EE^{5})
  + V^{1/2}\tilde{\Lambda}^{1/2}\left[(dz)^2 + ds^2(\EE^4)\right] \\
  - V^{-1/2}\tilde{\Lambda}^{-1/2}\left[\theta_1\omega^{23} 
  + \theta_2\omega^{45}+ V\left(\theta_3\omega^{67} 
  + \theta_4\omega^{89}\right)\right]^2~,
\end{multline*}
where we remind the reader that $\omega^{ij} := x^i dx^j - x^j dx^i$.
In addition the RR 1-form $A_{1}$, RR 6-form $H_6$ and dilaton $\Phi$ are 
non-trivial and listed below
\begin{footnotesize}
\begin{equation*}
  \begin{aligned}[m]
    A_1 &= \tilde{\Lambda}^{-1}\left\{\theta_1 \omega^{23} + 
    \theta_2 \omega^{45} + V\left(\theta_3 \omega^{67} + 
    \theta_4 \omega^{89}\right)\right\} \\
    H_6 &= -a\dvol\left(\EE^5\right)\wedge dV^{-1} \\ 
    \Phi &= \tfrac34 \log \left(\tilde{\Lambda}\cdot V^{-1/3}\right)~.
  \end{aligned}
\end{equation*}
\end{footnotesize}

The configuration depends on an scalar function $\tilde{\Lambda}$
which is defined in terms of the scalar function $\Lambda$ appearing
in the general discussion section, by
\begin{equation*}
\Lambda = V^{-1/3}\cdot \tilde{\Lambda}~,
\end{equation*}
and equals
\begin{footnotesize}
  \begin{multline*}
    \tilde{\Lambda} = -a^2 +
      (\theta_1)^2\left[(x^2)^2 + (x^3)^2\right] +
      \left(\theta_2\right)^2\left[(x^4)^2 + (x^5)^2\right] \\
    + V\left\{\left(\theta_3\right)^2\left[(x^6)^2 +(x^7)^2\right]
    +\left(\theta_4\right)^2\left[(x^8)^2 +(x^9)^2\right]\right\}~.
  \end{multline*}
\end{footnotesize}

\section{IIA/IIB discrete quotients and duality}
\label{sec:dual}

All the ideas and formalism developed so far apply equally well to any
background in type IIA/IIB supergravity having some isometry group
$G$.  In particular, D-brane backgrounds \cite{HorowitzStrominger}
allow an analogous description of the form \eqref{eq:elemint} and
adding the corresponding non-trivial profile for the dilaton.  Their
Killing spinors satisfy the same properties as the ones discussed in
Section~\ref{sec:branes}.  Thus we can again conclude that the action
of the symmetry group on the Killing spinors is induced by the action
of $\Spin(1,p)\times\Spin(9-p)$ on the asymptotic spinors.  Once more,
translations will act trivially on spinors, whereas Lorentz
transformations will impose certain constraints analogous to the
vanishing of \eqref{eq:susycons}.  It should conceptually be clear,
that we could classify the (not necessarily) supersymmetric
freely-acting spacelike isometries as we did for the M2-brane and
M5-brane configurations.  By constructing the discrete quotients
$\eM_{\text{Dp}}/\Gamma_0$, $\Gamma_0$ being some discrete subgroup of
the corresponding one-parameter subgroup $\Gamma$, such that
$\Gamma/\Gamma_0$ is compact, we would reach new supersymmetric smooth
type IIA/IIB configurations.

It is nevertheless well-known that the configurations described in the
previous sections are dual to the ones outlined above.  Consider an
M-theory background reduced along the orbits of $\xi_1=\partial_z +
\lambda$, with a further compact spacelike direction $x$, such that
translations along it, infinitesimally generated by
$\xi_2=\partial_x$, commute with the action generated by $\xi_1$.  The
claim is that the type IIA configuration obtained through
Kaluza--Klein reduction along the orbits of $\xi_1$ is equivalent to
the one obtained through Kaluza--Klein reduction along the orbits of
$\xi_2$, applying a T-duality transformation \cite{BHO} along the
orbits of $\xi_1$, plus an S-duality transformation in type IIB, and
finally, after a relabelling of coordinates, applying a T-duality back
to type IIA.  The mechanism just described is an obvious extension of
the well-known M-theory flip, when one of the orbits is
\emph{twisted}.

After all these preliminary remarks, it should be clear that most of
the results derived previously extend straightforwardly for
D$p$-branes, among other configurations in type IIA/IIB.  In
particular, it should be clear that we can construct time-dependent
backgrounds starting from D$p$-branes $(p\geq 3)$ and constructing the
quotient manifold associated with the discrete identification giving
rise to the ten-dimensional nullbrane.  Similar comments would apply
for the fluxbrane sector.  As a particular example, let us consider
D3-branes.  We shall concentrate on the non-trivial identifications
preserving $\nu=1/4$ of the spacetime supersymmetry generated by
$\xi=\partial_z + \lambda$, where $\partial_z$ generates translations
along the brane and $\lambda$ is either a rotation $\rho$ belonging to
an $\fsu(2)$ subalgebra (flux 5-brane construction) or a null rotation
$\nu$ belonging to an $\RR$ subalgebra (nullbrane construction).  Both
quotients survive the near horizon limit, even though they break the
superconformal symmetries, so that there is no supersymmetry
enhancement in this case.  It is natural to ask about the
corresponding gauge theory dual for type IIB in these configurations,
and it is natural to guess that it will be given in terms of the
corresponding orbifold constructions in $\mathcal{N}{=}4$
supersymmetric Yang--Mills.

Let us denote the matter fields transforming in the adjoint
representation of $\SU(N)$ by $\phi^i$ $i=1,\dots ,6$, whereas the
coordinates of the four-dimensional manifold where the field theory is
defined will be denoted by $(x^\mu,z)$ $\mu=0,1,2$ and $z$ standing
for the compact one.  Due to the isometries of the background, there
are two inequivalent $\fsu(2)$ constructions.  Indeed, one may
consider a rotation $\rho_\perp$ acting on the transverse directions
to the brane, or a rotation which acts both on $Z_1=\phi^1 + i\phi^2$
and $\omega=x^1+ix^2$.  In the first case, the compatibility of the
gauge structure of the theory with the orbifold requires the matrices
$Z_1$ and $Z_2=\phi^3 + i\phi^4$ to satisfy
\begin{equation*}
  \begin{aligned}
    Z_1(x^\mu,z+R) &= e^{i\theta}\Omega (x^\mu,z) Z_1(x^\mu,z)
    (\Omega (x^\mu,z))^{-1} \\ 
    Z_2(x^\mu,z+R) &= e^{-i\theta}\Omega (x^\mu,z) Z_2(x^\mu,z)
    (\Omega (x^\mu,z))^{-1}~,
  \end{aligned}
\end{equation*}
where $\Omega (x^\mu,z)$ is an $\SU(N)$ group element describing
a gauge transformation, whereas the remaining two adjoint matrices
satisfy the standard ones
\begin{equation*}
  \phi^i(x^\mu,z+R) = \Omega (x^\mu,z) \phi^i(x^\mu,z)
  (\Omega (x^\mu,z))^{-1} \quad i=5,6~.
\end{equation*}

In the second case, since the group also acts non-trivially on the
$\omega$ plane where the field theory is defined, one has four adjoint
matrices satisfying
\begin{equation*}
  \phi^i(x^0,e^{i\theta}\omega,z+R) = \Omega (x^\mu,z) \phi^i(x^\mu,z)
  (\Omega (x^\mu,z))^{-1} \quad i=3,4,5,6~,
\end{equation*}
whereas the $Z_1$ is twisted by a constant phase
\begin{equation*}
  Z_1(x^0,e^{i\theta}\omega,z+R) = e^{-i\theta}\Omega (x^\mu,z) 
  Z_1(x^\mu,z)(\Omega (x^\mu,z))^{-1}~.
\end{equation*}

Finally, in the null rotation identification, all adjoint matrices
$\phi^i$ satisfy that for all $i$,
\begin{equation*}
  \phi^i(\tilde{x}^\mu,z+R) = \Omega (x^\mu,z) \phi^i(x^\mu,z)
  (\Omega (x^\mu,z))^{-1}~,
\end{equation*}
where $\tilde{x}^\mu$ stands for the image of $x^\mu$ under a (not
infinitesimal) null rotation. 

\section*{Acknowledgments}

We started this work while participating in the programme
\emph{Mathematical Aspects of String Theory} at the Erwin Schrödinger
Institute in Vienna, and it is again a pleasure to thank them for
support and for providing such a stimulating environment in which to
do research.  JMF's participation in this programme was made possible
in part by a travel grant from PPARC.  The work was finished while JMF
was visiting the IHÉS, whom he would like to thank for support.
JS would like to thank the School of Mathematics of the University
of Edinburgh for hospitality during the final stages of this work.
JMF is a member of EDGE, Research Training Network HPRN-CT-2000-00101,
supported by The European Human Potential Programme.  The research of
JMF is partially supported by the EPSRC grant GR/R62694/01.  JS is
supported by a Marie Curie Fellowship of the European Community
programme ``Improving the Human Research Potential and the
Socio-Economic Knowledge Base'' under the contract number
HPMF-CT-2000-00480, and in part by a grant from the United
States--Israel Binational Science Foundation (BSF), the European
Research Training Network HPRN-CT-2000-00122 and by Minerva.

\appendix

\section{Group theory and spinors}
\label{sec:groups}

In this appendix we collect some facts about how the spinor
representation of $\Spin(1,10)$ decomposes under certain subgroups.
These results are useful in determining the supersymmetric
Kaluza--Klein reductions of the M2 and M5-brane solutions.

Let us start by recalling a few facts about the irreducible
representations of $\Spin(1,10)$ and of the Clifford algebra
$\Cl(1,10)$.  The Clifford algebra $\Cl(1,10)$ is isomorphic (as a
real associative algebra) to $\Mat(32,\RR) \oplus \Mat(32,\RR)$, where
$\Mat(n,\RR)$ is the algebra of $n\times n$ real matrices.  This means
that there are two inequivalent irreducible representations: real and
of dimension $32$.  They are distinguished by the action of the centre
which is generated by the volume form
\begin{equation*}
  \dvol(\EE^{1,10}) := \Gamma_{01\cdots\natural}~,
\end{equation*}
which squares to the identity.

The condition \eqref{eq:PM2} translates into an eight-dimensional
chirality condition on the spinor.  Indeed, decomposing
$\dvol(\EE^{1,10})$ into a product
\begin{equation*}
  \dvol(\EE^{1,10}) =   \dvol(\EE^{1,2})\dvol(\EE^8) = \Gamma_{012}
  \Gamma_{34\cdots\natural}
\end{equation*}
of two commuting operators, we obtain
\begin{equation*}
  \dvol(\EE^{1,2}) \varepsilon = \dvol(\EE^8)
  \dvol(\EE^{1,10})\varepsilon = \pm \dvol(\EE^8) \varepsilon~,
\end{equation*}
where the sign depends on the action of $\dvol(\EE^{1,10})$,
equivalently on the choice of irreducible representation of
$\Cl(1,10)$.  Let us assume that a choice has been made once and for
all and let $S_{11}$ denote the corresponding irreducible
representation.  This is an irreducible representation of
$\Spin(1,10)$.  Under the natural $\Spin(10)$ subgroup, $S_{11}$
remains irreducible as a real representation, even though its
complexification is reducible.  This is because $\dvol(\EE^{10})$ is a
complex structure and to diagonalise it requires complexifying the
spinors.  Indeed, we have
\begin{equation}
  \label{eq:complex}
  S_{11} \otimes \CC = S_{10} \oplus \Bar S_{10}~,
\end{equation}
where $S_{10}$ consists of those complex spinors $\varepsilon$ such
that
\begin{equation*}
  \dvol(\EE^{10}) \cdot \varepsilon = i \varepsilon~,
\end{equation*}
and $\Bar S_{10}$ is the complex conjugate.  We will abbreviate
equation \eqref{eq:complex} with the notation
\begin{equation*}
  S_{11} = [\![S_{10}]\!]~.
\end{equation*}
In other words, the double brackets indicate the underlying real
representation of the representation obtained by adding to a complex
representation its complex conjugate, which has a natural real
structure.  Notice that $\dim_\RR [\![S_{10}]\!] = 2 \dim_\CC S_{10}$
(which in this case is $32$), as the notation tries to suggest.  We
are interested in how $S_{11}$ breaks under the natural $\Spin(2)
\times \Spin(8)$ subgroup of $\Spin(10)$.  Since the volume element of
$\EE^2$ is a complex structure, whereas that of $\EE^8$ squares to the
identity, we see that
\begin{equation*}
  S_{10} = (S_2 \otimes S_8^+) \oplus (\bar S_2 \otimes S_8^-)~,
\end{equation*}
where $S_8^\pm$ are the half-spin representations of $\Spin(8)$, and
$S_2$ is the one-dimensional complex irreducible representation of
$\Spin(2)$ with weight $1$; that is, $S_2$ is the ``half-spin''
representation of $\Spin(2)$.  Since the representations $S_8^\pm$ are
real,
\begin{equation*}
  S_{11} = [\![S_2]\!] \otimes (S_8^+ \oplus S_8^-)~,
\end{equation*}
whence the subspace of $S_{11}$ consisting of spinors which obey
\eqref{eq:PM2} transforms under $\Spin(2) \times \Spin(8)$ as
$[\![S_2]\!] \otimes S_8^\pm$, for some choice of sign.

The problem of determining which Kaluza--Klein reductions of the
(delocalised) M2 brane preserve some supersymmetry comes down to
determining which elements in (the Lie algebra of a maximal torus of)
$\Spin(2) \times \Spin(8)$ (or $\Spin(2) \times \Spin(7)$ for the
delocalised M2 brane) preserve a spinor in $[\![S_2]\!] \otimes
S_8^\pm$.  As explained in Section~\ref{sec:method}, one way to do
this is to simply determine the weight decomposition of
$[\![S_2]\!] \otimes S_8^\pm$ under the maximal torus.  We will
work infinitesimally, hence we will decompose $[\![S_2]\!] \otimes
S_8^\pm$ under a Cartan subalgebra of $\fso(2) \times \fso(8)$.

Up to conjugation, a typical element in $\fso(2) \times \fso(8)$ can
be written as
\begin{equation*}
  \theta_1 R_{12} +  \theta_2 R_{34} + \theta_3 R_{56} +
  \theta_4 R_{78} +  \theta_5 R_{9\natural}
\end{equation*}
where the infinitesimal rotations $R_{ij}$ generate a Cartan
subalgebra in $\fso(2) \times \fso(8)$, with $R_{12}$ spanning
$\fso(2)$ and the rest spanning a Cartan subalgebra of $\fso(8)$.  In
the case of the delocalised M2-brane, we must restrict to a $\fso(7)$
subalgebra, which means setting $\theta_5 = 0$, say.  Relative to a
basis for the root space canonically dual to the $R_{ij}$, the weights
of the representation $[\![S_2]\!] \otimes S_8^\pm$ are easy to
work out.  First of all $[\![S_2]\!]$ has weights $\pm 1$, whereas
$S_8^\pm$ has weights $(\pm1,\pm1,\pm1,\pm1)$ where the signs are
uncorrelated, but where their product is $\pm1$ for $S_8^\pm$
respectively.  Putting these two results together we find that the
weights of $[\![S_2]\!] \otimes S_8^\pm$ are $(\pm1, \pm 1,
\pm1, \pm1,\pm1)$ where the signs are uncorrelated but where the
product of all but the first sign is $\pm 1$ for $[\![S_2]\!]
\otimes S_8^\pm$, respectively.  In other words, we have
\begin{multline}
  \label{eq:weightsM2}
  \text{weights}\left([\![S_2]\!] \otimes S_8^\pm\right) = \\
  \left\{ (\mu_1, \mu_2, \mu_3, \mu_4, \mu_5) \biggl| \mu_i^2 =
    1~\text{and}~\prod_{i=2}^5 \mu_i = \pm 1 \right\}~.
\end{multline}

For the (delocalised) M5-brane we have to break $S_{11}$ into
irreducible representations of $\Spin(1,5) \times \Spin(5)$ or
$\Spin(1,5) \times \Spin(4)$.  The situation is analogous.  We can
decompose the volume element $\dvol(\EE^{1,10})$ into a product
\begin{equation*}
  \dvol(\EE^{1,10}) = \dvol(\EE^{1,5})\dvol(\EE^5) =
  \Gamma_{01\cdots5} \Gamma_{67\cdots\natural}
\end{equation*}
of commuting operators which both square to the identity.  Again
a choice for the value of $\dvol(\EE^{1,10})$ on the
representation $S$ translates the condition \eqref{eq:PM5} into a
five-dimensional chirality condition on the spinor:
\begin{equation*}
  \dvol(\EE^{1,5}) \varepsilon = \dvol(\EE^5)
  \dvol(\EE^{1,10})\varepsilon = \pm \dvol(\EE^5) \varepsilon~,
\end{equation*}
where the sign depends on the choice of irreducible representation
$S_{11}$.  The low-dimensional isomorphisms
\begin{equation*}
  \Spin(1,5) \cong \SL(2,\HH) \qquad\text{and}\qquad
  \Spin(5) \cong \Sp(2)
\end{equation*}
tell us that the irreducible representations of these groups are
two-dimensional quaternionic; equivalently, four-dimensional complex
with a quaternionic structure.  Let $S_6^\pm$ denote the positive and
negative chirality half-spin representations of $\Spin(1,5)$ and let
$S_5$ denote the half-spin representation of $\Spin(5)$.  The
complexified spinors break up as
\begin{equation*}
  S_{11} \otimes \CC = (S_6^+ \oplus S_6^- )\otimes S_5~.
\end{equation*}
The right-hand side also has a natural real structure, since the
tensor product of two quaternionic representations is real.  We can
summarise this relationship by
\begin{equation*}
  S_{11} = [(S_6^+ \oplus S_6^-) \otimes S_5]~,
\end{equation*}
where the single brackets denote the underlying real representation of
a complex representation admitting a real structure.  (Notice that
$\dim_\RR [V] = \dim_\CC V$, which agrees with $\dim_\RR S_{11} = 32$
as it should.)  The subspace of spinors satisfying \eqref{eq:PM5}
transforms as $[S_6^\pm \otimes S_5]$, where the sign depends on the
choice of $S_{11}$.  In order to determine which Kaluza--Klein
reductions of the (delocalised) M5 brane preserve some supersymmetry,
we need to determine the weights decomposition of $[S_6^\pm \otimes
S_5]$ under a Cartan subalgebra of $\Spin(5) \times \Spin(5)$.  Both
half-spin representations $S_6^\pm$ of $\Spin(1,5)$ are isomorphic to
$S_5$ as representations of $\Spin(5)$.  Therefore we are interested
in the first instance in the weight decomposition of $S_5$ under a
Cartan subalgebra of $\Spin(5)$ and then in that of $[S_5\otimes S_5]$
under a Cartan subalgebra of $\Spin(5) \times \Spin(5)$.  A typical
element of the Cartan subalgebra of $\Spin(5)$ can be written as
\begin{equation*}
  \theta_1 R_{12} +  \theta_2 R_{34}
\end{equation*}
in the same notation as that used above.  Relative to a basis for the
weight system which is canonically dual to the $R_{ij}$, the weights
of the representation $S_5$ are given by $\{(\pm 1, \pm 1)\}$ with
uncorrelated signs for a total of four weights.  Similarly, the weight
decomposition of $[S_5\otimes S_5]$ is given by
\begin{equation}
  \label{eq:weightsM5}
  \text{weights}\left([S_5 \otimes S_5]\right) =  \left\{ (\mu_1,
    \mu_2, \mu_3, \mu_4) \biggl| \mu_i^2 = 1 \right\}~.
\end{equation}

\bibliographystyle{utphys}
\bibliography{AdS,ESYM,Sugra,Geometry,CaliGeo}

\end{document}